\documentclass[acmsmall]{acmart}

\AtBeginDocument{%
  \providecommand\BibTeX{{%
      \normalfont B\kern-0.5em{\scshape i\kern-0.25em b}\kern-0.8em\TeX}}}

\setcopyright{acmcopyright}
\acmJournal{TOMM}
\acmYear{2022} 
\acmVolume{X} 
\acmNumber{Y} 
\acmArticle{Z} 
\acmMonth{0} 
\acmPrice{15.00}\acmDOI{10.1145/3507917}

\usepackage{subfigure}
\usepackage{gensymb}
\usepackage{balance}

\usepackage[ruled,noend,linesnumbered]{algorithm2e}

\newcommand{\commentout}[1]{}

\usepackage[normalem]{ulem}
\newcommand{\revrev}[2]{#2}

\newcommand{\revtwo}[2]{#2}

\newcommand{\revthree}[2]{#2}

\newcommand{\revfour}[2]{#2}

\newcommand{\revfive}[2]{#2}
\newcommand{\revNot}[2]{#1}
\newcommand{\cutICPE}[1]{#1}
\newcommand{\revsix}[2]{#2}

\newcommand{\revAA}[2]{#2}

\newcommand{\revBB}[2]{#2}

\usepackage{fancyhdr}
\usepackage{extramarks}

\begin{document}

\title[Cross-user Similarities in Viewing Behavior for 360$\degree$ Video and Caching Implications]{\revsix{Had You Looked Where I'm Looking? Cross-user}{Cross-user} Similarities in Viewing Behavior for 360$\degree$ Video and Caching Implications}
\thanks{A preliminary version of this work appeared in ACM/SPEC ICPE 2020~\cite{CaEa20a}.}

\author{Niklas Carlsson}
\affiliation{\institution{Link\"oping University}
  \country{Sweden}}
\author{Derek Eager}
\affiliation{\institution{University of Saskatchewan}
  \country{Canada}}

\begin{abstract}

The demand and usage of 360$\degree$ video services are expected to increase.  However, despite these services being
highly bandwidth intensive, not much is known about the potential value
\revrev{of}{that}
basic bandwidth saving techniques such as
server or
\revrev{network-side}{edge-network on-demand}
caching (e.g., in a CDN) could have when used for delivery of such
\revrev{technologies.}{services.}
This problem
is both important and complicated as client-side solutions
\revrev{are}{have been}
developed that split the full 360$\degree$ view into
multiple tiles, and adapt the quality of the downloaded tiles based on the
\revrev{client's}{user's}
expected viewing direction and
bandwidth conditions.
\revsix{To better understand the potential bandwidth savings that caching-based techniques may offer
for this context, this paper presents the first characterization of the similarities in the viewing directions of
\revrev{clients}{users}
watching the same 360$\degree$ video, the
overlap in viewports of these
\revrev{clients}{users}
(the area of the full 360$\degree$ view they actually see),
and the potential cache hit rates for different video
\revsix{\revfive{categories, network conditions, and
  accuracy levels in the prediction of future viewing direction when prefetching.}{categories and network conditions.}}{categories,
  network conditions, and the prediction accuracy of the future viewing direction when prefetching.}}{This paper presents
  new trace-based analysis methods that incorporate users' viewports
  (the area of the full 360$\degree$ view the user actually sees),
  a first characterization of the cross-user similarities of the users' viewports,
  and a trace-based analysis of the potential bandwidth savings that caching-based techniques may offer under different conditions.
  Our analysis takes into account differences in the time granularity over which viewport overlaps can be beneficial
  for resource saving techniques, compares and contrasts differences between video categories,
  and accounts for uncertainties in the network conditions and the prediction of the future viewing direction when prefetching.}
The results provide substantial insight into the conditions under which overlap can be considerable and caching effective, 
\revBB{and can}{and} 
inform the design of new caching system policies tailored for 360$\degree$ video.
  
\end{abstract}



\maketitle

\fancypagestyle{firststyle}
               {
                 \fancyhf{}
                 \fancyfoot[CF]{\tiny
                   \copyright ACM (2021). This is the author's version of the work (as accepted). It is posted here by permission of ACM for your personal use. Not for redistribution. 
                   \\The definitive version will be published in 
                   {\em  ACM Transactions on Multimedia Computing, Communications, and Applications (TOMM)},
                   accepted Dec. 2021,
                   \url{https://doi.org/10.1145/3507917}.}
                   }
               \thispagestyle{firststyle}

\section{Introduction}

\revthree{Interactive streaming and}{Interactive streaming~\cite{vengat181,SoJR18,MaMF18,ToFr17,GRE+16,CEKP17,AAK+18} such as}
360$\degree$ video put the users in control of their viewing
\revrev{experience}{direction}
and have the opportunity to revolutionize
what users expect from their viewing experiences.  Already today, popular services such as Facebook and YouTube offer large catalogues of 360$\degree$ content. With rapidly increasing 360$\degree$ content catalogues and the introduction of 
\revBB{relatively inexpensive user interfaces (ranging from smartphone-based solutions to dedicated head mounted displays),}{inexpensive, easy-to-use interfaces,} 
the demand for
\revrev{these services are}{360$\degree$ streaming services can only be}
expected to increase.

With 360$\degree$ streaming services being highly bandwidth intensive, identifying and understanding bandwidth
saving opportunities in the wide-area delivery of 360$\degree$ video is therefore becoming an increasingly
important problem.  Perhaps the most popular bandwidth saving opportunity studied in the research literature
is based on the observation that, with 360$\degree$ video, only a limited fraction of the full view (called the {\em viewport})
is displayed at each point in time.
Motivated by this observation,
\revtwo{}{to reduce the bandwidth usage and to improve the expected playback quality given a fixed bandwidth,}
different streaming delivery
techniques have been studied that allow alternative playback qualities to be delivered for each candidate
viewing direction~\cite{vengat181,vengat179,SoJR18,SoJR18b,CSDC17,LFL+17}.

\revrev{One approach to do so is with the use of tiles.  In this case, the full 360$\degree$ view is split into
multiple tiles, each of which is encoded at different qualities, allowing the client to adapt the quality
downloaded of each tile independently.  An alternative approach is to create multiple 360$\degree$ versions
that differ both by the direction with the highest encoded quality and the overall encoding quality across
all directions, and then have the client request the versions that they are most likely to benefit the
most from.  Furthermore, regardless of encoding technique, the video is typically split into chunks
(e.g., 2-5 second in duration), allowing the clients to adapt their playback quality based on current
network conditions, for example, and typically try to build up a buffer to protect against stalls caused by
future bandwidth variations, for example.  Therefore, download decisions of each chunk
(consisting of a set of tiles) must therefore be done well
ahead of the time a chunk is being played, resulting in a prefetch-aggressiveness problem~cite{AAK+18}that must balance}{With
  video delivery systems using HTTP-based Adaptive Streaming (HAS), a video is split into chunks (e.g., 2-5 seconds in duration) that
  \revthree{can each be}{are each}
  encoded at multiple quality levels, allowing clients to adapt their playback quality based on current network
  conditions, for example,
  \revthree{and typically}{and}
  to build up a buffer to protect against stalls that may be caused by future bandwidth
  variations.  With 360$\degree$ video, each chunk can further be split into multiple tiles, each corresponding to a portion of the
  360$\degree$ view.  This division into tiles complicates prefetching, since now, when prefetching data from a future chunk,
  the client player needs to determine which tiles from the chunk to prefetch and a quality level for each. The prefetching
  policy must address a prefetch-aggressiveness tradeoff~\cite{AAK+18} and balance}
the use of a larger buffer (to protect against stalls) against making prefetching
decisions closer to the time of playback (improving predictions of future viewing directions).
To address this problem various head-movement prediction techniques have been proposed and evaluated~\cite{XiZG18,vengat178,vengat179}.
However, prior work
\revrev{have ignored the implications that the use of tiled 360$\degree$ video combined with such adaptive techniques
  may have on caching opportunities within the network and at the servers.}{has not considered the implications of tiling
  and associated quality-adaptive prefetching techniques for 360$\degree$ video on the performance of content caches.}

\revsix{\revfour{Naturally, proxy caches, server-side caches, and other network- and server-side aggregation techniques that try to
reduce the amount of
\revrev{computation or data}{data}
being transferred from
\revrev{disk, for example, is}{disk or over a network are}
most efficient
if there is a significant overlap in the data being requested.
To better understand the potential bandwidth savings
of caching-based techniques for
\revrev{this context, in this paper, we therefore}{360$\degree$ video delivery, in this paper we}
characterize the similarities in the viewing directions and viewports
\revtwo{}{(i.e., the area of the full 360$\degree$ view that each user sees)}
of
\revrev{clients}{users}
watching the same video,
and then analyze and discuss the implications these findings may have on caching
\revrev{of tiled 360$\degree$ video.}{performance.}
To the best of our knowledge, this is the first paper to analyze the
viewport overlaps between users watching the same video and the implications that
\revtwo{similarities in head movements}{such overlaps}
have on the caching of tiled 360$\degree$ video.}{Content caches will be more effective
  the greater the overlaps in the data requested by clients.  In this paper we carry out the first
  analysis, to the best of our knowledge, on the similarities in the viewing directions and viewports
  (i.e., the area of the full 360$\degree$ view that each user sees) of users watching the
  same 360$\degree$ video, and then analyze and discuss the implications these findings may have on caching performance.}}{This paper presents
  new trace-based analysis methods that incorporate users' viewports
  (the area of the full 360$\degree$ view the user actually sees),
  a first characterization of the cross-user similarities of the users' viewports,
  and a trace-based analysis of the potential bandwidth savings that caching-based techniques may offer under different conditions.}
The paper has three main parts, with the second and third parts building
  on the prior parts.  Furthermore, each part includes both novel methodological contributions
and a trace-based characterization or 
\revBB{analysis.}{analysis providing insights that can help guide the design of more effective caching policies and other related system solutions.}

\revsix{First,
we present
\revtwo{an}{a general}
analysis of the
\revrev{instantaneous viewing similarities;}{similarities in viewing direction among different
  \revrev{clients}{users}
when at identical playback points within the same 360$\degree$ video;}
e.g., as measured by angular differences
of the viewing directions, overlap in viewports, and how
\revrev{these overlaps increase as a server/cache sees
  more and more clients watching the same video.}{the viewport's overlap with the aggregate view cover from prior
  \revrev{client}{user}
  views increases with the number of such
  \revrev{clients.}{users.}}}{First,
  we focus on similarities in the viewports among different users when at identical playback points within the same 360$\degree$ video.
  Here, we define basic similarity metrics such as the overlap in viewports for two or more users
  and study how the
  overlap between a viewport and
  the aggregate view cover from prior user
  views increases with the number of such users and how this differs between different video categories.}
This analysis provides insight into inherent similarities in viewing behavior,
  \revsix{using measures that}{and the metrics}
  are not affected by the details of how video data is 
  \revAA{delivered to users.}{delivered.} 

\revfour{Second, we extend the
  analysis to
  \revtwo{take into account}{incorporate}
  \revrev{that the video typically is chunked,}{chunking of the video content,}
  \revtwo{}{and the chunk granularity in particular,}
  \revtwo{evaluating}{allowing us to evaluate}
  \revtwo{aspects such as the impact that chunk durations have on some of the observations from
  \revrev{our instantaneous analysis.}{our identical playback point analysis.}
  \revtwo{}{This is important to understand the impact that different time granularities
  have on chunk-based video delivery (e.g., as in the case of
  HTTP-based Adaptive Streaming (HAS)) and other prefetch-dependent techniques that
  may choose to operate at different time granularities.}}{its impact on the insights in the first part of the paper.
  This analysis is important with respect to understanding how similarities in viewing direction would impact caching
  performance.  For example, consider the case where two users have significantly different viewing directions
  at a particular time instant, and yet over the time duration of a chunk have essentially the same viewport cover.}}{Second,
  we extend the
  \revsix{}{metrics and}
  analysis to
evaluate the impact of chunk granularity on the insights from the first part of the paper.
  This analysis is important
to understand
  how similarities in viewing direction would impact caching
performance.  For example, consider the case where two users have significantly different viewing directions
at a particular time instant, and 
\revBB{yet}{yet,} 
over the time duration of a 
\revBB{chunk}{chunk, they}
have essentially the same viewport cover.}

Finally, we present 
\revBB{}{simulation results using}
a novel simulation 
\revBB{model that}{model.  The model} 
captures steady-state performance of 
\revBB{a large number of}{many} 
independent sessions,
  while using only a limited number of 
\revAA{traces, and a trace-based analysis in which we simulate a proxy cache and}{traces.  We simulate a proxy cache and}
evaluate the cache hit rates observed
when using prefetching algorithms that attempt
  to adaptively select which tiles to download and the quality of each
  to optimize the user's
  quality of experience.  The observed cache hit rates 
  \revBB{}{(analyzed in Section~\ref{sec:simulation})}
  reflect not only viewing direction
  similarities, both at identical time instants 
  \revBB{(as analyzed in the first part of the paper)}{(analyzed in Section~\ref{sec:characterization})}
  and over the time duration of a chunk 
  \revBB{(as analyzed in the second part of the paper),}{(analyzed in Section~\ref{sec:chunk-based}),}
  but also differences in chunk quality selections caused by bandwidth variations and
  uncertainties in viewing direction.
\revBB{}{The simulation results provide quantitative example comparisons and deliver insights into how viewing direction similarities and potential caching performance depend on the nature of the 360$\degree$ video content.}

\revBB{Throughout our analysis we use head-movement traces
collected for different categories of 360$\degree$ video~\cite{AAK+18}, allowing us to provide
quantitative example comparisons and deliver
insights into
how viewing direction similarities and potential caching performance depend on the nature of the 360$\degree$ video content.
  For evaluation of cache performance, we combine our traces with
  previously collected network measurements capturing a wide range of network conditions~\cite{RVGH13,anon16}.}{Throughout our analysis we place particular focus on insights regarding differences seen when comparing categories of 360$\degree$ videos.  For this analysis, we use head-movement traces collected for different 360$\degree$ categories~\cite{AAK+18}. For evaluation of cache performance, we combine the use of these traces with previously collected network measurements capturing a wide range of network conditions~\cite{RVGH13,anon16}.  Of particular interest are insights into how the different categories are affected by bandwidth variations and other uncertainties that may impact the client's quality of experience.}

\revrev{In general, the results provided in Sections 3-5 help
highlight substantial differences in the pairwise viewport overlaps
observed between different video categories and the impact that they have on caching of these videos.
The results}{The results provide substantial insight into the conditions under which overlap can be considerable and caching effective.
  Particularly noteworthy perhaps are the substantial differences observed between different video categories and, in some cases,
  playback positions within the
  \revNot{video, and the impact of prefetching accuracy on caching effectiveness.}{video.}
  \revtwo{}{For example,
    the category of
    videos for which ``the main focus of attention
    is deemed to always be at the same location in the video''~\cite{AAK+18} appears to provide
    the greatest
    opportunities, among the categories we consider.  However,
    this is not the case until 20-30 seconds into these videos,
    as viewers of these videos often have an initial exploratory phase during which viewing similarities are smaller
    compared
    to the
    \revsix{\revfive{category of
    videos in which the ``object of attention is moving across the 360$\degree$ sphere''
    or the category in which
    \revfive{``the users take a virtual ride in which the camera is moving forward at a high speed''.
Caching effectiveness is greatest when the viewing direction predictions used in prefetching are accurate and
    bandwidth conditions are relatively stable.}{``the users take a virtual ride in which the camera is moving forward at a high speed''.}}{
        ``moving focus'' category of videos or the ``rides'' category.}}}}{category of videos in which the
  ``object of attention is moving across the 360$\degree$ sphere'' or the category in
  which ``the users take a virtual ride in which the camera is moving forward at a high speed''. Caching effectiveness
  is greatest when the viewing direction predictions used in prefetching are accurate and
  bandwidth conditions are relatively stable.}
  
Our
\revsix{measurement}{characterization}
and analysis results can inform the design of new caching system policies tailored for 360$\degree$ video.
\revBB{}{For example, 
our results suggest that
selective edge-cache insertion 
policies~\cite{MaSi15,GaVa16,CaEa17,CaEa21}
should consider the video category and when in a video a chunk occurs.}
\revtwo{}{Our results
  \revsix{may also}{also}
  have implications for other policies.  For example, cache hit rate may
    benefit from cap-based network solutions that stabilize the bandwidth seen
    by individual clients (e.g.,~\cite{KrCH18}).}
\revtwo{may also have implications for other contexts than caching.
  For example, our characterization clearly highlights that}{With respect to prefetching policies, our results show how}
the value of using
\revrev{other clients viewing direction}{the viewing directions of previous \revrev{clients}{users}}
for viewport prediction
\revtwo{may vary substantially between}{varies among different}
\revrev{}{360$\degree$ video}
\revrev{categories.}{categories and, in some cases,
  \revtwo{the}{also depends on the}
  playback position.}

The remainder of the paper is organized as follows.
Section~\ref{sec:background} presents
\revrev{a brief overview}{background}
and introduces the head-movement dataset used here.
Section~\ref{sec:characterization} presents our
\revsix{}{metrics and}
\revrev{instantaneous analysis of the pairwise viewing differences,
viewport overlaps, and how the viewport overlap with the aggregate view cover of prior clients increase
with the number of prior clients.}{analysis of viewing direction similarities between pairs of
  \revrev{clients}{users}
  at identical playback points, pairwise viewport overlaps, and viewport overlaps with aggregate view covers from different numbers of prior
  \revrev{clients.}{users.}}
Section~\ref{sec:chunk-based} extends
\revsix{this}{these metrics and}
analysis to
\revtwo{\revrev{consider chunk-based head movements,}{take into account chunking,}}{take into account the chunk granularities used,}
before Section~\ref{sec:simulation} presents our trace-based simulations of cache performance
under different network bandwidth conditions and uncertainties in
\revrev{the}{viewing direction} prediction accuracies.
Finally, Section~\ref{sec:related} presents related 
\revBB{work and Section~\ref{sec:conclusions} concludes the paper}{work, Section~\ref{sec:insights} summarizes design insights, and Section~\ref{sec:conclusions} concludes the paper.}


\section{Background and Dataset}\label{sec:background}

\begin{figure}[t]
  \begin{minipage}[t]{0.28\textwidth}
  \centering
\vspace{8pt}
\includegraphics[trim = 0mm 4mm 0mm 4mm, clip, width=0.94\textwidth]{{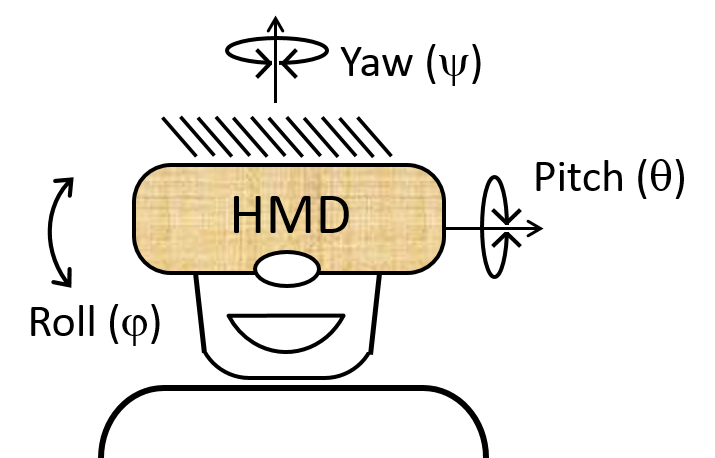}}
  \vspace{-10pt}
  \caption{Head-movement coordinates: Yaw, pitch, and roll.}
  \label{fig:directions}
  \vspace{-12pt}
\end{minipage}
\hfill
\begin{minipage}[t]{0.7\textwidth}
\centering
  \subfigure[All time instances and pairs]{
    \includegraphics[trim = 2mm 8mm 4mm 6mm, width=0.48\textwidth]{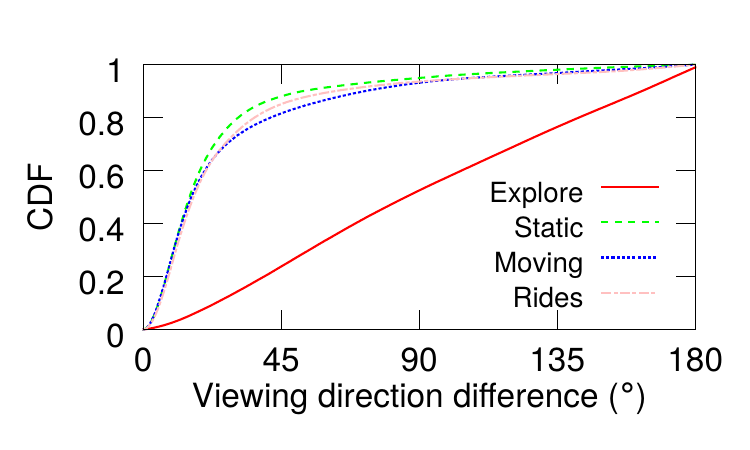}}
  \subfigure[Averages per session pair]{
    \includegraphics[trim = 2mm 8mm 4mm 6mm, width=0.48\textwidth]{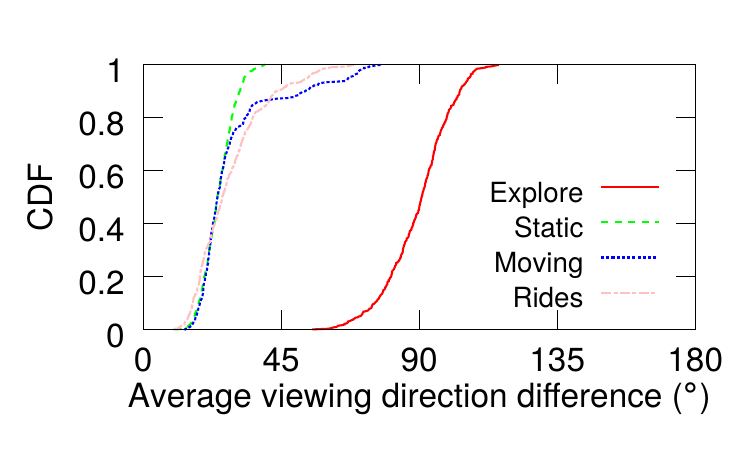}}
  \vspace{-16pt}
  \caption{CDFs of pairwise viewing direction \revthree{differences for the}{differences;} representative videos.}
  \label{fig:cdf-angle}
  \vspace{-10pt}
\end{minipage}
\end{figure}

360$\degree$ videos
capture the view in all directions
  and allow users to look in any direction at each point during playback;
e.g., by moving their head while wearing a head mounted display (HMD).  While 360$\degree$ videos also can be
viewed in the browser on PCs, on smartphones, or on tablets, for the work presented here we
assume use of an HMD.  
\revAA{Figure~\ref{fig:directions} shows an example user wearing an HMD and defines the viewing angles (i.e., yaw, pitch, and roll) used in our work.}{As illustrated in Figure~\ref{fig:directions}, we characterize viewing directions using yaw, pitch, and roll.}
\revAA{}{Yaw ($\pm 180\degree$) measures sideways rotations (relative to a 0$\degree$ line corresponding to the initial viewing direction as set in the video), the pitch ($\pm 90\degree$) vertical head rotations (relative to a horizontal plane), and the roll ($\pm 90\degree$) rotations of the head (relative to holding the head straight).}
All angles are measured in degrees and normalized so that two users will have the same recorded viewing direction at a given point during their viewing of the same video whenever their viewports completely overlap,
regardless of original head positioning.

For our analysis we use a dataset collected by Almquist et al.~\cite{AAK+18}.  The dataset
consists of fine grained head-movement data collected when 32 people watched
360$\degree$ videos from a set of 30 such videos.
The videos were downloaded and played in 4K resolution,
were
1-5 minutes long (3 min. on average), and were (by the authors) split
across five categories~\cite[p.~260]{AAK+18}:
{\em exploration} (``no particular object or direction of special interest and the users are expected to explore the entire sphere throughout the video duration''),
{\em static focus} (``the main focus of attention is deemed to always be at the same location in the video''),
{\em moving focus} (``story-driven videos where there is an object of special interest that is moving across the 360$\degree$ sphere''),
{\em rides} (``the users take a virtual ride in which the camera is moving forward at a high speed, making users feel that they too are moving forward quickly''), and
{\em miscellaneous} (``includes videos that were deemed to have a mix of the characteristics of the other categories or had a hard-to-classify, unique feel, to them'').
\revAA{}{While alternative classifications would be possible using machine learning, we decided to use these predefined categories, since they come with easy-to-understand human labels that simplify the interpretation of the results.  For our purposes, this is desirable over classification clustering that may have higher coherent scores but may be more difficult to interpret.  Of course, for studies that would try to optimize prefetching and caching policies, other choices may be desirable.}

In total, the dataset includes head movements from 439 unique viewings
(totaling 21 hrs and 40 min).
The ``semi-random'' design of the user study ensured that all 32 users watched one ``representative''
video from each
category (these videos were named ``Zayed Road'', ``Christmas scene'', ``Christmas story'',
``F1'', and ``Hockey'' in the Almquist et al. paper),
while the other videos got between 8-13 views each.  In this paper,
we primarily focus
on the
representative videos for the first four (more well-defined) 
categories, 
\revAA{as these allow for a richer head-to-head comparison of the similarities and differences in viewing direction, and hence also of the caching opportunities, when multiple viewers watch the same video, but}{but} 
also report some summary results for the other videos.
Since Almquist et al. found that yaw movements dominate, followed by pitch, with only small roll movements, we focus only on yaw and pitch.

\revAA{}{There is very limited work on the topic of caching of tile-based 360$\degree$ video implemented using HAS.  However, like for regular HAS videos, we foresee clients to be directed through a proxy cache (e.g., within a CDN) where copies of tiles in different qualities can be stored after having been requested by prior clients. A client requesting a previously requested tile would hence be possible to serve from the cache, rather than the origin server.  Such cache hits result in bandwidth savings, reduce the load on the origin servers, and help improve client performance (e.g., due to shorter round-trip-times and faster download times). 
Since each client still needs to perform their own bit-rate adaptation, make their own chunk requests, and their own tile selection independently, it is important to understand cross-user similarities in 
viewing behavior and their caching implications.}



\section{Similarity Characterization}\label{sec:characterization}

In this section we present an initial characterization of the viewing
similarities and differences between users watching the same video.
For each video,
we calculate and report summary statistics based on the viewing directions
observed every 50ms.  To account for the timestamps not always aligning
perfectly between the traces,
we use interpolation and note that good accuracy is ensured by the
use of a measurement granularity of 10ms
\revAA{(i.e., 100 measurements per second) in}{in} 
the data collection.

\subsection{Pairwise viewing differences}

First, we consider the difference in viewing direction of
  two users at identical playback points within the same 
  \revAA{video, as measured by the angle between these directions.}{video.}
Figure~\ref{fig:cdf-angle} shows cumulative distribution functions (CDFs)
of the pairwise differences, 
\revAA{when combining}{as measured by the angle between the viewing directions of two users at identical playback points.  Here, we combine} 
 the differences in both yaw and pitch,
for all pairs of viewing sessions of each of the representative videos.
(For each of these videos we have 32 user traces and therefore 496 pairs.)
In particular, Figure~\ref{fig:cdf-angle}(a)
shows CDFs for the differences,
as measured every 50 ms throughout 
\revAA{the viewing sessions,}{every pair of viewing sessions,} 
and Figure~\ref{fig:cdf-angle}(b) shows CDFs
for the average of these differences for each session pair.
\revAA{}{For the average metric, we first calculate the average pairwise viewing difference for each possible session pair, and then report the set of average values (over all such pairs) as a CDF.}

As expected, the pairwise differences are substantially larger for the {\em explore} category than
for the other 
\revAA{categories (i.e., {\em static}, {\em moving}, and {\em rides}).}{categories.}
For example, the close-to-straight {\em explore} line in Figure~\ref{fig:cdf-angle}(a) suggests that
the viewing directions of
users
watching {\em explore} videos are close to independent.
In contrast, for the other categories the view angle differences are less than 45$\degree$
for 80\% of the time instances, showing that viewers of these 
\revAA{three video categories}{other videos} 
often are looking at the same parts of the video.

\begin{figure}[t]
  \begin{minipage}[t]{0.50\textwidth}
  \centering
\vspace{-6pt}
\includegraphics[trim = 0mm 8mm 0mm 0mm, width=1\textwidth]{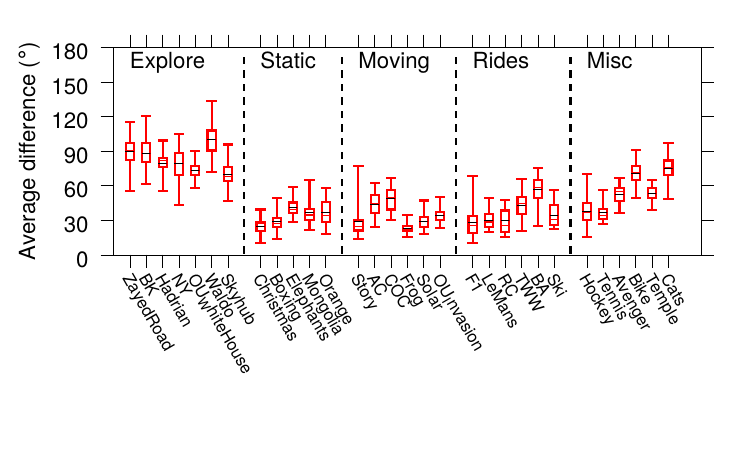}
  \vspace{-32pt}
  \caption{Pairwise average viewing direction differences for each video.}
  \label{fig:pairwise-wiskar-deltaTot}
  \vspace{-10pt}
\end{minipage}
\hfill
\begin{minipage}[t]{0.48\textwidth}
\centering
  \subfigure[Yaw only]{
\includegraphics[trim = 18mm 4mm 14mm 6mm, width=0.48\textwidth]{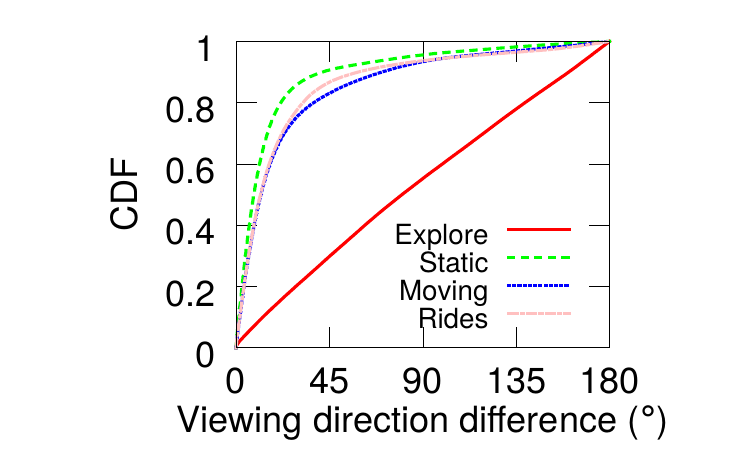}}
  \subfigure[Pitch only]{
\includegraphics[trim = 18mm 4mm 14mm 6mm, width=0.48\textwidth]{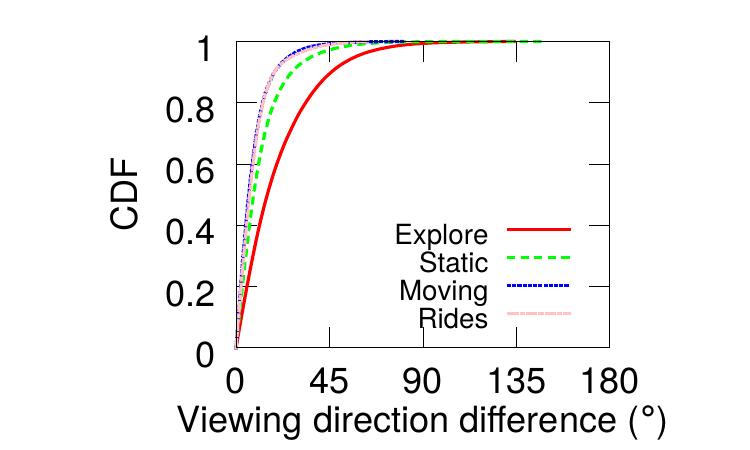}}
   \vspace{-12pt}
  \caption{\revBB{Breakdown of the pairwise}{Pairwise} 
  viewing direction differences, across all time instances and user pairs.}
  \label{fig:cdf-angle-breakdown}
  \vspace{-10pt}
\end{minipage}
\end{figure}

These significant differences
among the categories
are also
clearly
visible when considering
the viewing direction difference averaged over the entire playback duration
(Figure~\ref{fig:cdf-angle}(b))
and when considering the average differences 
\revAA{also for the other}{for the full set of} 
videos in the 
dataset.
\revAA{Figure~\ref{fig:pairwise-wiskar-deltaTot} shows
a box-and-whisker plot with results for all videos,
as categorized by Almquist et al.~\cite{AAK+18}.
For each video the figure}{Figure~\ref{fig:pairwise-wiskar-deltaTot}} 
shows the minimum over
  all pairs of sessions of the average viewing direction difference
(bottom marker), 25-percentile (bottom of box),
median (middle red marker), 75-percentile (top of box),
maximum (top marker), and average 
\revAA{(black marker).}{(black marker) for each of the videos in the Almquist et al. dataset.}

While the above results are based on the total directional
differences across both yaw and pitch,
the observations (and values) are very similar when
focusing on yaw only.
\revBB{In general}{One reason for this is that}
\revAA{}{the head movements are much smaller along pitch,
with angular differences (again) most noticeable for the {\em explore} category.} 
This is illustrated in Figure~\ref{fig:cdf-angle-breakdown},
which breaks down the angular differences observed in
Figure~\ref{fig:cdf-angle}(a) into yaw ($|\psi_A-\psi_B|$) and pitch ($|\theta_A-\theta_B|$).
\revBB{}{The 
total directional differences (yaw + pitch) are also highly correlated with the differences seen for yaw only.  For example, 
for the four representative videos the
Pearson correlation coefficients of the two per-session metrics are 0.981, 0.923, 0.994, and 0.989 for the pairwise comparisons.
(The corresponding correlations when using every per-instance measurement of the two differences are 0.977, 0.951, 0.987, and 0.984.)}

\subsection{Viewport-based metrics}

\begin{figure}[t]
  \centering
  \subfigure[Yaw + pitch]{
    \includegraphics[trim = 0mm 12mm 0mm 12mm, width=0.42\textwidth]{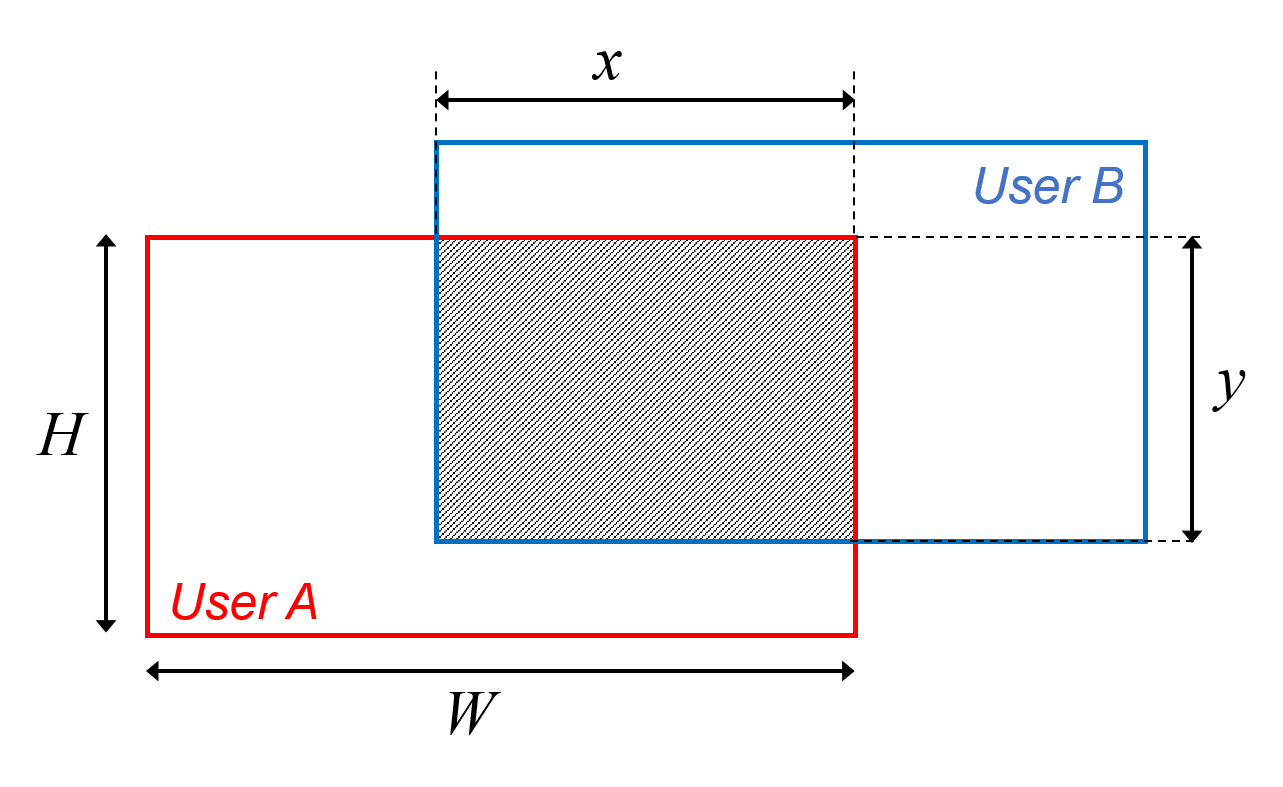}}
  \subfigure[Yaw only]{
    \includegraphics[trim = 0mm 0mm 0mm 12mm, width=0.28\textwidth]{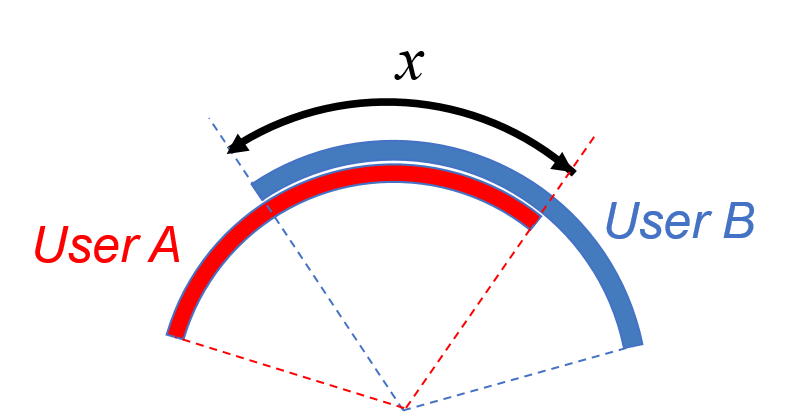}}
  \vspace{-16pt}
  \caption{Definition of the pairwise viewport overlap metric.
  (Handling of wraparound effects described in text.)}
  \label{fig:metric-overlap}
  \vspace{-12pt}
\end{figure}

\revBB{Similarities in what content clients download and what they actually watch depend not only on the users' viewing directions, but more importantly on their viewports.}{Similarities in what content clients download and watch depend not only on the users' viewing directions but more importantly on their viewports.}
\revAA{}{Here, it is important to note that the overlap between two user's viewports 
is determined by a combination of the viewing direction differences and the size of their viewports.}
\revAA{Here, we}{We} 
consider two types of viewports.
First, we consider the 2D area of the viewing field being
\revNot{displayed (i.e., the area of immediate interest to the user).}{displayed.}
\revNot{Let $W \times H$ define this area,
where $W$ and $H$ are the width and height of the viewport of consideration (as measured in degrees), respectively.
Second,}{Second,}
motivated by most head movements being along the yaw angle,
we consider a {\em sliced version},
in which we ignore the pitch and only consider the yaw angle.
\revNot{While each user typically
\revrev{only would watch part of such}{would only watch part of such a}
slice at each time instance,
this abstraction better matches systems that
\revrev{select}{opt}
to use vertical tiles
(e.g., with higher quality
\revrev{in the typically pitch angle}{in the range of typical pitch angles}
and somewhat lower quality downwards and upwards in the viewfield).}{In both cases, we report overlaps normalized by the total viewport size.
  Figure~\ref{fig:metric-overlap} illustrates the metrics.}

\cutICPE{
\revsix{\revtwo{}{{\bf Metrics:}}}{{\bf Pairwise viewport overlap:}}
Consider the viewports of two arbitrary users A and B at the same
\revrev{playpoint}{playback point}
$t$ of the video.
Figure~\ref{fig:metric-overlap}(a) shows
\revrev{this case}{the viewport overlap}
when taking into account both the yaw and the pitch angle,
whereas in Figure~\ref{fig:metric-overlap}(b) we consider the yaw angle alone (this time observing the viewports ``from above'').
\revfour{\revrev{Again, given}{Given}
the small variations in pitch observed in prior studies,
this second metric can be of great interest on its own.
In both figures,}{In both figures}
the viewports of users A and B are shown in red and blue, respectively,
and $x$
\revrev{shows}{denotes}
the overlap along the yaw angle.
Furthermore, in Figure~\ref{fig:metric-overlap}(a),
$y$
\revrev{shows}{denotes}
the overlap in pitch and
the shaded area ($x \times y$) shows the overlap when accounting for both angles.
In the following, we report the normalized overlap, equal to $\frac{xy}{WH}$ and $x/W$, respectively.
In our calculations we account for wraparound effects along the yaw angle
(using $x = \max(0,W-\min(|\psi_A-\psi_B|, 360- |\psi_A-\psi_B|))$,
where $\psi_A$ and $\psi_B$ are the yaw angles for the two users)
but do not consider overlaps due to users catching a glimpse of what is behind them due to pitch angles outside the range $\pm 90$
(using $y=\max(0,H-|\theta_A-\theta_B|)$, where $\theta_A$ and $\theta_B$ are the pitch angles for the two users).
}

\revsix{}{{\bf Multi-user viewport cover and overlap:}
\revrev{Thus far we have focused on the viewport overlap of two arbitrary
  clients watching the same video.  Clearly, the overlap (and also the hit rate)
  increases as more user view the same video.  To quantify this increase,
  we next evaluate how the normalized overlap of a client's viewport and
  that of all $N$ prior clients watching the same video.}{\revfour{To gain further insight into viewport overlaps
    and implications for potential cache performance,}{To gain insight into how the potential cache performance may be impacted
    by the number of users having watched a video,}
  we also look beyond pairwise viewport overlaps
  and consider overlaps among larger sets of
  \revrev{clients.}{users.}  Specifically, we evaluate how the
  overlap between a viewport and
  the aggregate view cover from prior
  \revrev{client}{user}
  views increases with the number $N$ of such
  \revrev{clients.}{users.}}}

\begin{figure}[t]
 \begin{minipage}[t]{0.64\textwidth}
    \centering
    \vspace{10pt}
\subfigure[Example sequence]{
      \includegraphics[trim = 4mm 0mm 4mm 0mm, width=0.58\textwidth]{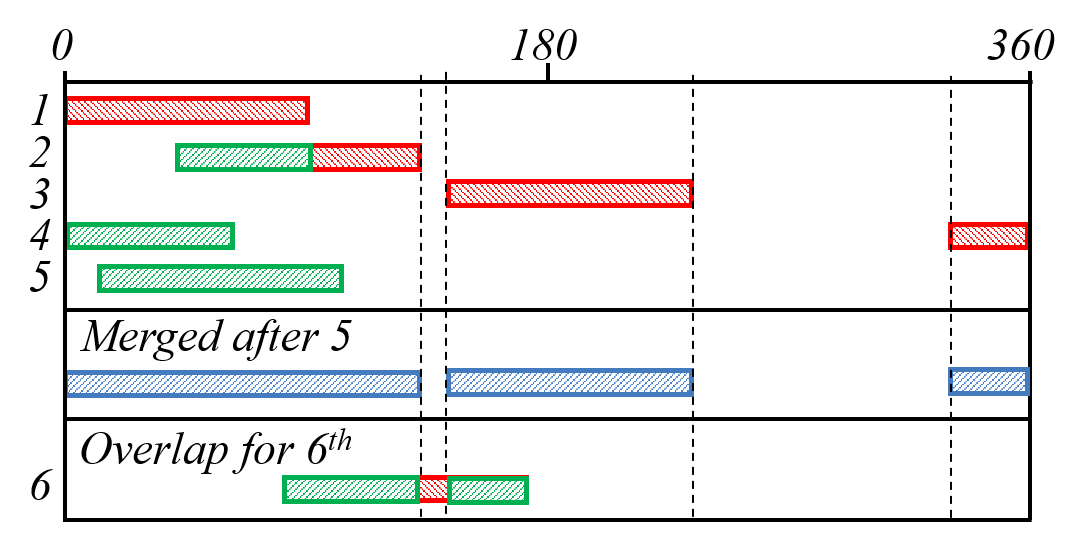}}
    \hspace{-12pt}
    \subfigure[Overlap cases]{
      \includegraphics[trim = 6mm 0mm 6mm 0mm, width=0.40\textwidth]{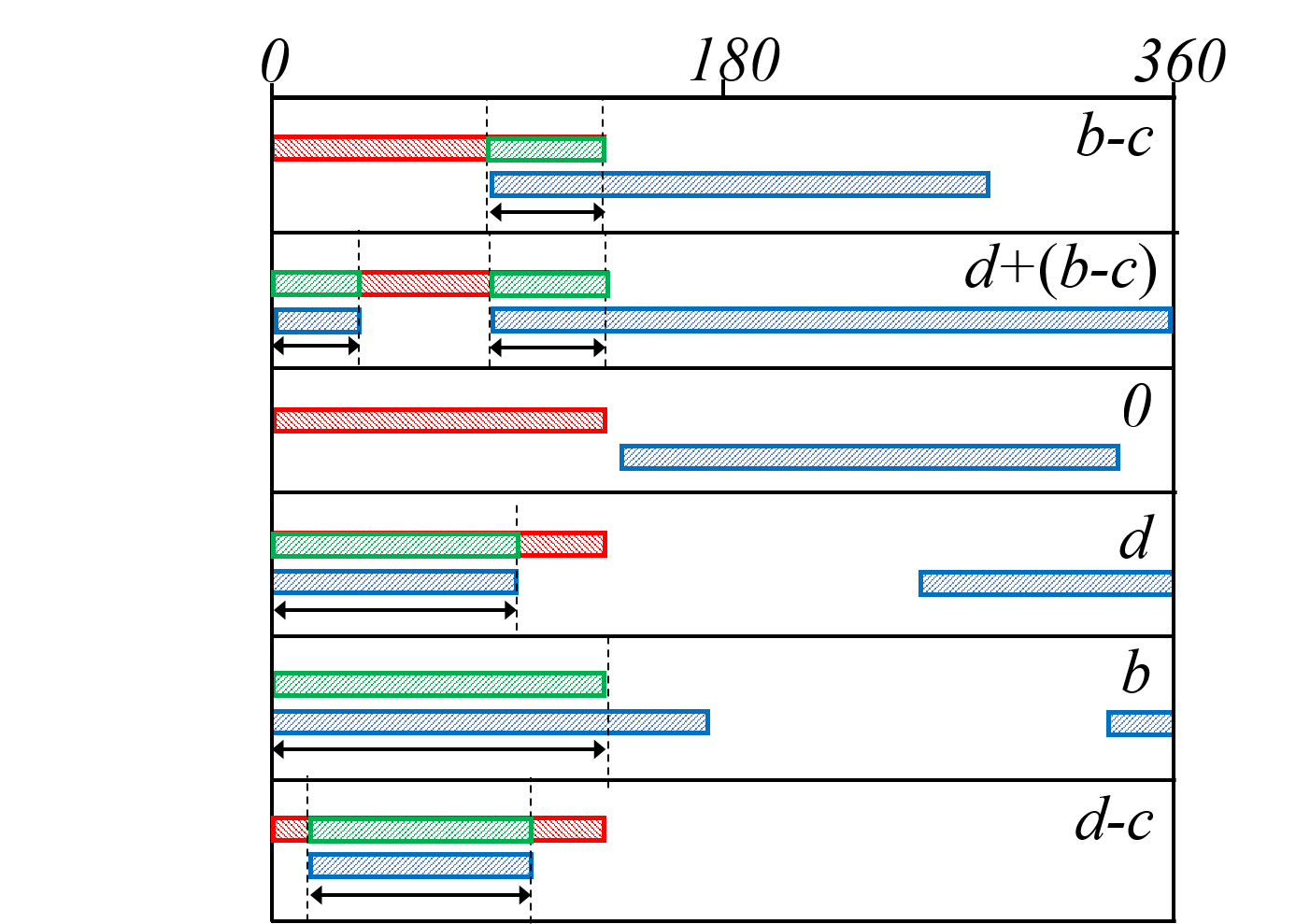}}
    \vspace{-8pt}
    \caption{Examples illustrating the combined multi-user viewport coverage and a user's overlap with prior users. \revAA{}{Sub-figure (b) shows the six intersection cases when assuming that the most recent viewport goes from $a$$=$$0$ to $b$ and the (merged) interval due to prior user sessions is a single interval from $c$ to $d$ (modulus 360).}}
    \label{fig:multi-user-cover-overlap}
    \vspace{-10pt}
\end{minipage}
\hfill
\begin{minipage}[t]{0.34\textwidth}
  \centering
  \subfigure[All time instances and pairs]{
    \includegraphics[trim = 2mm 8mm 4mm 6mm, width=0.96\textwidth]{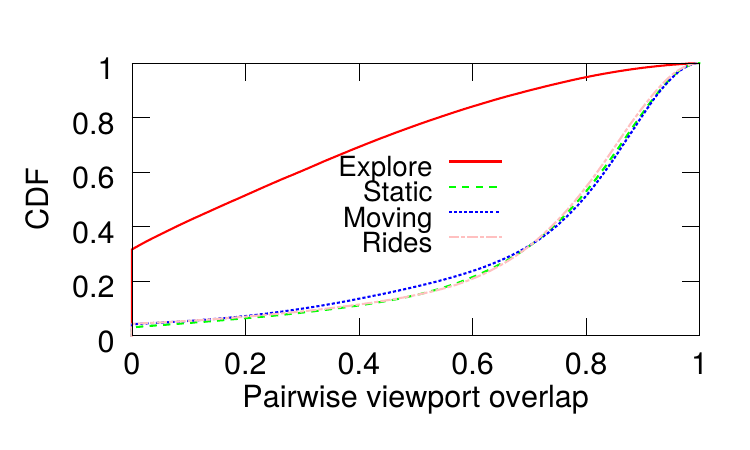}}
  \subfigure[Averages per session pair]{
    \includegraphics[trim = 2mm 8mm 4mm 14mm, width=0.96\textwidth]{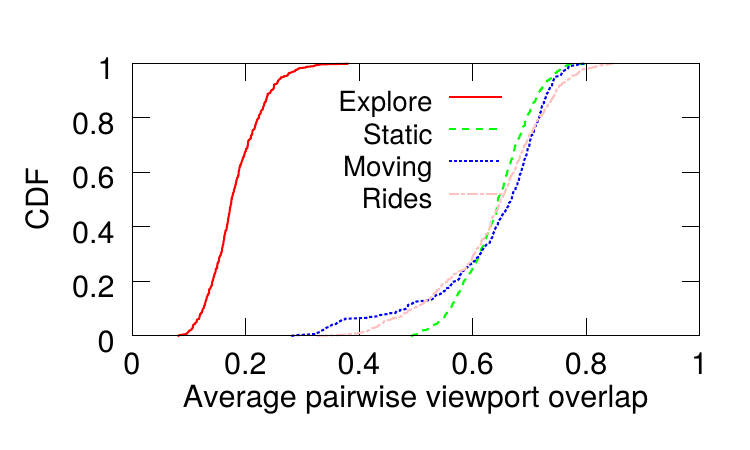}}
  \vspace{-12pt}
  \caption{CDF normalized pairwise viewport overlap.}
  \label{fig:pairwise-cdf-all}
  \vspace{-10pt}
\end{minipage}
\vspace{-6pt}
\end{figure}

For this analysis, 
\revAA{we ignore pitch (i.e., use vertically sliced viewports) and for}{we use vertically sliced viewports.  For} 
each time instance and session sequence, we first merge the viewport coverage
of all $N$ prior
users
\revBB{into a number of}{into} 
non-overlapping (merged) viewport
areas
(as represented by the blue rectangles in the example shown in Figure~\ref{fig:multi-user-cover-overlap}(a)).
Then,
we calculate the overlap
\revAA{of these non-overlapping (merged) viewport intervals with}{with}
the current
user's
viewport (overlap represented by the bottom green rectangles in Figure~\ref{fig:multi-user-cover-overlap}(a)),
before adding this
user's viewport to the merged intervals
and repeating the calculations for the next
user  in the sequence.
\revAA{}{In summary, for each user, we simply repeat the overlap calculations (last row) and merge step (second last row) illustrated in Figure~\ref{fig:multi-user-cover-overlap}(a) to calculate the overlap and combined coverage respectively.}

  By keeping track of the list of non-overlapping intervals
    that have been merged thus far (including updating the list for each new client),
    we can calculate the intersections that the viewport of the latest client
    has with the intervals associated with all prior clients
    using an efficient one-pass algorithm in which we simply
    consider the clients in the order that they arrive to the system.
    \revAA{}{(This is implemented using a structure that keeps track of all intervals merged thus far and considering the viewport of one client at a time.)}
    For each such client, both
  the calculations needed to merge intervals and to calculate intersections of intervals
  require a significant number of cases to consider.
    Figure~\ref{fig:multi-user-cover-overlap}(b) illustrates the six intersection cases that arise assuming
  that the most recent viewport goes from $a$$=$$0$ to $b$,
  and the (merged) interval
 from prior user sessions
 \revAA{}{results in a single interval that}
  goes from $c$ to $d$
  (modulus 360, accounting for wraparound).


\subsection{Pairwise viewport overlap}

{\bf Results for representative videos:}
Figure~\ref{fig:pairwise-cdf-all} shows CDFs of the normalized pairwise overlap
for the representative videos when using a 120$\times$67.5 viewport.
Here, Figure~\ref{fig:pairwise-cdf-all}(a) shows
CDFs for the pairwise overlap at identical playback points
\revAA{(as measured every 50 ms)}{(as measured every 50 ms for every possible viewport pair)}
and Figure~\ref{fig:pairwise-cdf-all}(b) shows
CDFs for the average of these overlaps for each session pair.
\revAA{}{(For the average metric, we first calculate the average pairwise viewport overlap for each possible session pair, and then report this set of average values as a CDF.)}
As before we observe significant differences when comparing the {\em explore} category
with the other categories.
For example, with the {\em explore} video,
more than 35\% of the time there is no pairwise overlap,
whereas for the other categories there is at least a 50\%
overlap in more than 80\% of the instances.  (See Figure~\ref{fig:pairwise-cdf-all}(a).)
\revNot{Furthermore, considering}{Considering}
the average normalized pairwise viewport overlap (Figure~\ref{fig:pairwise-cdf-all}(b)),
no pair of {\em explore} sessions had an average overlap of more than 40\%,
\revrev{while such low (40\%) average overlap was observed for less}{while less}
than 6.5\% of the {\em moving} session pairs, less than 1\% of the {\em rides} session pairs,
and none of the {\em static} session
\revrev{pairs.}{pairs had an average overlap that did not exceed 40\%.}
\revtwo{}{In fact, for these three categories,
more than 70\% of the sessions see an average overlap of at least 60\%.}

\begin{figure}[t]
  \begin{minipage}[t]{0.48\textwidth}
  \centering
  \includegraphics[trim = 0mm 12mm 0mm 4mm, width=0.98\textwidth]{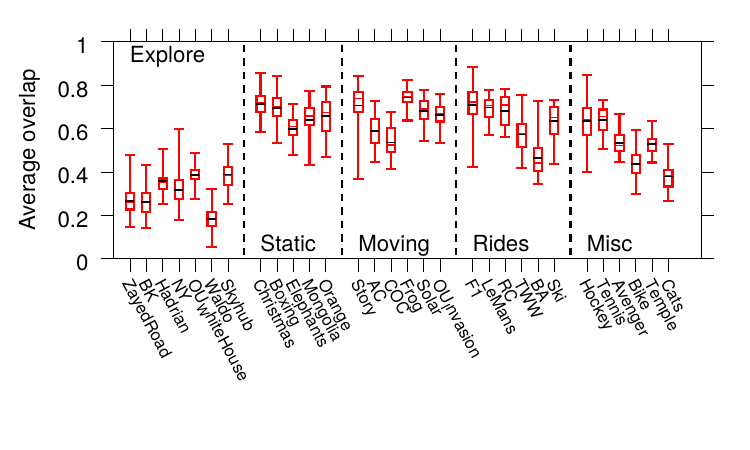}
  \vspace{-10pt}
  \caption{Average normalized pairwise overlap for all videos. (Viewport size 120$\times$67.5.)}
  \label{fig:pairwise-wiskar-relOverlap}
  \vspace{-10pt}
\end{minipage}
  \hfill
  \begin{minipage}[t]{0.48\textwidth}
  \centering
  \includegraphics[trim = 0mm 8mm 6mm 12mm, width=0.98\textwidth]{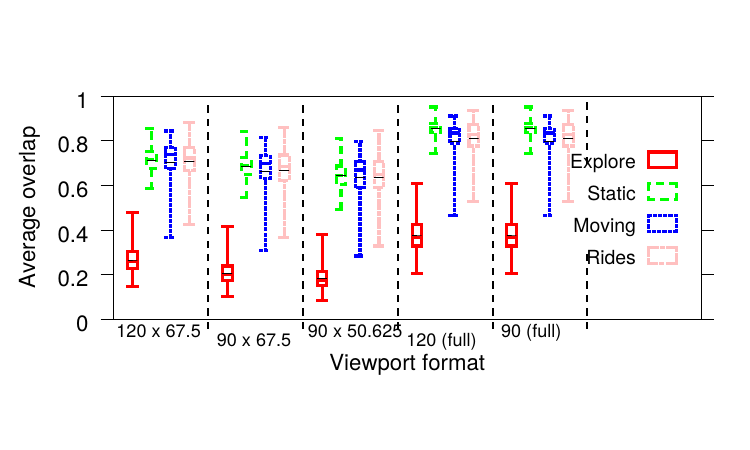}
  \vspace{-10pt}
  \caption{Average normalized pairwise overlap for representative videos, when using different viewports.}
  \label{fig:pairwise-wiskar-viewport}
  \vspace{-10pt}
  \end{minipage}
\end{figure}

\revtwo{}{{\bf Results for other videos:}}
Similar large differences
\revtwo{were also}{were}
observed when calculating the
average normalized pairwise overlap
\revfour{for other videos.
Figure~\ref{fig:pairwise-wiskar-relOverlap}
\revrev{shows a per-video wiskar plot}{shows per-video box-and-whisker plots}
for the average normalized
\revrev{viewport}{pairwise}
overlap when considering a 120$\times$67.5 viewport.
Although there are significant differences between the videos
in each category, these results clearly show that there are substantial
\revrev{average pairwise viewport overlaps}{overlaps}
for all videos in the {\em static}, {\em moving}, and {\em rides} categories.
\revtwo{}{The above observations suggest that there may be substantial
  caching opportunities
with these video
categories.}}{for other videos, as seen in Figure~\ref{fig:pairwise-wiskar-relOverlap},
  which shows per-video box-and-wisker plots for a 120$\times$67.5 viewport.
  Figures~\ref{fig:pairwise-cdf-all} and~\ref{fig:pairwise-wiskar-relOverlap} suggest that there may be substantial caching opportunities for
  videos in the {\em static}, {\em moving}, and {\em rides} categories. }

\revtwo{}{{\bf Impact of viewport:}}
Figure~\ref{fig:pairwise-wiskar-viewport} shows
\revNot{similar summary
statistics for the representative
\revrev{videos of each category}{videos}}{summary statistics}
for five alternative viewports
(the last two ignoring differences in \revthree{the pitch angle}{pitch}).
\revNot{}{For each class we show the minimum over
  all pairs of sessions of the average viewing direction difference
  (bottom marker), 25-percentile (bottom of box),
  median (middle colored marker), 75-percentile (top of box),
maximum (top marker), and average (black marker).}
We note that as the viewports become
\revrev{larger (e.g., 120$\times$67.5) and/or ignores pitch (e.g., 120 full and 90 full),
the overlaps increase compared to the smaller viewports (e.g., 90$\times$50.625).}{larger, the overlaps increase.}

\begin{figure}[t]
  \centering
  \subfigure[120$\times$67.5]{
    \includegraphics[trim = 4mm 8mm 4mm 0mm, width=0.36\textwidth]{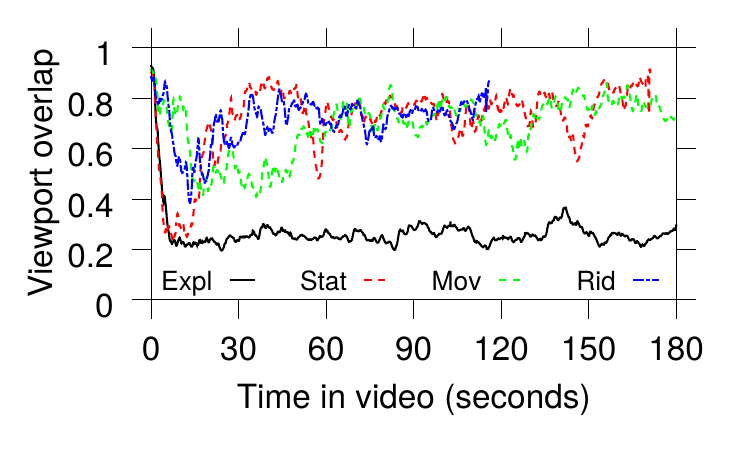}}
  \subfigure[90 full]{
  \includegraphics[trim = 4mm 8mm 4mm 0mm, width=0.36\textwidth]{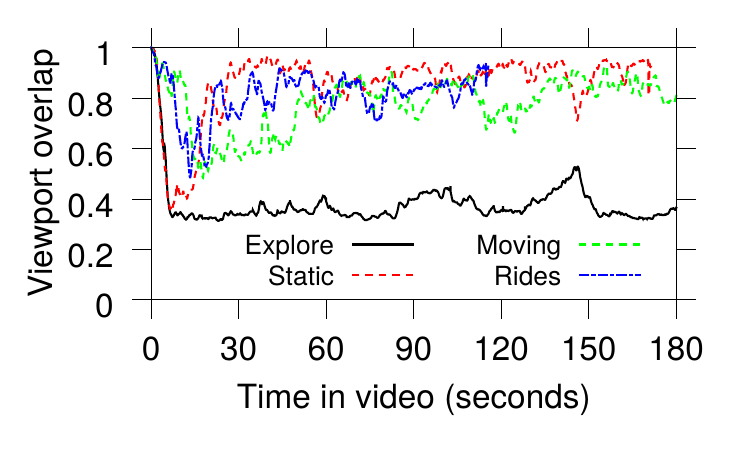}}
  \vspace{-14pt}
  \caption{Time-line plot of the normalized pairwise overlap.}
  \label{fig:timePairs-relOverlap-Representative}
  \vspace{-12pt}
\end{figure}

\revtwo{}{{\bf Longitudinal playpoint dependencies:}}
\revrev{Naturally,}{Note that}
\revrev{the relative differences are not constant over the playback session.}{pairwise overlaps vary over the playback duration.}
For example, all clients start with the same viewing direction and prior work~\cite{AAK+18}
has shown that with {\em static} videos
\revrev{often have}{there is often}
an initial exploration phase.
Figures~\ref{fig:timePairs-relOverlap-Representative}(a) and~\ref{fig:timePairs-relOverlap-Representative}(b)
show the
\revrev{average normalized pairwise overlap as function of time
  (where averages here are calculated over all pairs every 50 ms)}{overlap averaged over all session pairs as a function of the time from the start of the video,}
for two example 
\revAA{viewports; the first (120$\times$67.5) taking pitch into consideration and the second (90 full) ignoring pitch.}{viewports.}
In addition
\revrev{to a brief exploration phase for {\em static},
  which are represented by initially smaller overlaps,}{to smaller initial average
  overlaps for the {\em static} video, resulting from initial exploration,}
we also observe a somewhat smaller
\revrev{overlap in}{average overlap at}
the beginning of the
videos in the {\em rides} and {\em moving} categories than towards the end of those videos.
This
\revrev{may suggests}{suggests}
\revfour{that there may be
\revrev{viewing behavior related optimization opportunities
that also may allow hit rates of high-quality chunks associated with the viewports}{opportunities for cache hit rates}
to improve over the duration of a video session.}{that cache hit rates may improve over the duration of many video sessions.}

\begin{figure}[t]
  \centering
  \subfigure[Explore]{
    \includegraphics[trim = 2mm 12mm 4mm 8mm, width=0.24\textwidth]{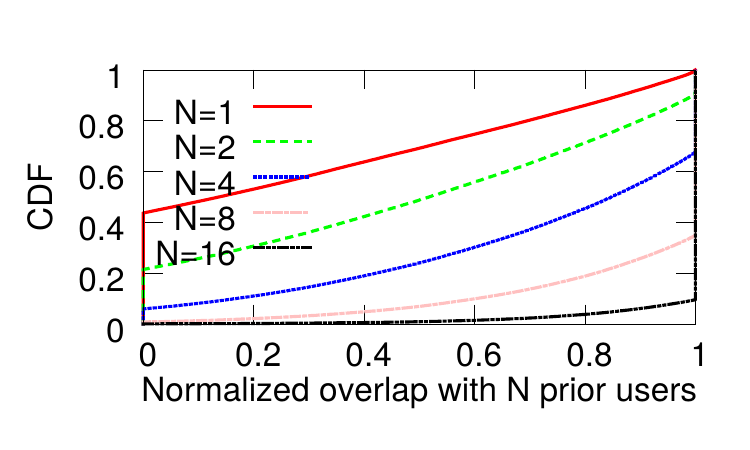}}
  \hspace{-6pt}
  \subfigure[Static]{
    \includegraphics[trim = 2mm 12mm 4mm 8mm, width=0.24\textwidth]{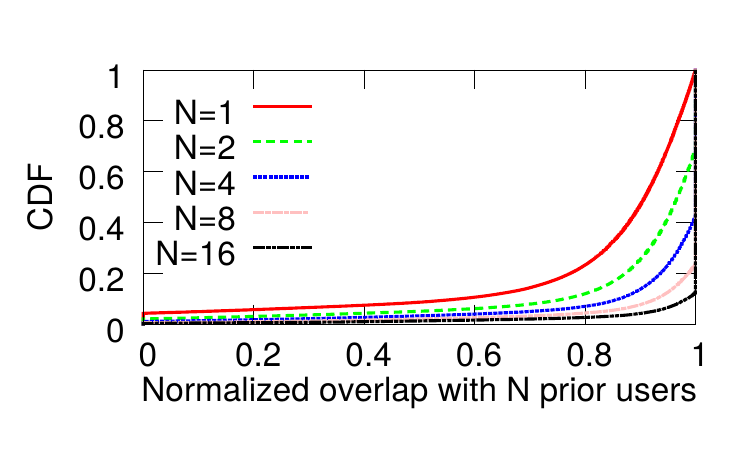}}
  \hspace{-6pt}
  \subfigure[Moving]{
    \includegraphics[trim = 2mm 12mm 4mm 18mm, width=0.24\textwidth]{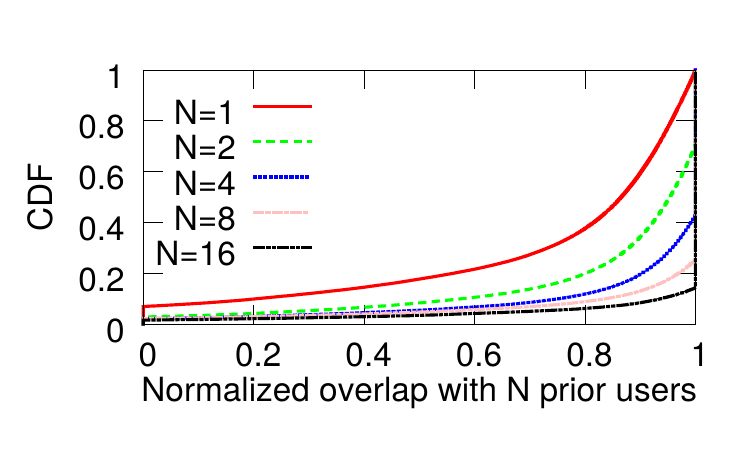}}
  \hspace{-6pt}
  \subfigure[Rides]{
    \includegraphics[trim = 2mm 12mm 4mm 18mm, width=0.24\textwidth]{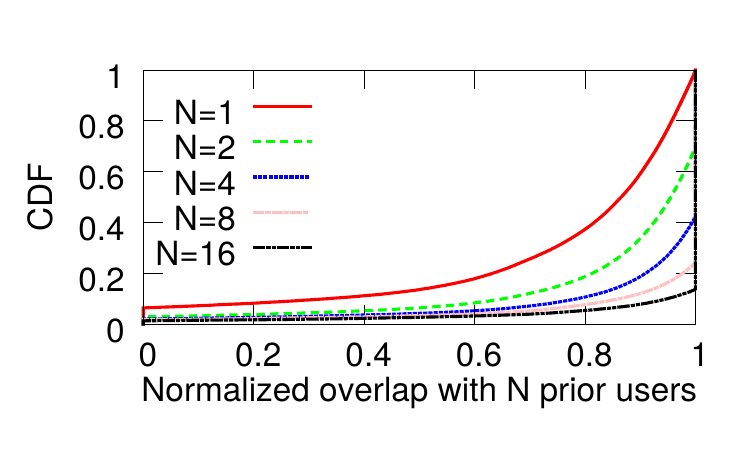}}
  \vspace{-14pt}
  \caption{CDF of normalized viewport overlap with $N$ prior \revrev{clients.}{users.} (Viewport size $W$=90.)}
  \label{fig:sequence-cdf-90}
  \vspace{-10pt}
\end{figure}


\subsection{Viewing sequence analysis}

\revsix{}{Consider next each client's viewport overlap with the aggregate view cover from prior
  user views as a function of the number $N$ of such users.
    For each representative video,
    we created 1,000 random orderings of the 32 viewing sessions recorded in the dataset for that video,
    and for each sequence and viewing session, evaluated
    the overlap at identical playback points between the respective user's viewport and the aggregate
    viewing area covered by all prior users
    in that viewing sequence.}

\begin{figure}[t]
  \centering
  \subfigure[Smaller ($W$=60), explore]{
    \includegraphics[trim = 2mm 12mm 4mm 8mm, width=0.24\textwidth]{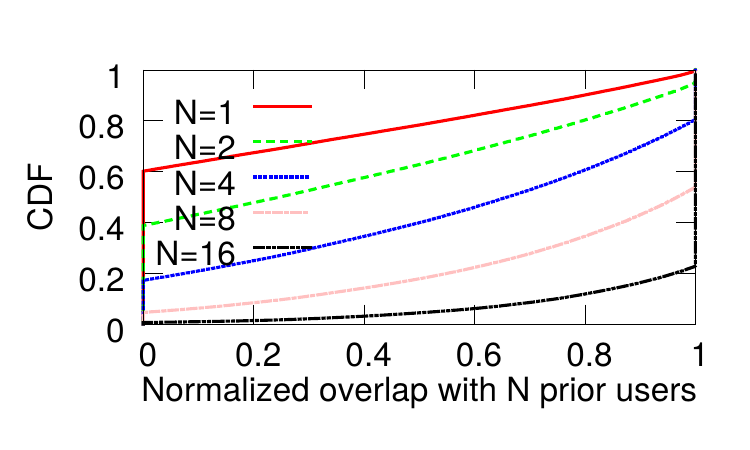}}
  \hspace{-6pt}
  \subfigure[Larger ($W$=120), explore]{
    \includegraphics[trim = 2mm 12mm 4mm 8mm, width=0.24\textwidth]{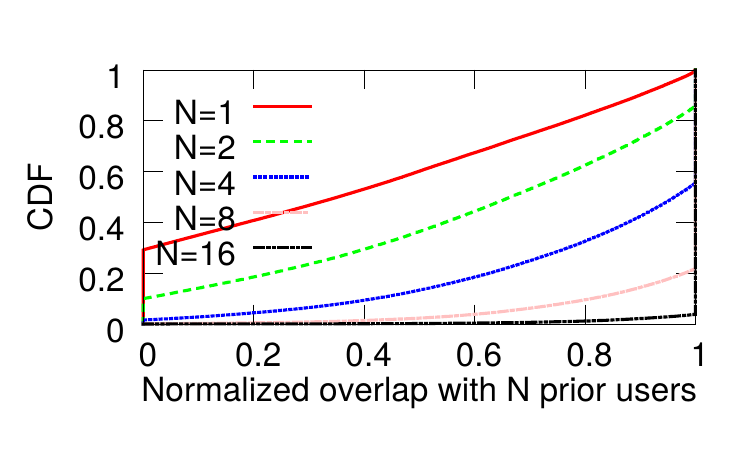}}
  \hspace{-6pt}
  \subfigure[Smaller ($W$=60), static]{
    \includegraphics[trim = 2mm 12mm 4mm 18mm, width=0.24\textwidth]{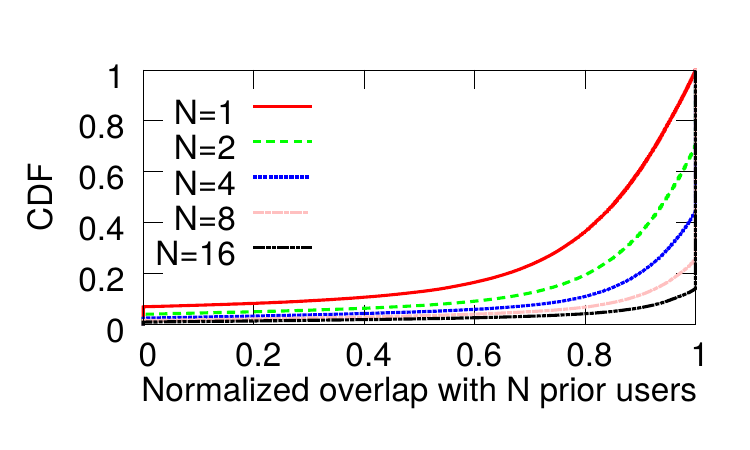}}
  \hspace{-6pt}
  \subfigure[Larger ($W$=120), static]{
    \includegraphics[trim = 2mm 12mm 4mm 18mm, width=0.24\textwidth]{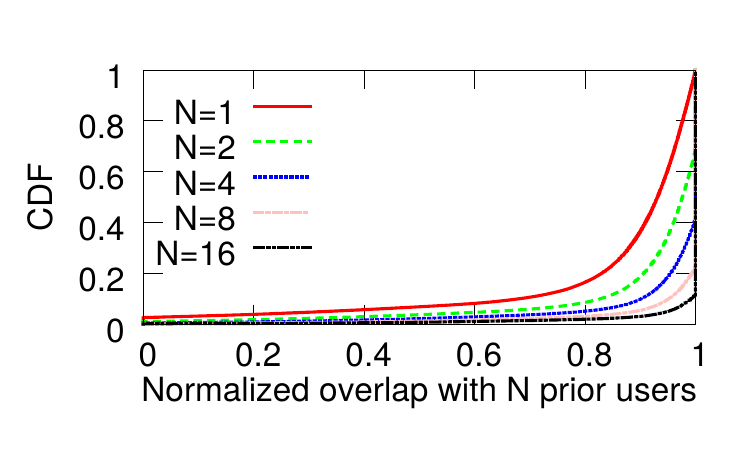}}
    \vspace{-16pt}
  \caption{\revrev{Examples showing the impact of viewport size.}{Impact of viewport size on normalized overlap.}}
  \label{fig:sequence-cdf-W}
  \vspace{-14pt}
\end{figure}

{\bf Category and viewport dependencies:}
Figure~\ref{fig:sequence-cdf-90} shows CDFs of the normalized viewport overlap
for the representative videos and different numbers
of prior users
$N$, where the CDFs are each over all 1,000 random sequences and all playback points at a granularity of 50 ms.
Here we used a sliced viewport with width $W$$=$$90$.
\revAA{}{As expected, looking at the extremes, for small $N$, we note a big point mass in the distribution at a normalized overlap of 0 for the {\em explore} video, and for large $N$ we observe a big point mass at a normalized overlap of 1 for all videos.  These cases result in big steps in the CDFs taken at the x-values of x=0 and x=1, respectively.}
\revAA{We}{More generally, we} 
note that there is a substantial increase in the normalized overlap as $N$ increases,
but with diminishing returns with each doubling of $N$.
\revAA{}{(For example, the lines are for the most part
spaced closer and closer to each other with each doubling of $N$.)}
\revAA{Note}{We also note} 
that as $N$ increases and the CDFs approach the ideal case with all probability mass concentrated at a normalized overlap of 1, the differences between the results for the representative videos diminish.
  In fact, for $N$$=$$16$ and
  $W$$=$$120$,
  the distribution of
  the normalized overlap with the {\em explore} video (Figure~\ref{fig:sequence-cdf-W}(b))
  has greater mass on larger values than that for the other representative videos
(as exemplified by the {\em static} video in Figure~\ref{fig:sequence-cdf-W}(d)).
More generally,
differences between categories,
  as seen in
  Figure~\ref{fig:sequence-cdf-W}, are larger for smaller viewport sizes and for smaller $N$.

\begin{figure}[t]
  \centering
  \subfigure[Explore]{
    \includegraphics[trim = 3mm 12mm 3mm 4mm, width=0.24\textwidth]{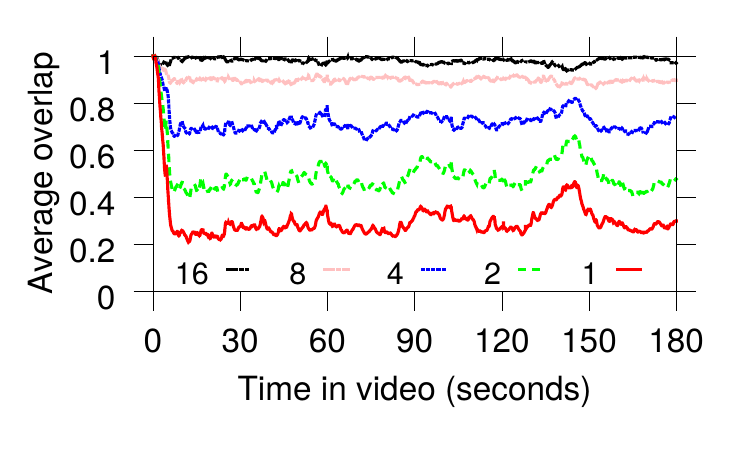}}
  \hspace{-6pt}
  \subfigure[Static]{
    \includegraphics[trim = 3mm 12mm 3mm 4mm, width=0.24\textwidth]{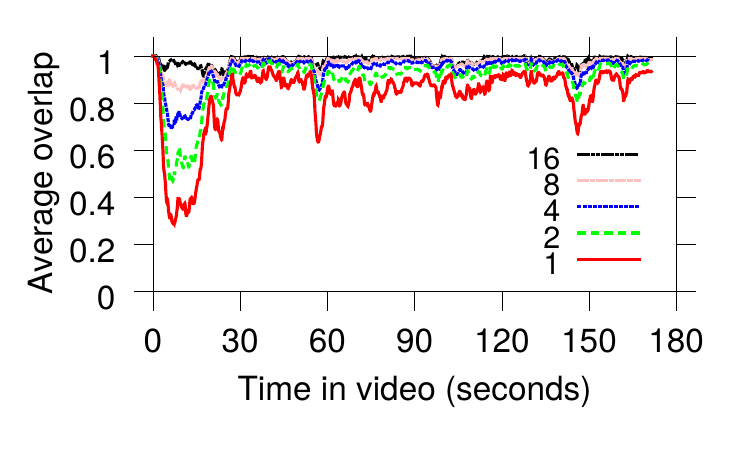}}
  \hspace{-6pt}
  \subfigure[Moving]{
    \includegraphics[trim = 3mm 12mm 3mm 16mm, width=0.24\textwidth]{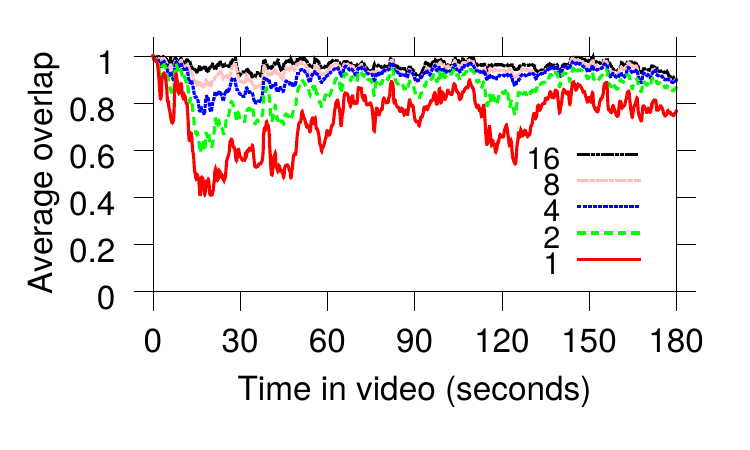}}
  \hspace{-6pt}
  \subfigure[Rides]{
    \includegraphics[trim = 3mm 12mm 3mm 16mm, width=0.24\textwidth]{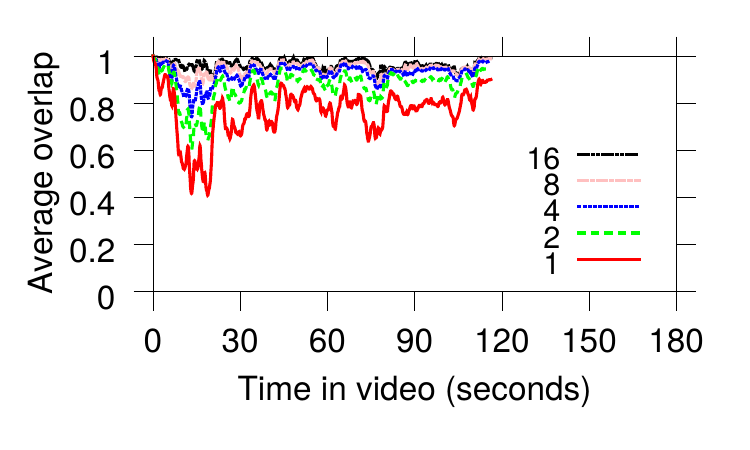}}
  \vspace{-14pt}
  \caption{Average normalized viewport overlap as function of time.
    \revrev{Example videos when using wiewport size $W=90$.}{(Viewport size $W=90$.)}}
  \label{fig:sequence-time-average-90}
  \vspace{-14pt}
\end{figure}

\begin{figure}[t]
\begin{minipage}[t]{0.42\textwidth}
\centering
  \subfigure[Explore]{
    \includegraphics[trim = 3mm 12mm 3mm 4mm, width=0.58\textwidth]{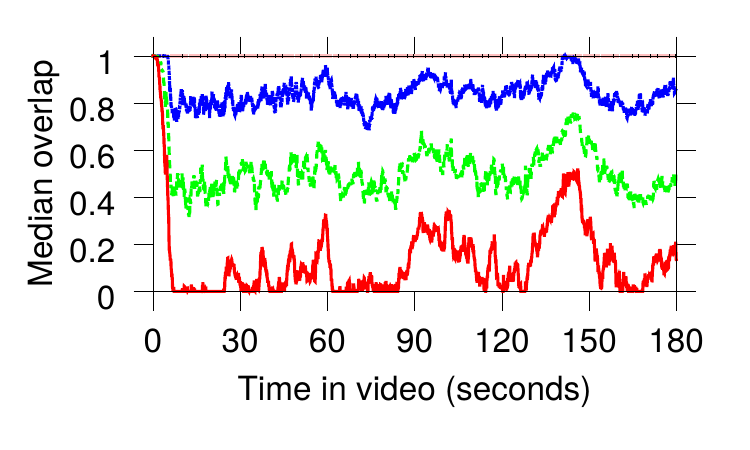}}\\
  \vspace{-12pt}
  \subfigure[Static]{
    \includegraphics[trim = 3mm 12mm 3mm 4mm, width=0.58\textwidth]{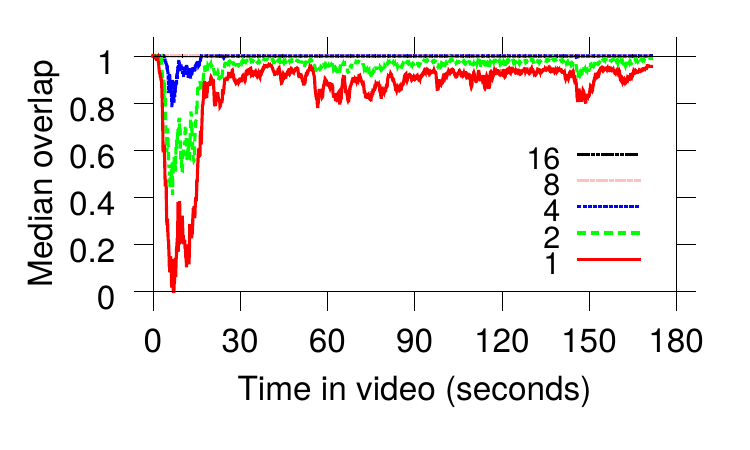}}
  \vspace{-14pt}
  \caption{Median normalized viewport overlap as function of time. (Viewport size $W=90$.)}
  \label{fig:sequence-time-median-90}
  \vspace{-10pt}
\end{minipage}
  \hfill
  \begin{minipage}[t]{0.54\textwidth}
    \centering
    \vspace{1pt}
    \includegraphics[trim = 0mm 0mm 0mm 0mm, width=0.92\textwidth]{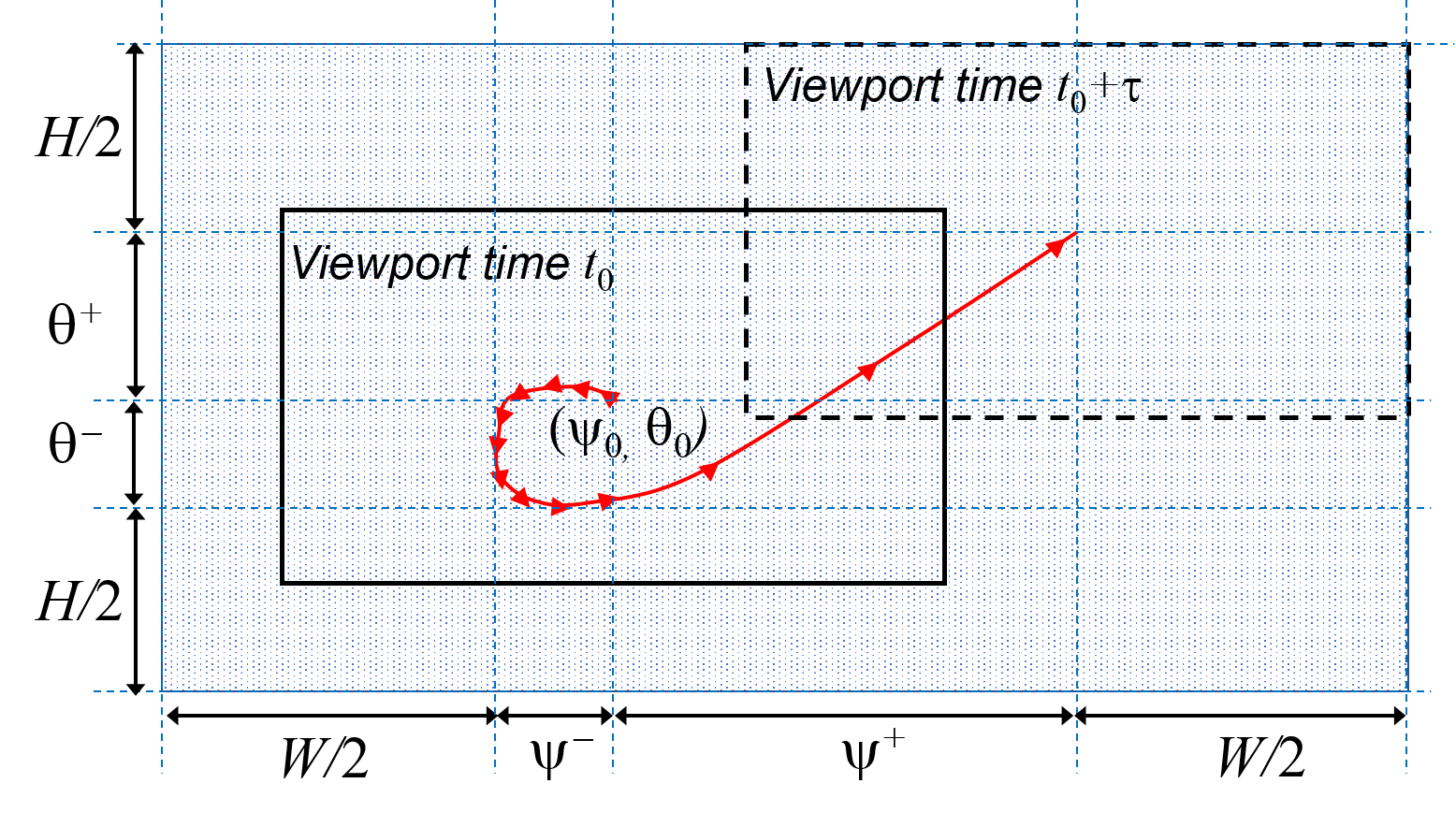}
    \vspace{-12pt}
    \caption{Bounding the maximum
      change in viewing direction
      and 
      \revAA{the relative}{the} 
      viewing field covered by the viewport.}
    \label{fig:metric-cover}
    \vspace{-10pt}
\end{minipage}
\end{figure}

\revNot{\revtwo{}{{\bf Longitudinal evaluation:}}}{}
\revNot{The above observations
\revrev{are also consistent}{also hold}
when considering the normalized overlap observed over time.
\revNot{Figures~\ref{fig:sequence-time-average-90} and~\ref{fig:sequence-time-average-W} show}{Figure~\ref{fig:sequence-time-average-90} shows}
the corresponding}{Figure~\ref{fig:sequence-time-average-90} shows}
timeline plots of the
\revrev{average}{average (over the 1,000 random orderings of viewing sessions)}
normalized viewport overlap.
\revrev{We again note}{Note}
that the
\revrev{largest benefits of}{benefits of}
\revrev{additional requests being visible during exploration}{\revrev{having more}{more} prior video viewings \revrev{occur}{increase} when there is more variability in where users are looking}
(e.g., {\em explore} videos or the beginning of the {\em static} video).
Interestingly, the improvements are even larger
\revrev{when looking at}{for}
\revrev{the median}{median}
overlap, as
\revrev{exemplified}{seen}
in Figure~\ref{fig:sequence-time-median-90}.
The larger median improvements
  show that 
  \revBB{the majority of the}{most} 
  sessions quickly see significant benefits from each additional
  prior
client.
  For example,
  with just four prior clients, in the case of the {\em static} video,
  \revBB{the majority of}{most} 
  clients have 100\% overlap from roughly the 15 second mark.

\cutICPE{
\begin{figure}[t]
  \centering
  \subfigure[Smaller ($W$=60), explore]{
    \includegraphics[trim = 3mm 12mm 3mm 4mm, width=0.24\textwidth]{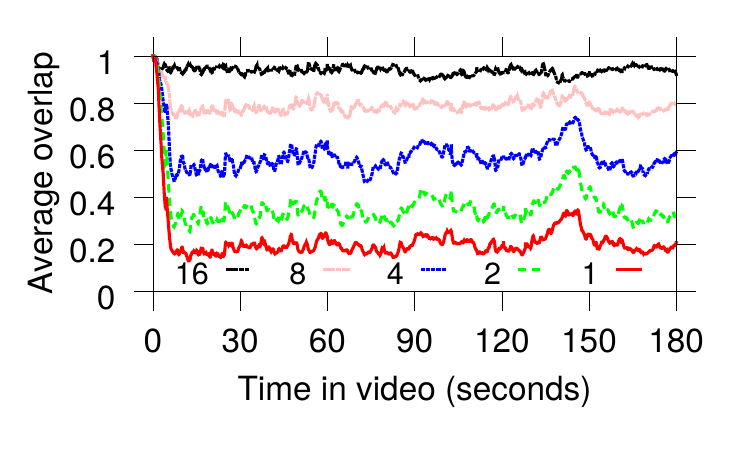}}
  \hspace{-6pt}
  \subfigure[Larger ($W$=120), explore]{
    \includegraphics[trim = 3mm 12mm 3mm 4mm, width=0.24\textwidth]{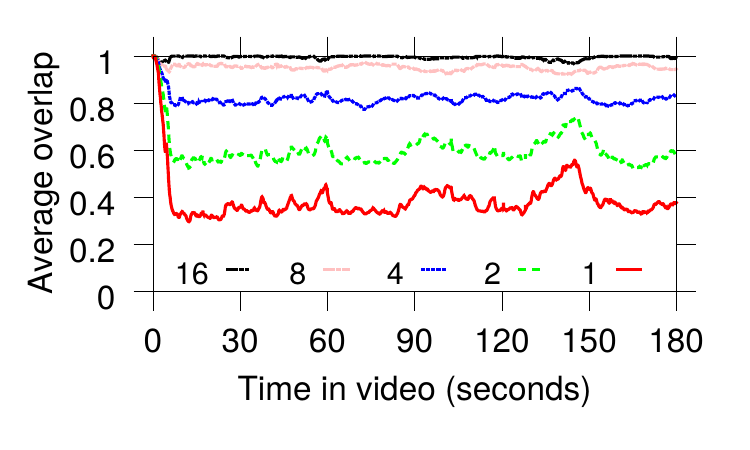}}
  \hspace{-6pt}
  \subfigure[Smaller ($W$=60), static]{
    \includegraphics[trim = 3mm 12mm 3mm 16mm, width=0.24\textwidth]{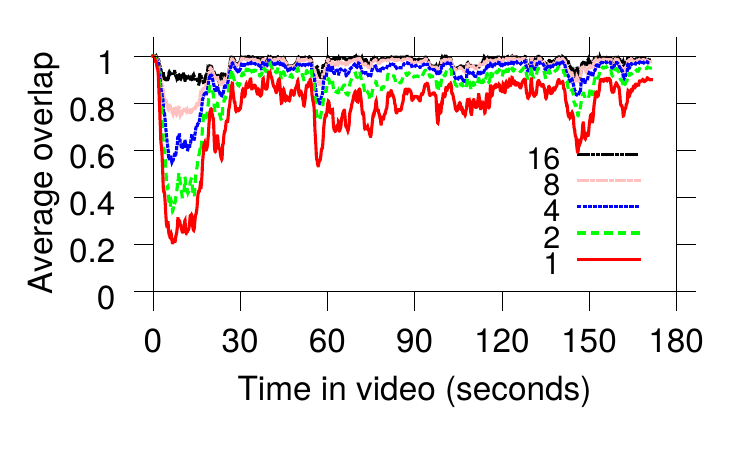}}
  \hspace{-6pt}
  \subfigure[Larger ($W$=120), static]{
    \includegraphics[trim = 3mm 12mm 3mm 16mm, width=0.24\textwidth]{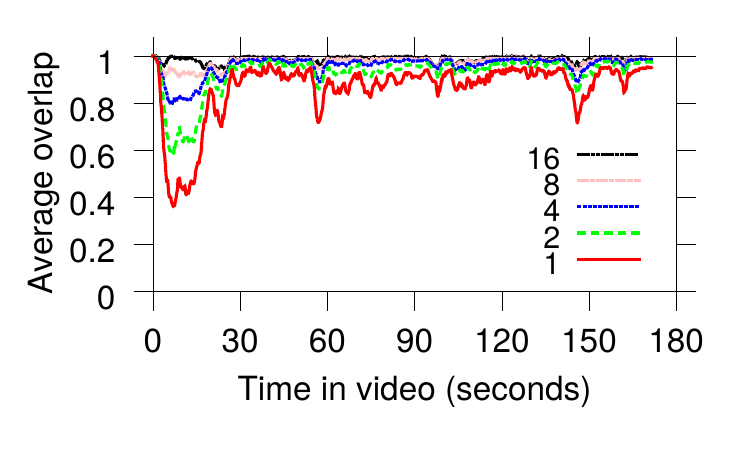}}
  \vspace{-16pt}
  \caption{Average normalized viewport overlap as function of time, with different viewport sizes.}
  \label{fig:sequence-time-average-W}
  \vspace{-12pt}
\end{figure}
}

\section{\revtwo{Impact of Chunking}{Chunk granularity analysis}}\label{sec:chunk-based}

\revrev{When \revrev{comparing caching opportunities,}{considering potential cache performance,} it}{It}
is important to remember that caching
(and video delivery itself) typically is done on a per-chunk basis.
\revrev{Clearly, the}{The}
\revAA{viewport}{viewing direction (and hence also the tiles seen within the viewport)} 
may change during the playback duration of
a chunk,
  resulting in a larger per-chunk viewport cover 
  \revAA{}{(defined next)}
  than the viewport
  at an individual playback point.
  The overlaps between per-chunk viewport covers and those of prior clients are important in caching.  We next
study and report on per-chunk
\revfour{statistics such as the angular
changes in viewing direction
during the
playback duration
of a chunk,
\revrev{\revrev{and the relative viewing area covered by the instantaneous viewport due to these head movement changes.}{per-chunk viewport cover sizes, and finally per-chunk viewport cover overlaps.}}{and per-chunk viewport cover sizes and overlaps.}}{statistics.}

\subsection{\revrev{View direction changes}{Changes in viewing direction}}

\cutICPE{
\revrev{To bound the maximum change in the viewing direction and the total viewing field
covered during the playback of a chunk,
we track all directional changes in the viewing during the playback of each chunk.
This is illustrated in Figure~\ref{fig:metric-cover}.
For simplicity, let us assume}{Figure~\ref{fig:metric-cover} illustrates how the total viewing field covered during the playback period of a chunk,
  and the maximum changes in viewing direction during this time period,
  are determined from the fine-grained head movement data in our dataset.}
\revthree{\revrev{}{We assume in this figure}
  that the playback of the chunk starts
at time $t_0$ and lasts for $\tau$ seconds.}{Here, $t_0$ denotes the time at which the chunk starts playback, and $\tau$ its duration.}
\revrev{Over}{For}
this time period,
\revrev{we then keep track of the small head movement changes in $\psi(t)$ and $\theta(t)$
  that happens at a 10ms granularity, and sum up}{the small head movements that were captured at a 10ms granularity in the trace data are used to calculate}
the maximum accumulated changes to the left
(i.e., $\psi^-$), to the right (i.e., $\psi^+$), upwards (i.e., $\theta^+$), and
downwards (i.e., $\theta^-$),
\revrev{all measured relative}{relative}
to the original viewing direction ($\psi_0,\theta_0$) at time $t_0$.
\revrev{By tracking all viewing direction changes at a 10ms granularity, we avoid missing wraparound effects
and any errors in one particular data points is cancelled by an equally large error in the opposite direction
in the following $\psi(t+\delta)-\psi(t)$ and $\theta(t+\delta)-\theta(t)$ values, where $\delta$ is 10 ms.footnote{To account for wraparound effects
  in yaw, we added (or subtracted 360) to the 10ms changes whenever the change $\psi(t+\delta)-\psi(t)$
  was larger than -180 (and +180) degrees.  If we call each of these change $\Delta\psi_{n}$ and assume that there are $N$ measurements during a chunk,
  then we can easily calculate $\psi^- = \min_{n}[\sum_{i=1}^n \Delta\psi_{i}]$ and $\psi^+ = \min_{n}[\sum_{i=1}^n \Delta\psi_{i}]$.}}{Using fine-grained measurements
  allows us to keep track of wraparound effects and ensures accurate calculation of these values.}
}

\begin{figure}[t]
  \centering
  \subfigure[Explore]{
    \includegraphics[trim = 2mm 12mm 4mm 8mm, width=0.24\textwidth]{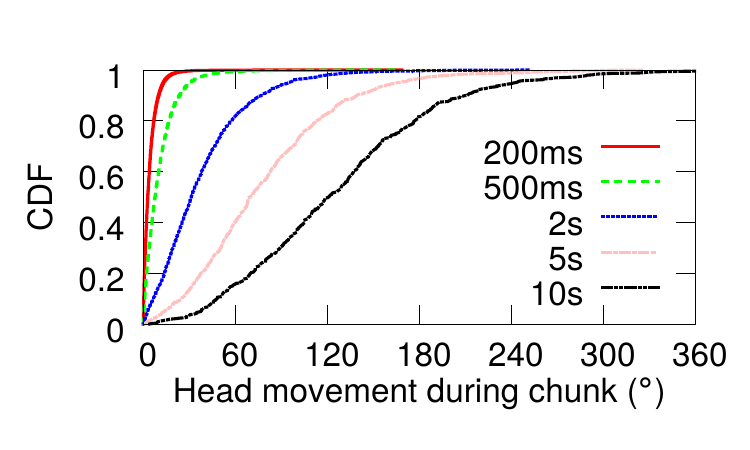}}
  \hspace{-6pt}
  \subfigure[Static]{
    \includegraphics[trim = 2mm 12mm 4mm 8mm, width=0.24\textwidth]{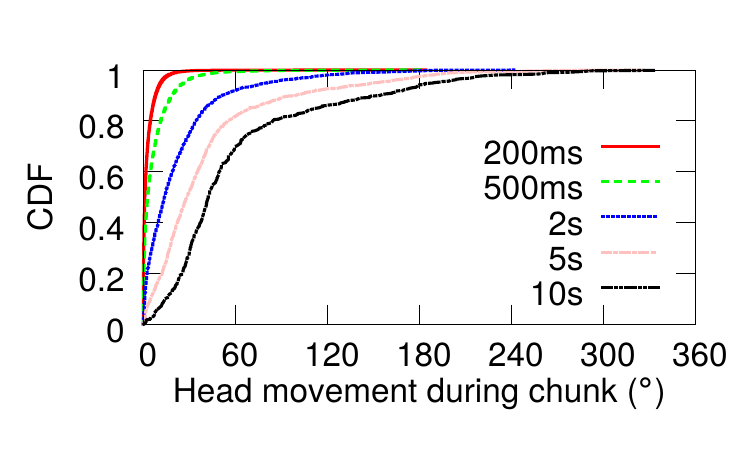}}
  \hspace{-6pt}
  \subfigure[Moving]{
    \includegraphics[trim = 2mm 12mm 4mm 18mm, width=0.24\textwidth]{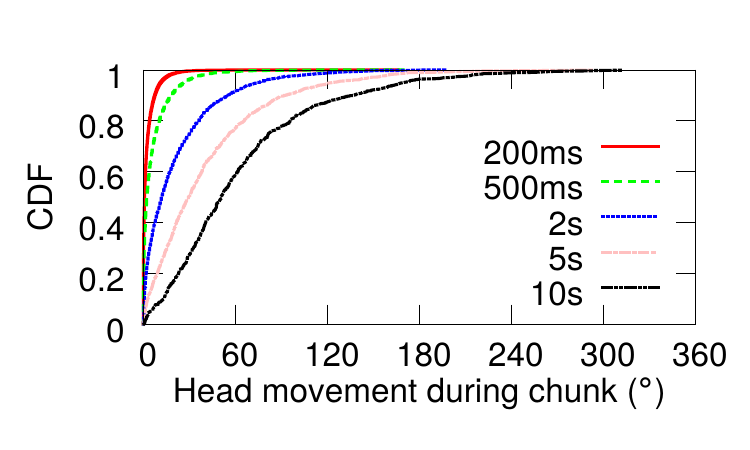}}
  \hspace{-6pt}
  \subfigure[Rides]{
    \includegraphics[trim = 2mm 12mm 4mm 18mm, width=0.24\textwidth]{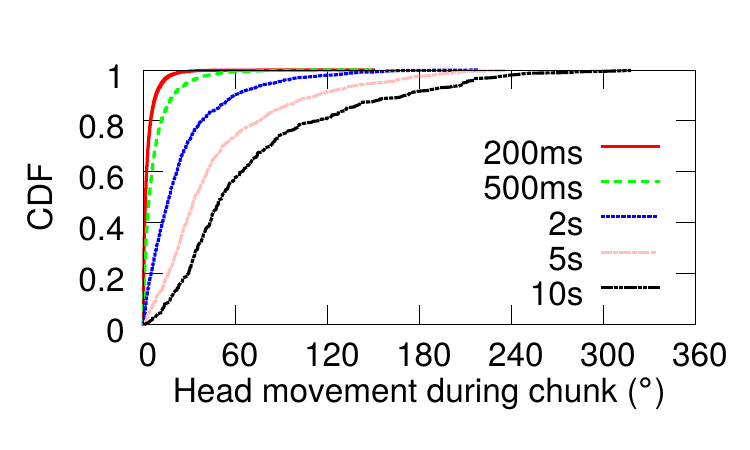}}
  \vspace{-14pt}
  \caption{Impact of chunk duration on the change in viewing angle for
    \revrev{representative example}{the representative}
    videos.}
  \label{fig:chunk-cdf-delta-time}
  \vspace{-12pt}
\end{figure}

\begin{figure}[t]
  \centering
  \subfigure[Yaw + pitch]{
    \includegraphics[trim = 2mm 12mm 4mm 8mm, width=0.24\textwidth]{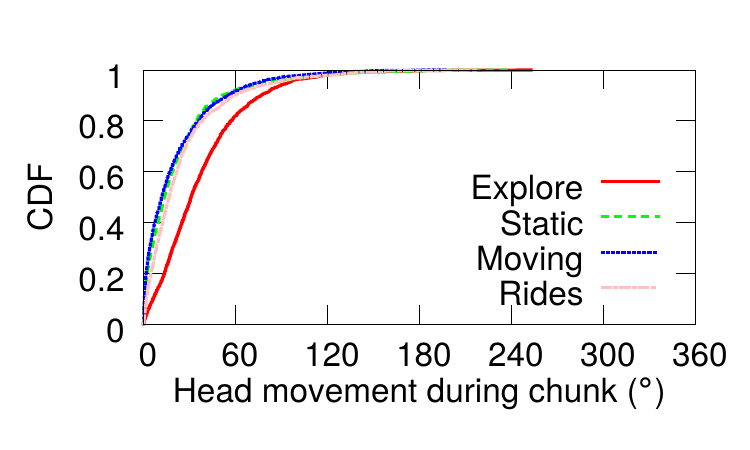}}
  \hspace{-6pt}
  \subfigure[Yaw only]{
    \includegraphics[trim = 2mm 12mm 4mm 8mm, width=0.24\textwidth]{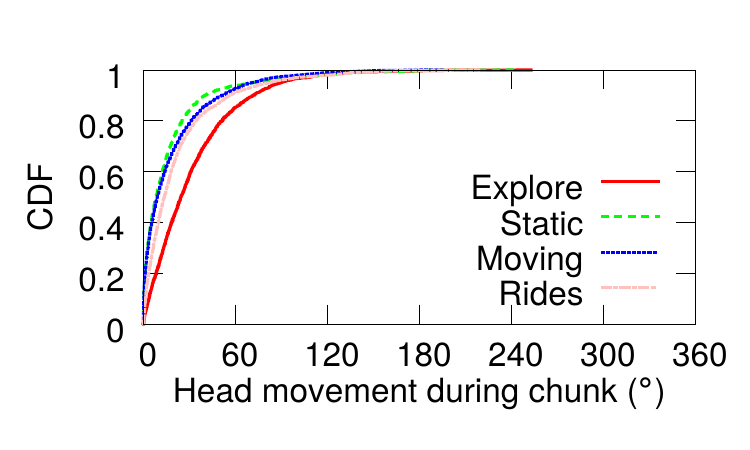}}
    \hspace{-6pt}
  \subfigure[\revAA{}{Pitch only}]{
    \includegraphics[trim = 2mm 12mm 4mm 8mm, width=0.24\textwidth]{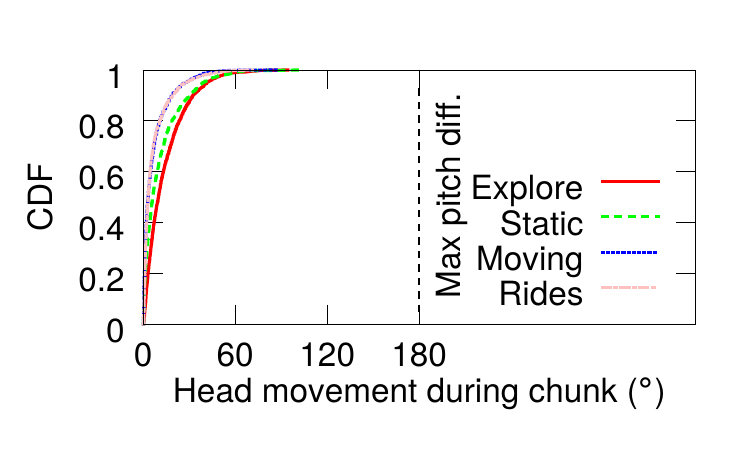}}
   \vspace{-14pt}
  \caption{\revrev{Comparison of which view angle is used for the results shown in
      Figure~\ref{fig:chunk-cdf-delta-time} when using 2s chunks.}{Impact of using only yaw angle rather than yaw + pitch.  (2s chunks.)} \revAA{}{As a reference point we also include pitch (only) movement results.}}
  \label{fig:chunk-cdf-delta-measure}
  \vspace{-12pt}
\end{figure}

\cutICPE{\revtwo{}{{\bf Impact of chunk granularity:}}}
Figure~\ref{fig:chunk-cdf-delta-time} shows CDFs of
\revNot{the}{a}
bound on the maximum viewing direction
\revNot{\revfour{change}{change over a chunk duration,}
given by $\sqrt{(\psi^{+} + \psi^{-})^2 + (\theta^{+} + \theta^{-})^2}$,}{change over a chunk duration, as calculated using the fine-grained measurements in our dataset~\cite{CaEa20-arxiv},}
for the representative
\revrev{example videos. Here, each line represent a different chunk size,
  ranging from small chunk durations of 200ms to very long chunk durations of 10s.}{videos and a range of chunk
  \revfour{durations, from very small (200ms) to very large (10s).}{durations (200ms-10s).}}
As before, the {\em explore} category stands out, with much larger head movements.
However,
\revrev{it is encouraging to see}{note}
that for intermediate chunk durations (e.g., 2s),
the head movements still only cover a small fraction of the view field.
For example,
\revrev{for 80\% of the chunks the maximum head movements for each representative video is}{for the representative videos the maximum viewing direction changes for 80\% of the chunks are}
upper bounded
\revNot{(as per the above formula) by}{by}
57.7$\degree$, 34.5$\degree$, 36.3$\degree$, and 38.7$\degree$,
\revfour{\revrev{The differences are consistent (and even more extreme for the {\em static} video) when only considering the yaw angle.   For example,}{When considering only the yaw angle,}
the corresponding 80\% values are: 51.5$\degree$, 25.3$\degree$, 31.3$\degree$, 34.5$\degree$.
\revrev{Figure~\ref{fig:chunk-cdf-delta-measure}(a) and~\ref{fig:chunk-cdf-delta-measure}(b)}{Figure~\ref{fig:chunk-cdf-delta-measure}}
presents a direct comparison of
the CDFs 
when using both yaw and pitch
(as in Figure~\ref{fig:chunk-cdf-delta-time}) and when using only yaw,
\revrev{respectively, for}{for}
the case of
\revrev{a 2s chunk size.}{2s chunks.}
\revrev{As noted, the differences in the CDFs of {\em static} an {\em explore} are substantial.
However, since pitch movements in general are small,}{As can be seen,}
the results are
\revrev{generally relatively consistent}{similar in nature}
regardless of whether only the maximum yaw change is used
or the bounding angle (across both directions) is used.}{As shown in Figure~\ref{fig:chunk-cdf-delta-measure} for the case of 2s chunks,
the results are similar regardless of whether both yaw and pitch are considered (as in Figure~\ref{fig:chunk-cdf-delta-time}) or only the maximum yaw change.}
\revAA{}{Here, we break down the total movement (Figure~\ref{fig:chunk-cdf-delta-measure}(a)) into yaw only (Figure~\ref{fig:chunk-cdf-delta-measure}(b)) and pitch only (Figure~\ref{fig:chunk-cdf-delta-measure}(c)). The pitch movements are again more restricted and contribute significantly less to the total movements.}
\revBB{}{There is 
a high Pearson correlation between
the total head movement over a chunk duration
and the yaw movement only
(e.g., 0.961, 0.957, 0.978, 0.979 for the four videos when using 2s chunks).}

\subsection{Viewport-based metrics}

\revsix{\revNot{\revtwo{}{{\bf Metric:}}}{}}{{\bf Per-chunk viewport cover:}}
\revrev{\revrev{Naturally, the}{The}
observed head movements during playback of a chunk
will impact the total viewing area that is covered during this playback duration.
To measure how much this inflates the area that actually is viewed of a chunk,
we use the relative viewports covered during the playback of a chunk.
More specifically,
referring back to Figure~\ref{fig:metric-cover},}{To measure the total viewing area that is included within a
  \revrev{client's}{user's}
  viewport for at least some portion of a chunk's playback period,}
\revsix{\revfive{we calculate a bounding box of the head movements
\revrev{during the playback of the chunk}{during this time period}
  using the maximum changes in each direction
  \revrev{(i.e., $\psi^-, \psi^+, \theta^-, \theta^+$)
and upper bound the potential viewing area that
  has been covered during this time period as:}{(i.e., $\psi^-, \psi^+, \theta^-, \theta^+$).
  We define this bounding box as the {\em per-chunk viewport cover}, with size given by:}
  $\max[360, W + \psi^- + \psi^+] \times \max[180, H + \theta^- + \theta^+]$.
  Finally, we calculate the normalized
  \revrev{{\em per-chunk viewport cover}}{per-chunk viewport cover size}
  by dividing
  \revrev{this area by}{by}
  $W \times H$.
  For the sliced viewport (ignoring pitch), the
  \revrev{calulation simply reduced to}{calculation reduces to simply}
  $\frac{1}{W}\max[360, W + \psi^- + \psi^+]$.}{we calculate a bounding box of the head movements
    during this time period, which we term the {\em per-chunk viewport cover}.
    In the following, we report per-chunk viewport covers normalized by the total size of the viewport.}}{we calculate
  a bounding box of the head movements during this time period using fine-grained measurements from our dataset.
  \revBB{In particular, we}{We} 
  define the bounding box as the {\em per-chunk viewport cover}, with size given by:
  $\max[360, W + \psi^- + \psi^+] \times \max[180, H + \theta^- + \theta^+]$, where $\psi^-, \psi^+, \theta^-, \theta^+$
  are the maximum changes in each direction over the full playback duration of the chunk.
  In the following, we report per-chunk viewport covers normalized by the total size of the viewport
  (i.e., we divide by $W \times H$).
  For the sliced viewport (ignoring pitch), the
  calculation reduces to simply $\frac{1}{W}\max[360, W + \psi^- + \psi^+]$.}

\begin{figure}[t]
    \centering
    \subfigure[Yaw + pitch]{
      \includegraphics[trim = 0mm 4mm 0mm 14mm, width=0.4\textwidth]{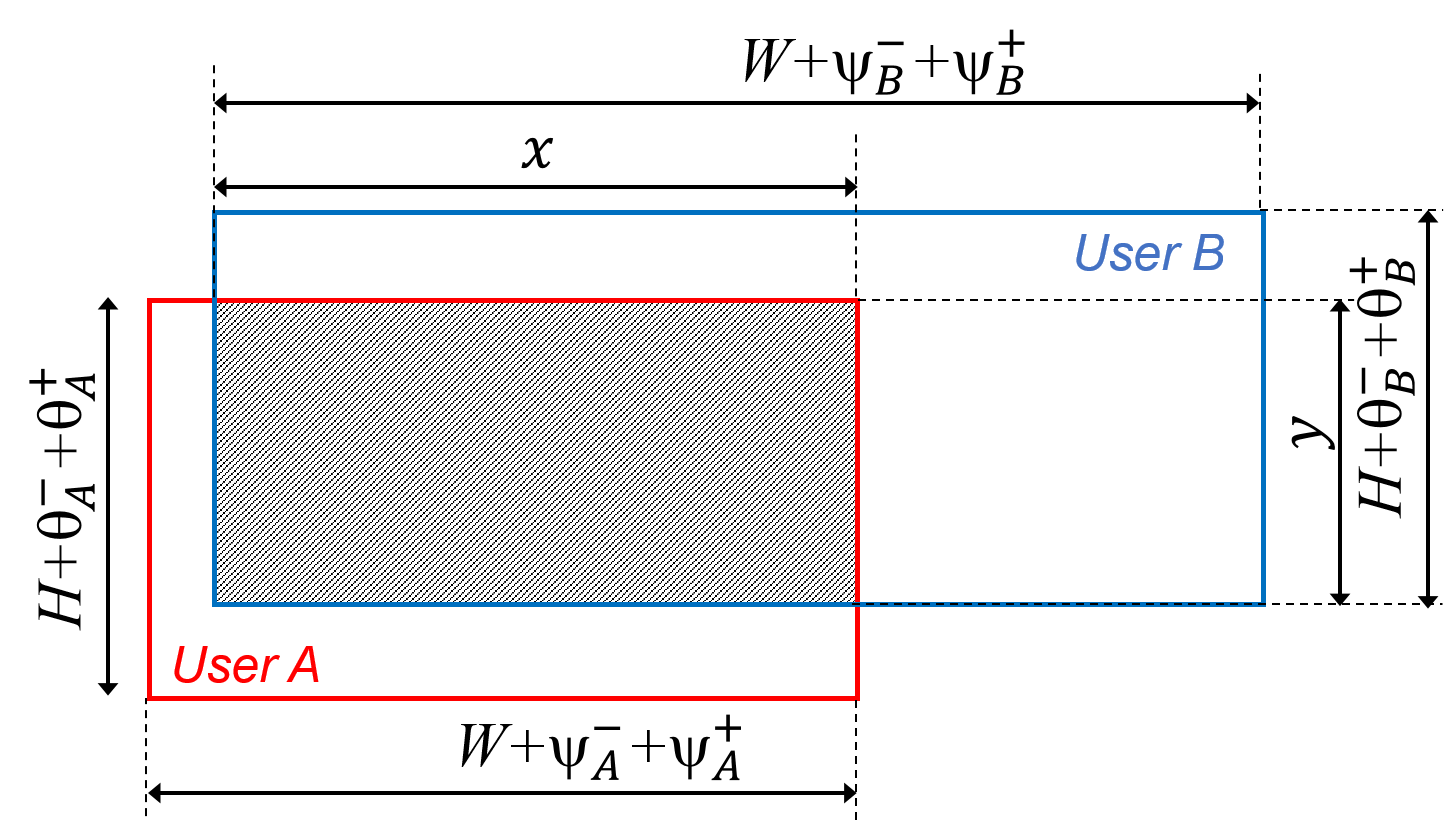}}
    \hspace{-12pt}
    \subfigure[Yaw only]{
      \includegraphics[trim = 0mm 4mm 0mm 14mm, width=0.3\textwidth]{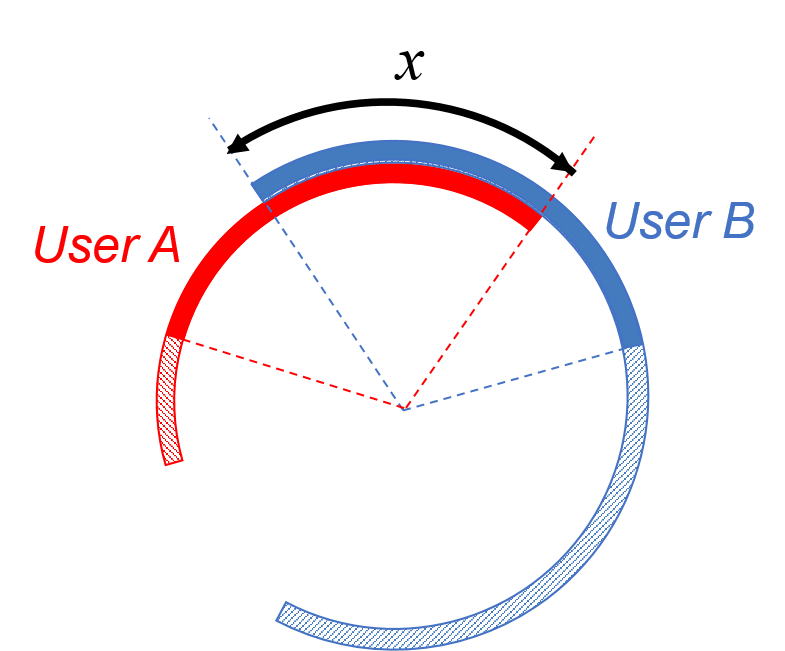}}
    \vspace{-14pt}
    \caption{\revfour{Notation for pairwise per-chunk cover overlap metrics.}{Pairwise per-chunk cover overlap.}}
    \label{fig:metric-cover-overlap}
    \vspace{-12pt}
\end{figure}

\revsix{}{{\bf Pairwise cover overlap:}
  We
next combine our
techniques for analysis of pairwise viewport overlap and for determining per-chunk viewport covers,
to measure the pairwise overlap in per-chunk viewport cover.
  Figure~\ref{fig:metric-cover-overlap} illustrates how this metric is calculated
  for
  two
  users  
  A and B for a particular chunk.
  Here, 
  \revBB{we have assumed that user}{user} 
  A has the
  per-chunk viewport cover
  illustrated by the red box,
  and user B's corresponding viewport cover
  is given
  by the blue box;
  both bounding boxes calculated as described above.
  To calculate the overlap $xy$ when 
  \revBB{taking into account}{considering} 
  both dimensions (Figure~\ref{fig:metric-cover-overlap}(a))
  or $x$ when 
  \revBB{taking into account}{considering} 
  only the yaw angle (Figure~\ref{fig:metric-cover-overlap}(b)),
  we
  extend the methodology used for calculating overlapping viewports to account for
  the two bounding boxes 
  \revAA{(capturing the individual viewport cover during the playback duration of the chunk) to be}{being} 
  of different size.
  A case-based analysis is used for this extension
  (similar to for the yaw-only case shown in Figure~\ref{fig:multi-user-cover-overlap}).}

\subsection{\revrev{Relative viewports covered}{Per-chunk viewport cover}}
  
\begin{figure}[t]
  \begin{minipage}[t]{0.34\textwidth}
  \centering
  \subfigure[120$\times$67.5]{
    \includegraphics[trim = 2mm 12mm 4mm 6mm, width=0.80\textwidth]{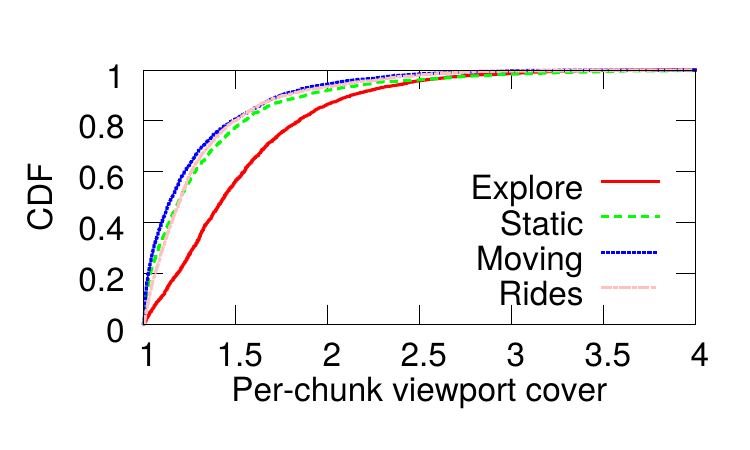}}
\subfigure[90 full]{
    \includegraphics[trim = 2mm 12mm 4mm 20mm, width=0.80\textwidth]{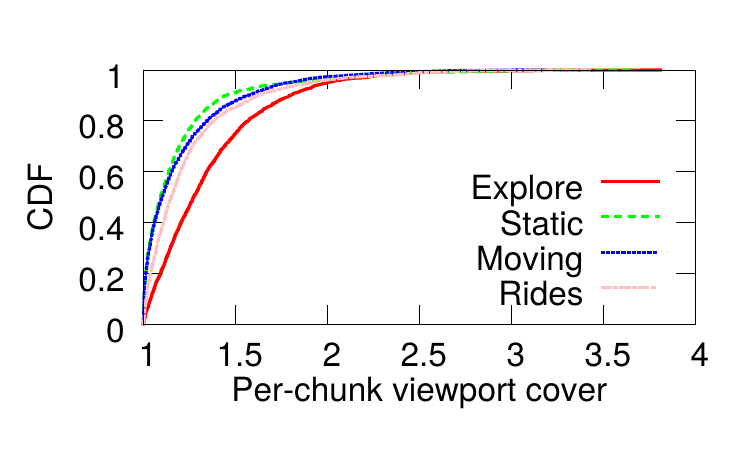}}
  \vspace{-14pt}
  \caption{\revrev{The normalized}{Normalized} per-chunk viewport
    \revrev{cover when using 2 second long chunks.}{cover size.  (2s chunks.)}}
  \label{fig:cover-cdf}
  \vspace{-10pt}
 \end{minipage}
  \hfill
  \begin{minipage}[t]{0.6\textwidth}
  \centering
\vspace{4pt}
  \includegraphics[trim = 0mm 16mm 0mm 0mm, width=1\textwidth]{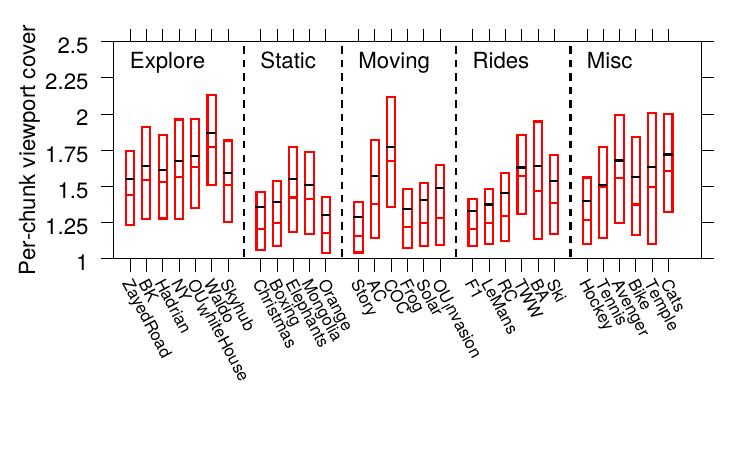}
  \vspace{-10pt}
  \caption{\revrev{The normalized}{Normalized} per-chunk viewport cover \revrev{}{size} for each video when using a 120$\times$67.5
    \revrev{viewport and 2 second long chunks.}{viewport.  (2s chunks.)}}
  \label{fig:cover-wiskar-relOverlap}
  \vspace{-10pt}
  \end{minipage}
\end{figure}

\revNot{\revtwo{}{{\bf Results for representative videos:}}}{}
Figure~\ref{fig:cover-cdf} presents CDFs of the normalized per-chunk viewport cover
\revrev{}{size}
for each of the four representative videos,
\revfour{\revrev{when using 2 second long chunks with four different viewports.}{for four different viewport sizes, and 2 second chunks.}
 When interpreting these results,
 \revthree{it should be noted}{note}
 that the
 maximum theoretic
 \revrev{cover during any chunk, when using each of these four viewports, are:}{normalized cover sizes with these viewport sizes are}
  8, 14.2, 3, and 4, respectively.
  While we do see a few extreme values close to the theoretic maximums
  for the two sliced viewports
  \revrev{(i.e., 120 full and 90 full in Figures~\ref{fig:cover-cdf}(c) and~\ref{fig:cover-cdf}(d)),}{(120 full and 90 full),}
  the coverage is typically much smaller.
  For example, with the sliced 90 full (with a theoretic maximum of 4),}{for 2 second chunks, and two viewport sizes (120$\times$67.5 and 90 full)
  with maximum theoretic normalized cover sizes of 8 and 4, respectively.
  Note that the coverage is typically much smaller than the theoretic maximum.  For example, with the sliced 90 full viewport,}
80\% of the chunks have a normalized cover size of
\revrev{}{at most}
1.57, 1.28, 1.35, and 1.38, respectively,
for the four
\revrev{video categories.}{representative videos.}
These small cover sizes suggest that tiles could 
\revAA{}{indeed}
fruitfully be prioritized 
\revAA{}{(by a client)}
on a per-chunk basis
since a significant portion of the potential viewing area is not viewed during the playback of a chunk.

\revtwo{}{{\bf Results for other videos:}}
The above observations are relatively consistent across the videos in each
\revfour{category.
Figure~\ref{fig:cover-wiskar-relOverlap} shows example distribution statistics for the
normalized per-chunk viewport cover size for each
\revrev{individual video}{video}
when using}{category, as shown in Figure~\ref{fig:cover-wiskar-relOverlap} for}
a 120$\times$67.5
\revrev{viewport. Again, remember that for this case the maximum theoretic per-chunk viewport cover is 8.}{viewport (for which the maximum theoretic normalized cover size is 8).}
To improve readability,
\revrev{in Figure~\ref{fig:cover-wiskar-relOverlap}, we leave out}{we omit whiskers for}
minimum (always 1) and
\revfour{maximum (larger than 3.5 for all videos except 3 {\em static} videos, 1 {\em moving} video, and 2 {\em ride} videos).}{maximum.}
While the relative differences  between the
\revrev{categories}{video categories in these results}
(e.g., looking at averages or medians)
are less apparent than 
\revAA{when looking at}{for} 
the pairwise differences in viewing directions
(e.g., Figures~\ref{fig:pairwise-wiskar-deltaTot} and~\ref{fig:pairwise-wiskar-relOverlap}),
we typically see the largest head movements associated with videos in the {\em explore} and {\em miscellaneous} categories
and the smallest in the {\em static} category (even when taking into account that these videos often have an initial exploration phase).
Furthermore,
we note that the normalized per-chunk viewport
\revrev{cover is less than two (viewports)}{cover size is less than two (i.e., less than double the viewport size)}
for more than 75\% of the chunks
\revfour{for almost all videos; Waldo ({\em explore}) and COC ({\em moving}) being the two exceptions.}{for all but two videos:
Waldo ({\em explore}) and COC ({\em moving}).}

\begin{figure}[t]
  \begin{minipage}[t]{0.49\textwidth}
  \centering
  \includegraphics[trim = 0mm 14mm 0mm 9mm, width=0.98\textwidth]{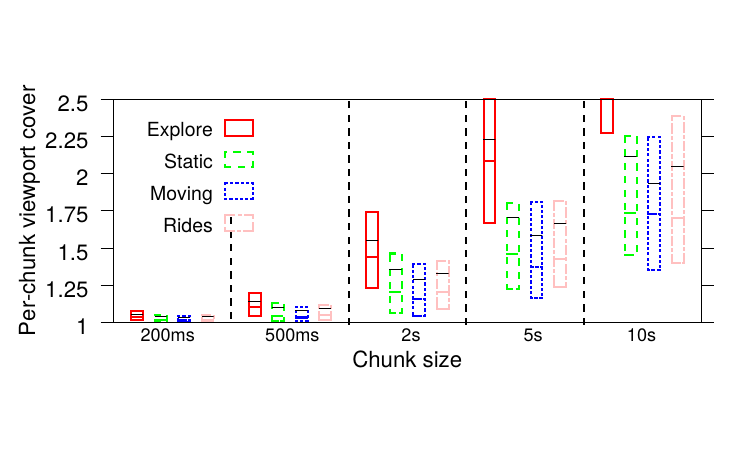}
  \vspace{-8pt}
  \caption{Impact of chunk duration on normalized per-chunk viewport cover size.  (Viewport size 120$\times$67.5.)}
  \label{fig:cover-wiskar-chunksize}
  \vspace{-10pt}
\end{minipage}
\hfill
\begin{minipage}[t]{0.49\textwidth}
  \centering
  \includegraphics[trim = 0mm 14mm 0mm 9mm, width=0.98\textwidth]{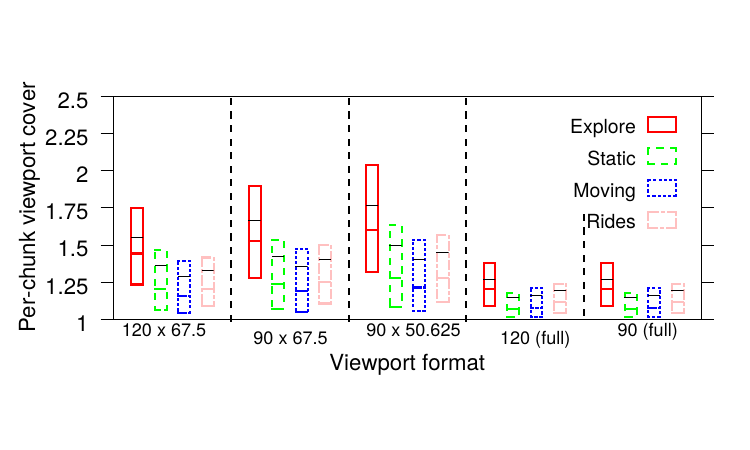}
  \vspace{-8pt}
  \caption{Impact of viewport format on normalized per-chunk viewport cover size. (Chunk duration of 2s.)}
  \label{fig:cover-wiskar-viewport}
  \vspace{-10pt}
\end{minipage}  
\end{figure}

\revsix{\revfive{\revfive{\revtwo{}{{\bf Chunk duration and viewport dependencies:}}}{{\bf Impact of chunk duration:}}}{}
\revfive{\revfour{Of course, the
\revrev{per-chunk viewport cover}{cover size}
depends on both the viewport of consideration and
the chunk durations.  Figures~\ref{fig:cover-wiskar-viewport}
and~\ref{fig:cover-wiskar-chunksize} show}{Figures~\ref{fig:cover-wiskar-viewport} and \ref{fig:cover-wiskar-chunksize} show}}{Figure~\ref{fig:cover-wiskar-chunksize} shows}
the impact of
\revfive{\revfour{these two factors}{the chunk duration and viewport}}{the chunk duration}
on the normalized
\revrev{per-chunk viewport cover, for the representative videos of each category.}{per-chunk viewport cover size for the representative videos.}
\revfour{For example, focusing}{Focusing}
on the 75-percentile values,
except for the
\revfive{three cases of
(i) {\em explore} using a small viewport of 90$\times$50.625,
(ii) {\em explore} using a chunk duration of 5 or more seconds,
  and (iii)}{cases of (i) {\em explore} using a chunk duration of 5 or more seconds, and (ii)}
extremely long duration chunks of 10 seconds,
the normalized per-chunk viewport cover size is again consistently
below 2.
In general,
\revfour{for the}{for}
   {\em static}, {\em moving}, and {\em rides}
\revfour{\revrev{sees substantially smaller head movements,}{videos the head movements are substantially smaller,}
suggesting that prioritizing of tiles may be most suitable for these
\revrev{video categories.}{categories.}}{this
  cover size
  is substantially smaller.}}{{\bf Impact of chunk duration:}
        Figure~\ref{fig:cover-wiskar-chunksize} shows the impact of the chunk duration on the normalized
        per-chunk viewport cover size for the representative videos.
        Focusing on the 75-percentile values,
        except for the cases of (i) {\em explore} using a chunk duration of 5 or more seconds, and (ii)
        extremely long duration chunks of 10 seconds,
        the normalized per-chunk viewport cover size is again consistently  below 2.
        In general,
        for {\em static}, {\em moving}, and {\em rides}
        this cover size is substantially smaller.
        Again, the small normalized per-chunk viewport cover values
suggest that prioritizing of tiles may be most suitable for these categories.
These observations are relatively consistent 
        \revAA{also across}{across} 
        different viewport 
        \revAA{formats.  Figure~\ref{fig:cover-wiskar-viewport} illustrates the impact
        of the viewport format.}{formats, as illustrated in Figure~\ref{fig:cover-wiskar-viewport}.}  
        \revAA{Here, we note}{Note} 
        that the 75-percentile values again remain
        below 2
        except for the representative {\em explore} video
        using the smallest of the
        considered viewports (i.e., 90$\times$50.625).}

\begin{figure}[t]
%
  \centering
  \subfigure[\revrev{Per per-chunk cover}{Relative to bounding box} (120$\times$67.5)]{
    \includegraphics[trim = 2mm 6mm 4mm 0mm, width=0.35\textwidth]{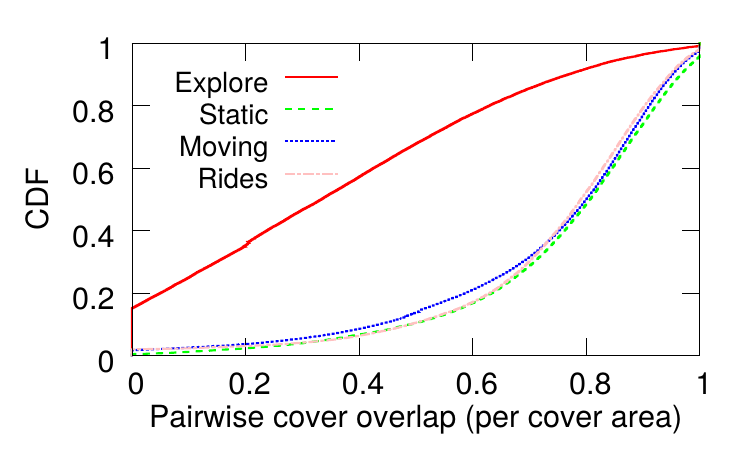}}
  \subfigure[\revrev{Per viewport}{Relative to viewport} (120$\times$67.5)]{
    \includegraphics[trim = 2mm 6mm 4mm 0mm, width=0.35\textwidth]{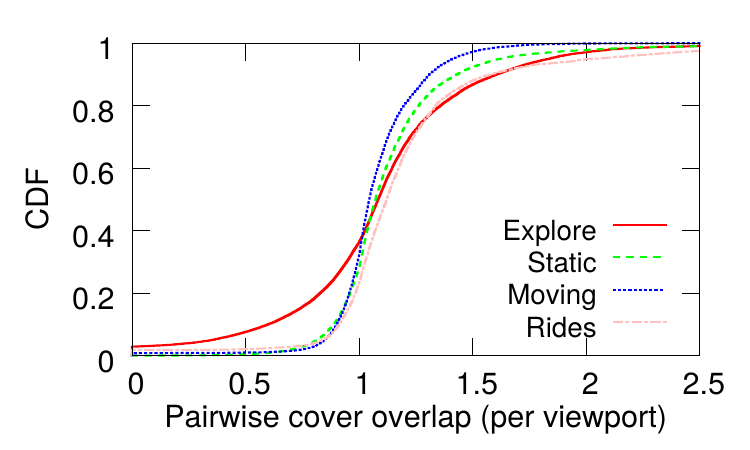}}
  \vspace{-10pt}
  \subfigure[Different viewports]{
    \includegraphics[trim = 0mm 15mm 0mm 19mm, width=0.62\textwidth]{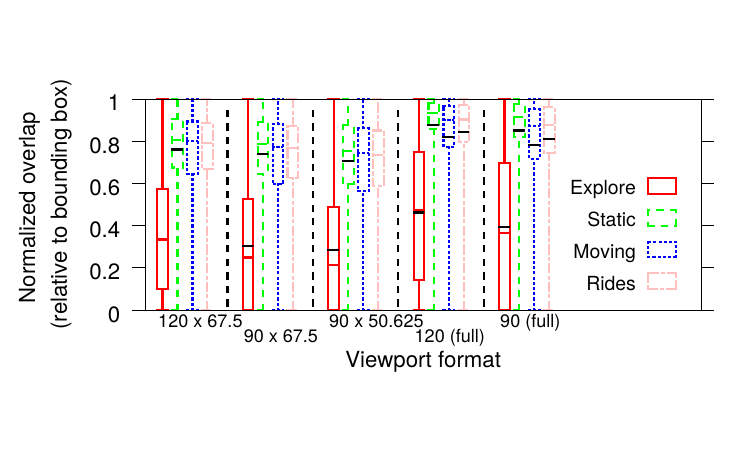}}
  \vspace{-6pt}
  \caption{Pairwise overlap in per-chunk viewport cover.}
  \label{fig:pairwise-chunk}
  \vspace{-10pt}
\end{figure}


\subsection{Pairwise cover overlap}

{\bf Results for representative videos and example viewports:}
The first two sub-figures in Figure~\ref{fig:pairwise-chunk} show the pairwise overlap in per-chunk viewport cover
normalized relative to the
size of the user's bounding box
(i.e., $\frac{xy}{(W+\psi^{+}+\psi^{-})(H+\theta^{+}+\theta^{-})}$) and
relative to the
viewport
size (i.e., $\frac{xy}{WH}$), respectively,
when using a 120$\times$67.5 viewport, and the third sub-figure shows
summary statistics (of the first kind) also for other viewports.
\revAA{}{The two metrics capture the pairwise similarities after accounting for the head movement variability seen over the playback duration of a chunk. The first metric uses the bounding box to normalize the overlaps against the head-movement variations themselves, while the second metric uses the viewport size itself (which always is the same size, regardless of head movements).  Per their definition, the first metric can have values no greater than 1, whereas the second metric can.}
We note that the {\em explore} category stands out even more than we have seen before,
when considering the overlap normalized
\revrev{per per-chunk cover}{relative to the bounding box size}
(Figures~\ref{fig:pairwise-chunk}(a) and~\ref{fig:pairwise-chunk}(c)).
For example,
referring to Figure~\ref{fig:pairwise-chunk}(a),
\revrev{while the three other categories observe a}{while there is at least a}
60\% overlap in cover for 79-83\% of the
\revrev{chunks}{chunks for the {\em static}, {\em moving}, and {\em rides} videos}
(83\%, 79\%, and 83\%, respectively),
the corresponding fraction of chunks is only 23\% for the {\em explore} video.
\revAA{}{(Note that these overlap coverage values are the complementary values to those shown in the figure.)}
\revrev{This captures that the}{This reflects the fact that the videos in the}
{\em explore} category typically
\revrev{sees}{have}
both larger head movements during a chunk
\revrev{duration and that the}{duration, and larger}
pairwise viewing direction differences
\revrev{(including during the chunk playback period) is larger.}{(including during the chunk playback period).}
Furthermore,
\revfour{due to the generally larger head movements associated with {\em explore}, the}{the}
variations in the absolute overlap
(e.g., as normalized relative to the viewport size, as in Figure~\ref{fig:pairwise-chunk}(b))
are much greater for the {\em explore} video,
and conversely, the variations are smallest for the {\em static} video.

\subsection{Request sequence analysis}

\begin{figure}[t]
\cutICPE{
  \centering
  \subfigure[Explore]{
    \includegraphics[trim = 2mm 12mm 4mm 8mm, width=0.24\textwidth]{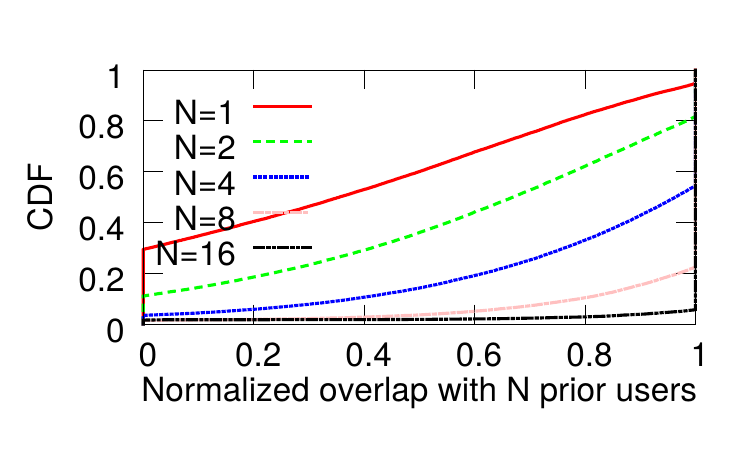}}
  \hspace{-6pt}
  \subfigure[Static]{
    \includegraphics[trim = 2mm 12mm 4mm 8mm, width=0.24\textwidth]{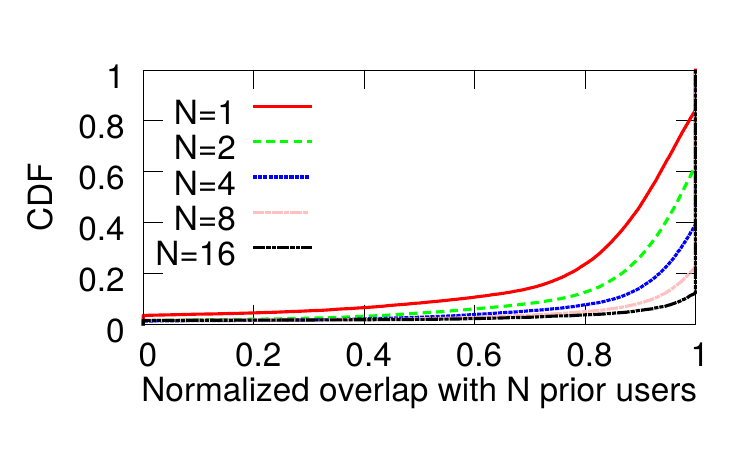}}
  \hspace{-6pt}
  \subfigure[Moving]{
    \includegraphics[trim = 2mm 12mm 4mm 18mm, width=0.24\textwidth]{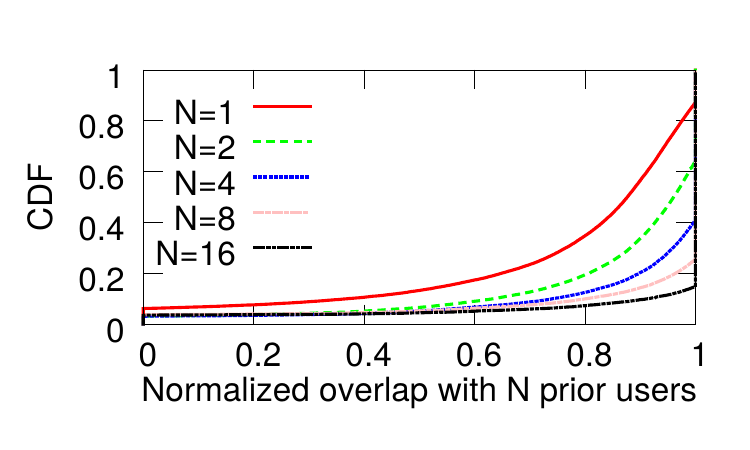}}
  \hspace{-6pt}
  \subfigure[Rides]{
    \includegraphics[trim = 2mm 12mm 4mm 18mm, width=0.24\textwidth]{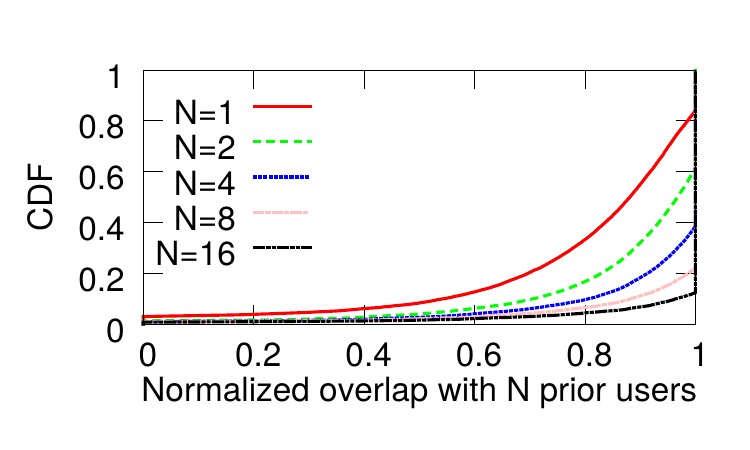}}
  \vspace{-14pt}
  \caption{Chunk-based CDFs of the normalized cover overlap. (Viewport size $W=90$.)}
  \label{fig:chunk-sequence-cdf-average-90}
  \vspace{-4pt}
  }
  \centering
  \subfigure[Explore]{
    \includegraphics[trim = 4mm 14mm 4mm 2mm, width=0.24\textwidth]{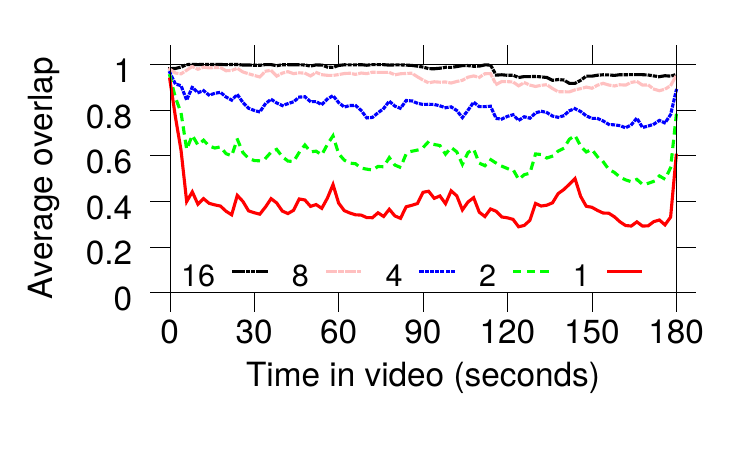}}
  \hspace{-6pt}
  \subfigure[Static]{
    \includegraphics[trim = 4mm 14mm 4mm 2mm, width=0.24\textwidth]{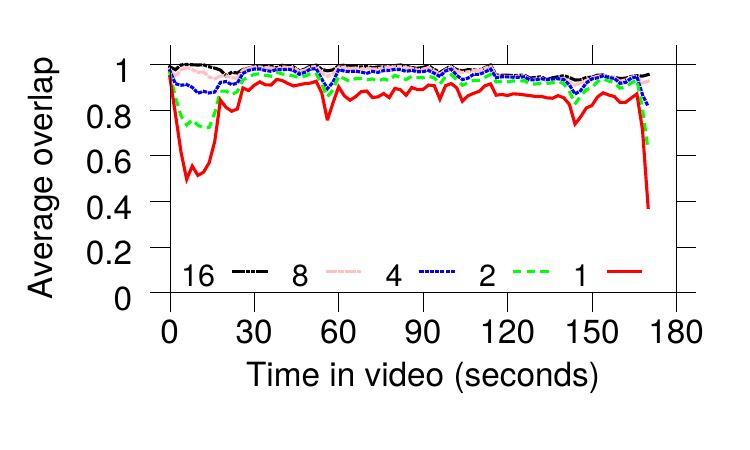}}
  \hspace{-6pt}
  \subfigure[Moving]{
    \includegraphics[trim = 4mm 14mm 4mm 16mm, width=0.24\textwidth]{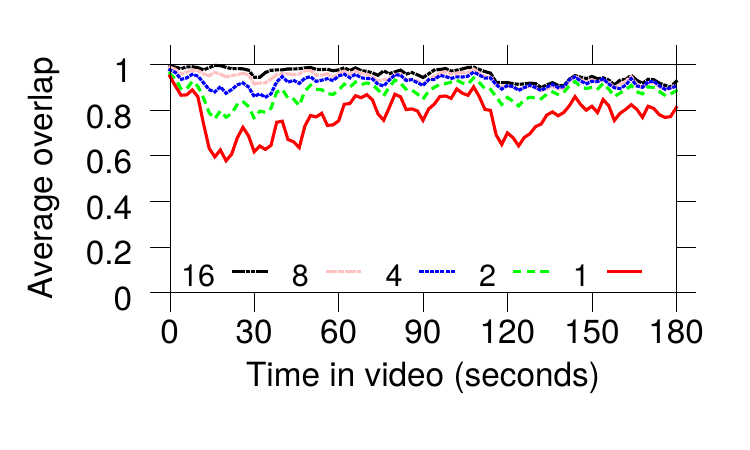}}
  \hspace{-6pt}
  \subfigure[Rides]{
    \includegraphics[trim = 4mm 14mm 4mm 16mm, width=0.24\textwidth]{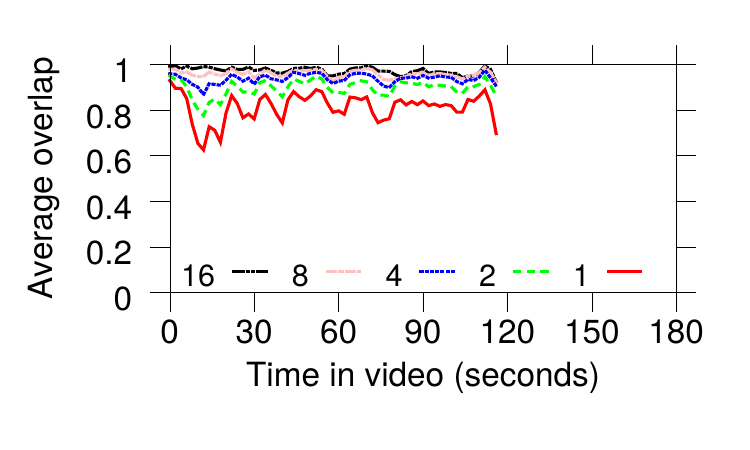}}
  \vspace{-14pt}
  \caption{Chunk-based time plot of the average normalized cover overlap. (Viewport size $W=90$.)}
  \label{fig:chunk-sequence-time-average-90}
  \vspace{-12pt}
\end{figure}

\revfour{Similar to the prior sub-section, we}{We}
next extend our analysis of the
\revrev{overlap that a client sees with $N$ prior clients watching the same video}{overlap with the aggregate
  view cover from $N$ prior
  \revrev{client}{user}
  viewings of the same video}
to account for chunk boundaries.
Throughout this section we use 2 second chunks,
a sliced 90$\degree$ viewport, and normalize the reported cover overlap
\revrev{with the}{relative to the size of the}
cover of the
\revrev{client}{user}
of consideration.
\revNot{In general, the distribution statistics of the overlap between the
 current
 \revrev{client's}{user's}
 per-chunk cover and prior
 \revrev{clients'}{users'}
 cover for the same chunk 
 (Figure~\ref{fig:chunk-sequence-cdf-average-90}(a))
 are similar to the corresponding
 \revrev{instantaneous overlaps}{statistics for individual playback points}
 (Figure~\ref{fig:sequence-cdf-90}).
 However,
 \revfour{we have observed that the}{the}
 larger head movements and bigger differences in viewing directions
 associated with the {\em explore} video result in even greater
 \revrev{pe-client gains in overlaps.}{gains in overlap as the number of prior
  \revrev{clients}{users}
  increases.}
In fact, with 16 prior user viewings
the overlap is greater than 99\% (of the user's cover)
for 94.7\% of the chunks for the {\em explore} video,
compared to 88.8\%, 86.0\%, and 88.5\% for the other videos.
\revAA{}{These results suggest that significant cache hit rates may be achievable already after a limited number of users, at least if they experience similar bandwidth conditions.}
 These observations are also apparent when considering the overlaps seen across the playback
 durations of the example
 videos (Figure~\ref{fig:chunk-sequence-time-average-90}).}{Figure~\ref{fig:chunk-sequence-time-average-90} shows
  the overlaps across the playback durations of the example videos.
  Note the larger overlaps compared to those in Figure~\ref{fig:sequence-time-average-90},
  although the qualitative differences among the results for the representative videos are quite similar.}
    \revsix{}{Note the larger overlaps compared to those in Figure~\ref{fig:sequence-time-average-90},
    although the qualitative differences among the results for the representative videos are quite similar.}
In fact, during the first 120 seconds of the {\em explore} video and
the initial explore phase of the {\em static} video the
average overlaps when there are $N$=8 and $N$=16 prior
\revrev{clients}{users}
are close to one.
In general, however, the overlaps when there are fewer prior
\revrev{clients}{users}
(e.g., $N$=1, $N$=2, and $N$=4 curves)
are greater when
\revrev{clients}{users}
are less exploratory (e.g., with {\em moving}, {\em rides},
and after the initial exploratory phase of the {\em static} video).
\revtwo{}{These chunk-level results again highlight important differences in the caching
  opportunities that different video categories
  present, and that videos of some categories
  (e.g., {\em static}) may require different optimizations
for the initial (exploratory) phase than the later parts of the videos.}

\section{Cache Performance Simulations}\label{sec:simulation}

Our trace-driven cache simulations 
\revBB{take into account}{are designed to take into account and study} 
multiple sources of uncertainty that impact prefetching and caching performance.
First, network bandwidth varies over time and clients do not know their future bandwidth.
Second,
with different scenes requiring different encodings, for example,
chunk sizes typically vary from chunk-to-chunk and across different parts of the same video.
These first two uncertainties result
in variable download times and buffer sizes,
as clients adapt the requested encodings so as
to try to maintain a relatively stable buffer and to avoid stalls.
Third, as seen here,
with 360$\degree$ video
there is a lot of variability and uncertainty in how
users
move their heads.
The client player can try to predict head movements and prefetch high quality tiles
only for some directions, but prediction accuracy will vary across 
\revAA{videos}{videos, prediction algorithm used,} 
as well as over time during video playback.

\subsection{Simulation model}

To better understand the impact that download 
\revAA{speed variability (caused by the first two uncertainties)}{time variability} 
and the view direction prediction accuracy
have on the cache efficiency under different quality selection algorithms,
we use a simple simulation model where we use
probability distributions
to capture each of the uncertainties.

In our model, we assume that client player $i$ makes its tile selection for each chunk $k$
based on a quality of experience (QoE)
optimization problem taking into account
(i) the capacity $C_{i,k}$ drawn from a distribution $P_C(C)$, and
(ii) the probability $P_n(n)$ that a specific tile $n$ will be
viewed (where the probability $P_n(n)$ depends on the class of videos considered
  and how far in advance of playback the client must make its tile selection for the chunk).
\revAA{}{We next present the optimized tile-selection algorithm used for our (default) simulations in which we assume a horizontally sliced viewport.}

\revAA{}{{\bf Optimized tile selection for sliced viewport:}}
\revAA{In particular, each}{Each} 
client greedily maximizes
the objective function proposed by Almquist et al.~\cite{AAK+18}:
\begin{align}
  & (1-\beta)\sum_{n=0}^{N-1} P_n(n)u(q_n)
  - \beta \sum_{n=0}^{N-1}\frac{P_n(n)+P_n(n+1)}{2}|u(q_n)-u(q_{n+1})|, \nonumber
\end{align}  
where $\beta$ is a weight factor giving more (or less) weight to the importance of small quality differences
between neighboring tiles in the 360$\degree$ space
versus high expected viewing quality,
and $u(q_n)$ is a concave utility
achieved when viewing at
quality $q_n$.
At each step of the simulation, each client maximizes
  this objective function given the capacity
constraint
that $\sum_{n=0}^{N-1} r(q_n) \le C_{i,k}$,
where $r(q_n)$ is the size of tile $n$.
To find the optimal solution for a given $C_{i,k}$ and $P_n(n)$ we solve the above
optimization problem using dynamic programming~\cite{AAK+18}.

\revAA{}{{\bf Greedy tile selection for general viewport:}
For the case when we have a 2D grid of tiles, we implemented a greedy approach based on the slightly modified objective function:
\begin{align}
    & (1-\beta)\sum_{n=0}^{N-1} P_n(n)u(q_n)
     - \beta \sum_{n=0}^{N-1}\sum_{m=0}^{N-1} \delta_{m,n} \frac{P_n(n)+P_n(m)}{4}|u(q_n)-u(q_{m})|, \nonumber
\end{align}
where $\delta_{m,n} = 1$ if tiles $m$ and $n$ are
direct neighbors (share a side, accounting for wrap-around effects) in the grid, and 0 otherwise.  
For the greedy allocation we simply started with a zero-bandwidth allocation for each tile, and then greedily allocated more and more bandwidth as long as there was free bandwidth to allocate from the total bandwidth budget $C_{i,k}$ (for client $i$, chunk $k$) and there existed at least one more feasible allocation that can be made.  Here, in each step, we selected to increase bandwidth (one quality level) for the tile that maximized the relative increase in utility per consumed bandwidth unit and that still fits within the capacity constraint.  In particular, in each step we select to increase the quality level (and bandwidth allocation) to the tile that maximizes the ratio $\frac{\Delta_n}{r(q_n^{new})-r(q_n^{old})}$, where $\Delta_n$ is the objective function if applying the change (``new" allocation) minus the objective function with the ``old" allocation.}

\revAA{}{{\bf Impact of view prediction accuracy:}}
Finally, to account for the third uncertainty,
the predicted viewing direction used when solving the 
\revAA{optimization}{optimization (or finding a good greedy allocation)}
is offset from the actual viewing direction at playback time by an angle $\psi_{i,k}^{\epsilon}$ 
\revAA{}{(or $\psi_{i,k}^{\epsilon}$+$\theta_{i,k}^{\epsilon}$)}
chosen by sampling from a probability distribution 
\revAA{$P_{\psi}(\psi)$.}{$P_{\psi}(\psi)$ (or $P_{\psi,\theta}(\psi,\theta)$).}

\revAA{}{{\bf Independent simulation steps for each chunk:}}
\revAA{}{To remove 
dependencies on the specific HAS algorithm in use (as these are still evolving),
head movement prediction algorithm used (as the quality of these differ and some quickly are improving), and the long-term effects of a cache miss and other 
factors
that may impact HAS performance (depending on algorithms used) over a longer period, we simulate each chunk of each video individually.}
\revAA{}{We next describe how our simulations are done for the sliced viewport simulations.  The general case naturally extends from this as per the above uncertainties and alternative chunk-selection algorithm.}

To obtain a hit rate estimate for a particular video and number of previous clients, we average results from 1,000 simulations, each with 32 
\revBB{randomly-ordered}{randomly ordered} 
users sequentially viewing the video.
Each client uses the user head movements recorded in our trace dataset for that
user when viewing the respective video.
For each chunk prefetch request within each viewing, we
(i) draw a random capacity $C_{i,k}$ from the distribution $P_C(C)$,
(ii) draw a random offset $\psi_{i,k}^{\epsilon}$ from the distribution $P_{\psi}(\psi)$,
(iii) use the actual viewing direction $\psi_{i,k}$ that the
user has at the start of the playback of the chunk
and $\psi_{i,k}^{\epsilon}$ to determine the center tile used for the optimization,
(iv) solve the above optimization problem using the $C_{i,k}$ and the distribution $P_n(n)$
(rotated by $\psi_{i,k}+\psi_{i,k}^{\epsilon}$), and 
(v) download the
qualities of tiles for the chunk that are determined by the optimization.
To emulate the behavior of a cache, we keep track of prior client requests for tiles of the same chunk.
For our simulation we
assume that the system always starts with an empty cache and measure how the hit rate (both in terms of tile objects and bytes delivered) changes as more and more users view the same video.

\revAA{}{{\bf Limitations discussion:}}
We do not model the buffer states of individual clients,
  correlations in the chunk qualities that individual clients may request for back-to-back chunks,
  or the correlations between the bandwidths that clients may observe during download of consecutive chunks.
  While these aspects may help model the quality of experience and performance of individual clients,
  they are not needed to capture the performance of a network or server-side cache.
\revAA{}{In fact, by treating each chunk individually and drawing independent bandwidth samples each time,
  we avoid having to make additional simplifying system assumptions about the clients and, most importantly,
  avoid introducing persistent biases (e.g., client A may have higher bandwidth than client B) that persists
  throughout each sample session of our longitudinal analysis.
  Instead, in our simulations, each sequence of requests for a chunk can be seen as an independent experiment,
  effectively increasing the statistical significance of our results (across all chunks), given the same number of simulations.}

\subsection{Parameters and example distributions}

\revNot{\revtwo{}{{\bf Bandwidth variations:}}}{}
\revrev{To capture different bandwidth conditions and head movement patterns
we use a combination of traces and data sources.
First, to capture the distribution $P_C(C)$
we draw random samples from two real-world datasets and two synthetic datasets:}{For the distribution $P_C(C)$ we use
  \revfour{two distributions}{distributions}
  obtained by drawing random samples from two real-world datasets, and two synthetic distributions.
  The real-world datasets are:}
(i) 10,000 download bandwidth measurements collected by
mobile 3G and 4G users of a dominant national speed testing service~\cite{anon16}
over a 
\revBB{19 hour}{19-hour} 
window on Feb. 15-16, 2015,
\revrev{}{and}
(ii) 10,000 sample points from ``bus'' commuter traces
collected in Norway by Riiser et al.~\cite{RVGH13} between Aug. 28, 2010, and Jan. 31,
\revrev{2011, (iii) a synthetic three-level trace }{2011.  The synthetic distributions we use are: (i) a distribution}
in which the bandwidth capacity $C$
varies across three different levels such that $C$ is equal to the
average bandwidth 40\% of the time,
twice the 
\revAA{average bandwidth}{average} 
20\% of the time, and
half the 
\revAA{average bandwidth}{average} 
40\% of the time,
and (ii) a constant bandwidth capacity.
To account for the fact that bandwidths have increased substantially since the
\revrev{original traces}{traces in the real-world datasets}
were collected 
\revAA{(2010-2011 and 2015, respectively)}{(2015 and 2010-2011, respectively)} and to
ensure a 
\revAA{more fair head-to-head}{fairer} 
comparison across the different
\revrev{bandwidth traces/datasets,}{distributions,}
we
\revrev{multiply all sample bandwidth with a factor (dependent on the trace) so to ensure
that the average bandwidth is the same for all four adjusted traces/datasets.
Furthermore, we normalize all reported average bandwidths to the (sample) bandwidth}{scale the bandwidths in the real-world datasets and choose parameters for the synthetic distributions so that the average bandwidth in each case is the same.  
\revAA{Furthermore, we}{We} 
use normalized units so that a normalized bandwidth of 1 corresponds to 
\revAA{the bandwidth}{that}}
needed to deliver all tiles at the maximum quality.

{\bf Head movements, their prediction uncertainties, and optimized quality selection:}
\revAA{}{For simplicity, consider the sliced viewport model.  (The general viewport model extends naturally as per the differences described in the previous section.)}
\revrev{To estimate the uncertainties in head movements associated with different categories}{To determine
  choices for the $P_{\psi}(\psi)$ and $P_n(n)$ distributions,}
\revrev{we measured the yaw movements during different time intervals
  and video categories and then plotted the CDFs.}{we
  \revfive{considered}{used}
  the yaw
  \revrev{movements during}{angle changes in the traces from the head-movement dataset over}
  different time intervals
  \revrev{and}{and for different}
  video
  \revNot{categories.}{categories~\cite{CaEa20-arxiv}.}}
With the averages close to zero and the CDFs following s-shaped distributions~\cite{AAK+18},
we decided to approximate
\revrev{the distribution}{yaw angle change distributions}
using normal distributions
and used these (or variations)
\revrev{to simulate the}{for the}
uncertainty $P_{\psi}(\psi)$ in yaw prediction
and the uncertainty in head movements $P_{n}(n)$ used for the optimizations.
Table~\ref{tab:variations} reports the standard deviations observed for
each video category and 
\revAA{three different time intervals $T=2,5,10$.}{four different time intervals.}
Clearly, the best $P_{\psi}(\psi)$ distribution to use here
would depend on the prediction techniques being used and
there can be both better and worse predictors of future head movements
than simply using the current viewing direction (as implicitly assumed here).
For this reason, we apply a scaling factor $f_{\psi}$
on these measured standard deviations,
with a factor $f_{\psi}$ smaller (greater) than one capturing a more (less)
accurate prediction of the future viewing direction.
Similarly, we use a factor $f_{n}$ to scale the
\revrev{distribution}{standard deviation}
used for the
\revrev{quality selection optimization problem,}{quality selection optimization distribution $P_n(n)$,}
with a factor $f_{n}$ smaller (greater) than one capturing a more (less) concentrated distribution.
As our default values we use the 10 second values of each category with $f_{\psi} = f_{n} = 1$.

Finally,
for the 
\revAA{example}{default} 
simulation results presented here, we consider a sliced 360$\degree$ video
with each 2-second chunk split into 6
tiles, each covering 60 degrees, and for which
the tile encoding rates are each
proportional to one of seven quality
levels: 0 
\revAA{(modelling the case in which the tile is not fetched),}{(tile is not fetched),}
and, in normalized units, 144, 268, 625, 1124, 2217, and 4198 (corresponding to the quality levels in an example YouTube video).
  For the dynamic programming optimization,
  we use these integers as the corresponding tile sizes
  together with a default average bandwidth $C$ of 12,000,
  resulting in a normalized average bandwidth of 12000 / (4198 $\times$ 6) = 0.476
  (allowing benefits from quality-adaptive tile delivery similar to those
  in prior works~\cite{vengat179,vengat181,QHXG18}).
For
the utility function we extend the
large-screen model by Vleeschauwer et al.~\cite{VVB+13} to include
a ``black-out penalty'' associated with a missing tile:
\begin{equation}\label{eqn:utility}
  u(q) = \left\{ \begin{array}{ll}
    b \cdot \frac{(q / \theta)^{1-a}-1}{1-a}, & \textrm{if}~q = 144, 268, 625, 1124, 2217, 4198\\
    - u(4198), & \textrm{if}~q = 0\\
  \end{array}\right.\nonumber
\end{equation}
where $a$, $b$, and $\theta$ are
  parameters
  with values chosen
  as 2, 10, and 200, respectively, to match their large-screen model,
and the negative
utility
when $q=0$ captures the black-out
penalty.

\subsection{Example results}

In this section we focus on the hit rate as a function of the number
of prior users that have watched the same video.  Each result for a particular video and scenario
is an average from 1,000 simulations, each using a random ordering of the 32
user sessions for that video from our trace dataset.

\begin{figure}
\begin{minipage}[t]{0.4\textwidth}
  \centering
  \vspace{-60pt}
  \captionof{table}{Head movement variations, as measured by yaw angle change,
      over 
      \revAA{a $T$ second interval ($T$ = 2, 5, or 10 seconds).}{(0.5, 2, 5, and 10 seconds.}}
  \vspace{-10pt}
         {\footnotesize
           \begin{tabular}{|l|r|r|r|r|}
             \cline{2-5}
             \multicolumn{1}{c|}{} & \multicolumn{4}{c|}{Standard deviation} \\\hline
             Category  & \revAA{}{500 ms} & 2 sec & 5 sec & 10 sec \\\hline
             Explore   & \revAA{}{17.24} & 50.77$\degree$     & 79.85$\degree$     & 94.09$\degree$ \\\hline
             Static    & \revAA{}{14.23} & 35.94$\degree$     & 46.32$\degree$     & 46.93$\degree$ \\\hline
             Moving    & \revAA{}{14.69} & 35.77$\degree$     & 48.10$\degree$     & 57.42$\degree$ \\\hline
             Rides     & \revAA{}{15.05} & 38.94$\degree$     & 50.02$\degree$     & 52.44$\degree$ \\\hline
         \end{tabular}}
         \label{tab:variations}
         \vspace{-8pt}
\end{minipage}
\hfill
\begin{minipage}[t]{0.28\textwidth}
\centering
  \includegraphics[trim = 6mm 6mm 6mm 6mm, width=1\textwidth]{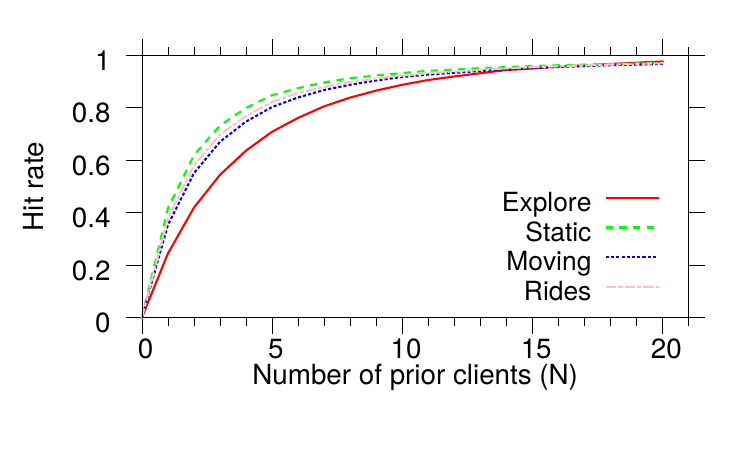}
  \vspace{-18pt}
  \caption{Object hit rate for trace-based simulations with fixed bandwidth.}
  \label{fig:staticBW-categories}
  \vspace{-10pt}
  \end{minipage}
  \hfill
  \begin{minipage}[t]{0.28\textwidth}
    \centering
  \includegraphics[trim = 6mm 6mm 6mm 0mm, width=1\textwidth]{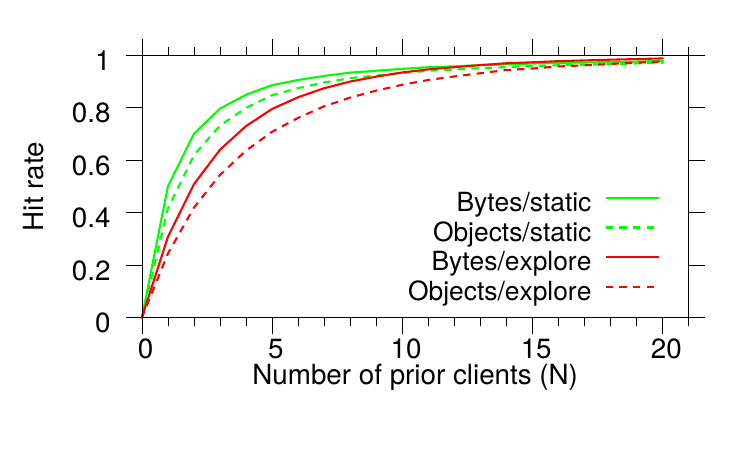}
  \vspace{-18pt}
  \caption{Object hit rate vs byte hit rate for trace-based simulations with fixed bandwidth.}
  \label{fig:staticBW-rates}
  \vspace{-10pt}
\end{minipage}
\vspace{-6pt}
\end{figure}

{\bf Results for representative videos:}
Figure~\ref{fig:staticBW-categories} shows a baseline
comparison of the tile object hit rates for the
  representative
  videos, using the 10 second values from Table~\ref{tab:variations} and $f_{\psi} = f_n = 1$.
This figure clearly
\revrev{shows that the hit rates quickly goes up for all video categories, but that}{illustrates that better cache performance is achieved with the}
{\em static}, {\em rides} and {\em moving}
\revrev{consistently outperform the {\em explore} category.}{videos compared to with the {\em explore} video.}
This observation is not surprising
given the results reported in previous sections, and is
\revrev{consistent also for other network traces and scaling
  factors of the uncertainties, but simply quantifies these advantages.}{also consistently seen with other distribution and parameter settings.}
For example, with four prior clients (i.e., $N=4$), the object hit
\revrev{rates for the first three categories are 0.75-0.80 and for {\em explore}}{rate for the {\em static}, {\em rides}, and {\em moving}
  videos ranges between 0.75-0.80, while for the {\em explore} video}
it is only 0.64.
\revAA{}{Higher hit rates can be directly translated into reduced loads on origin servers and the shorter round-trip-times to caches (than origin servers) can be translated into improved streaming performance for clients.}
\revAA{Note also that these differences can have a large impact on
  bandwidth requirements and cache write costs, for example, as}{However, these differences can also
  have a large impact on
  bandwidth requirements and cache write costs; both proportional to the miss rate.  For example, for $N$ = 4,}
  the object miss rate for {\em explore} is 80\% higher (a factor of 0.36/0.20) than for the {\em static} video.
In the reminder of this section we present results only for the two extreme cases of {\em static} and {\em explore} videos,
but note that the results for {\em rides} and {\em moving} are relatively similar to those of {\em static}.

\revtwo{}{{\bf Object vs byte hit rates:}}
\revfour{\revtwo{}{For our default settings,
  we have observed higher byte hit rates than object hit rates
  (Figure~\ref{fig:staticBW-rates}).  For example,
  with four prior clients,
  the byte hit rates for the four representative videos range from 0.73-0.85,
  substantially higher than the object hit rates.}}{Figure~\ref{fig:staticBW-rates} compares the byte and object
  hit rates for the {\em static} and {\em explore}
  \revNot{videos with our default parameter settings.}{videos.}
  The higher byte hit
  \revfive{rates (than object hit rates)}{rates}
  suggest even better cache benefits
  than suggested by the object hit rate results.
  The observed differences in byte hit rate (between classes) can have a large impact on bandwidth requirements.
  For example, with four prior clients, the byte hit rate for the {\em explore} video is 0.73
  while that for the {\em static} video is 0.85, implying an 80\% higher byte miss rate for the {\em explore} video.}

\begin{figure}[t]
\begin{minipage}[t]{0.64\textwidth}
\centering
  \centering
  \subfigure[Static]{
  \includegraphics[trim = 4mm 12mm 4mm 0mm, width=0.48\textwidth]{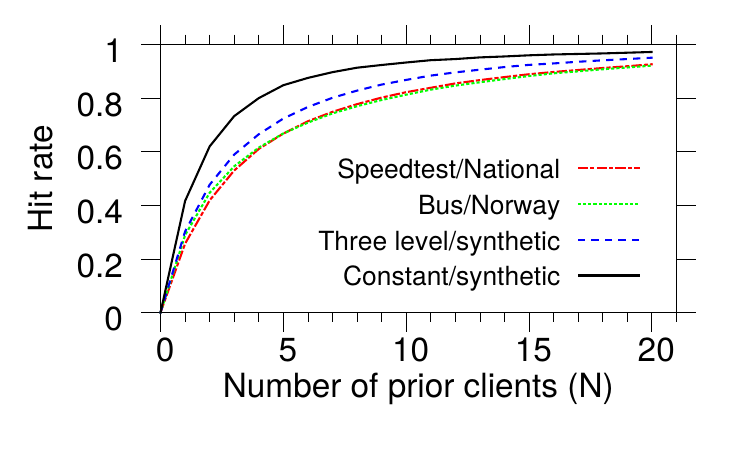}}
  \subfigure[Explore]{
  \includegraphics[trim = 4mm 12mm 4mm 0mm, width=0.48\textwidth]{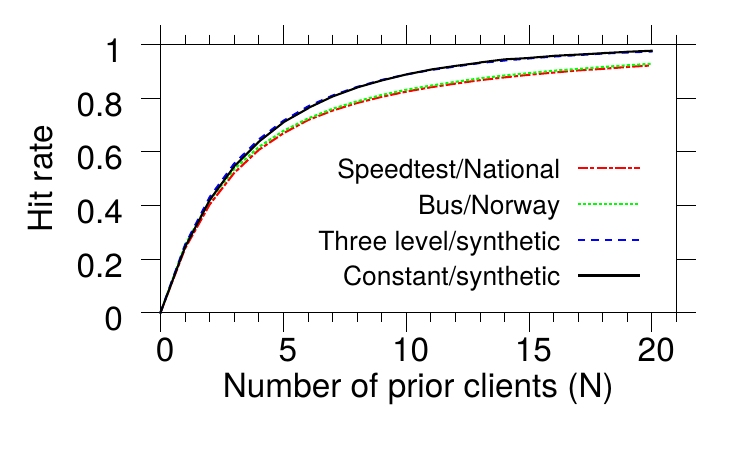}}
  \vspace{-16pt}
  \caption{Object hit rate for trace-based simulations using different network bandwidth profiles.}
  \label{fig:traces-constant}
  \vspace{-10pt}
  \end{minipage}
  \hfill
  \begin{minipage}[t]{0.32\textwidth}
    \centering
    \vspace{6pt}
      \includegraphics[trim = 4mm 6mm 4mm 0mm, width=0.98\textwidth]{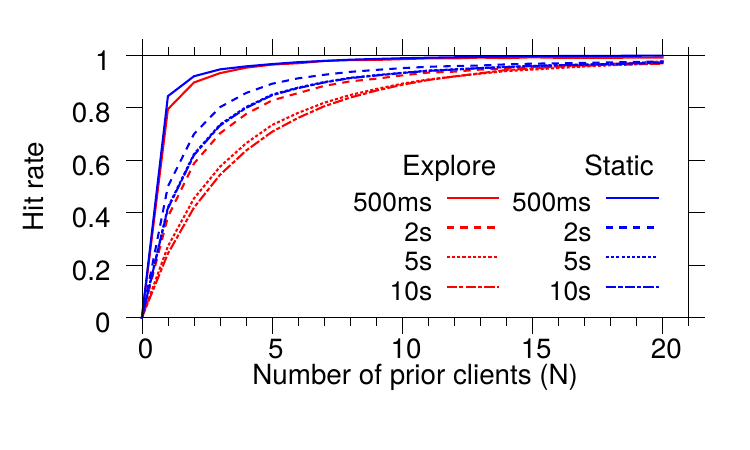}
  \vspace{-16pt}
  \caption{Impact of time threshold.}
  \label{fig:impact-of-T}
  \vspace{-10pt}
\end{minipage}
\vspace{-2pt}
\end{figure}

{\bf Impact of client's bandwidth variability:}
\revfour{\revtwo{\revrev{For our default settings we}{We}
have observed higher byte hit rates than object hit rates
(Figure~\ref{fig:staticBW-rates}) and that
\revthree{the hit}{hit}
rates typically reduce
\revrev{the more variability there are in the traces}{the greater the bandwidth variability}
(Figure~\ref{fig:traces-constant}).
For example,
\revrev{staying with the $N=4$ case, for which we reported object hit rates for the case
  when all clients have the same bandwidth,}{with four prior clients,}
the byte hit rates
\revrev{for the four video categories are 0.73, 0.85, 0.81, and 0.83,
and the object hit rates for this client ($N=4$)
when watching the {\em static} video under different
bandwidth variations are:}{for the four representative videos range from 0.73-0.85,
  substantially higher than the object hit rates.  With respect to the impact of bandwidth variability,
  note for example that for $N=4$ the object hit rates for
  the {\em static} video are}}{\revfour{We have observed that}{As seen in Figure~\ref{fig:traces-constant},}
  \revthree{the hit}{hit}
  rates typically reduce
  the greater the bandwidth
  \revfour{variability (Figure~\ref{fig:traces-constant}).}{variability.}
  For example,
    \revthree{we note that for $N=4$,}{for $N$=4,}
    the object hit rates for the {\em static} video are}
    0.60 with the national speed test service,
    0.61 with the Norwegian bus traces,
    0.66 with the three-level
    \revrev{synthetic traces,}{distribution,}
    and 0.80 when the bandwidth is constant.
    \revrev{Furthermore, comparing across bandwidth traces, we note that}{\revthree{We note}{Note}
      however that}{Note however that}}{As seen in Figure~\ref{fig:traces-constant}, hit rates typically reduce
  the greater the bandwidth variability. Note however that}
the relative impact of bandwidth variability
is smaller for the {\em explore}
\revrev{video (Figure~\ref{fig:traces-constant}(b))}{video}
than the {\em static}
\revrev{video (Figure~\ref{fig:traces-constant}(a)),}{video,}
showing that
\revrev{the higher}{higher}
uncertainty in viewing direction and bandwidth
do not contribute independently to
\revrev{the reduced}{reduced}
hit rates.

\revAA{}{{\bf Impact of time threshold T:}
Better head movement prediction is possible on shorter time scales, whereas better stall protection is achieved using larger buffers that account for chunk size variations, long round-trip-times (RTTs), or variations in the RTTs and the available bandwidths.  To glean 
\revBB{some insights}{insights} 
into the tradeoffs associated with how soon head-movement predictions are made, Figure~\ref{fig:impact-of-T} shows results for the {\em exploration} video (red) and {\em static} video (blue) for the head-movement uncertainties observed over four different time intervals: 500ms, 2s, 5s, and 10s.  As expected, the cache performance improves substantially 
as smaller buffer margins are used.  
However, 
it is important to note that most practical systems are likely to use a larger buffer to protect against unforeseen bandwidth variations and use larger chunks to allow more efficient encodings.}

\begin{figure}[t]
  \centering
    \subfigure[Categories comparison]{
  \includegraphics[trim = 4mm 6mm 4mm 6mm, width=0.32\textwidth]{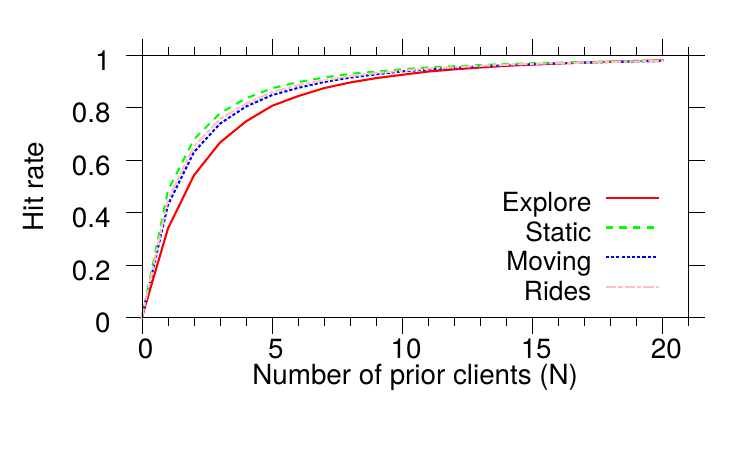}}
      \subfigure[Object vs byte hit rate]{
  \includegraphics[trim = 4mm 6mm 4mm 0mm, width=0.32\textwidth]{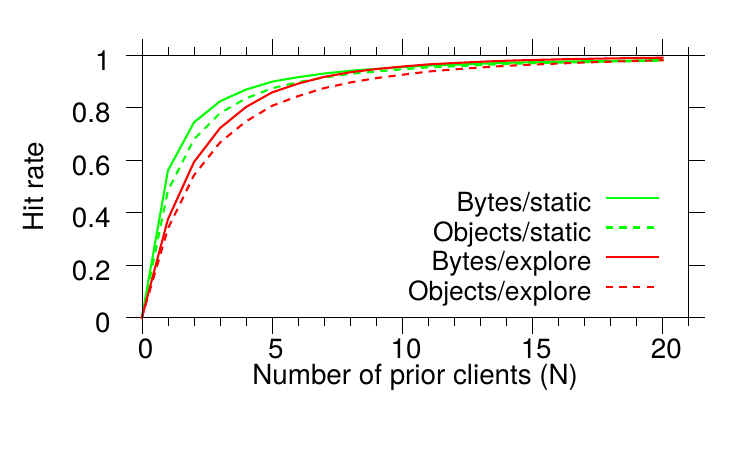}}
\subfigure[Bandwidth variability (Static)]{
  \includegraphics[trim = 4mm 12mm 4mm 0mm, width=0.32\textwidth]{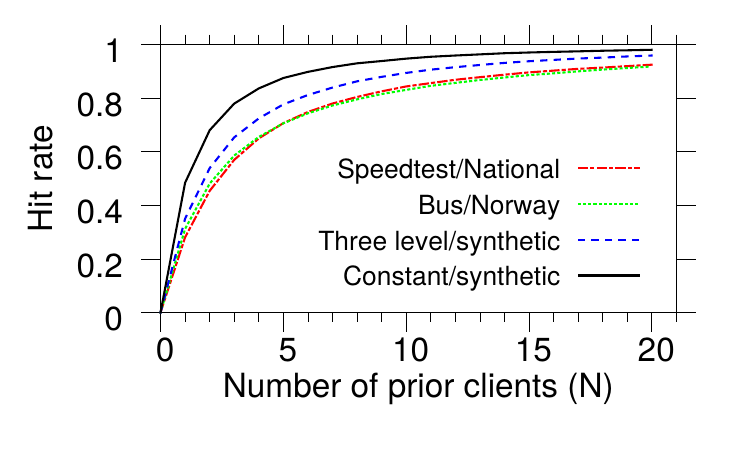}}
  \vspace{-16pt}
  \caption{\revAA{}{Simulation results using 4$\times$3 tiling.}}
  \label{fig:simulate-3x4-viewport}
  \vspace{-14pt}
\end{figure}

\revAA{}{{\bf Viewport and tiling scheme comparison:}
We have also run experiments with  4$\times$3 tiling. These results suggest similar tradeoffs between the different categories (Figure~\ref{fig:simulate-3x4-viewport}(a)), relationships between byte vs object hit rates (Figure~\ref{fig:simulate-3x4-viewport}(b)), and relative performance under different bandwidth variations (Figure~\ref{fig:simulate-3x4-viewport}(c)).}

\begin{figure}[t]
  \centering
  \subfigure[Error factor $f_{\psi}$ ~~~~~~~~~~~~.\newline Constant bandwidth]{
    \includegraphics[trim = 8mm 12mm 8mm 0mm, width=0.32\textwidth]{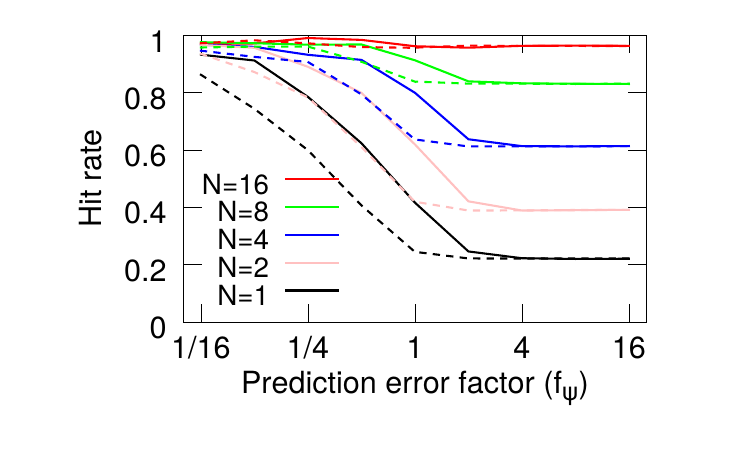}}
  \hspace{-4pt}
  \subfigure[Uncertainty factor $f_n$\newline Constant bandwidth]{
    \includegraphics[trim = 8mm 12mm 8mm 0mm, width=0.32\textwidth]{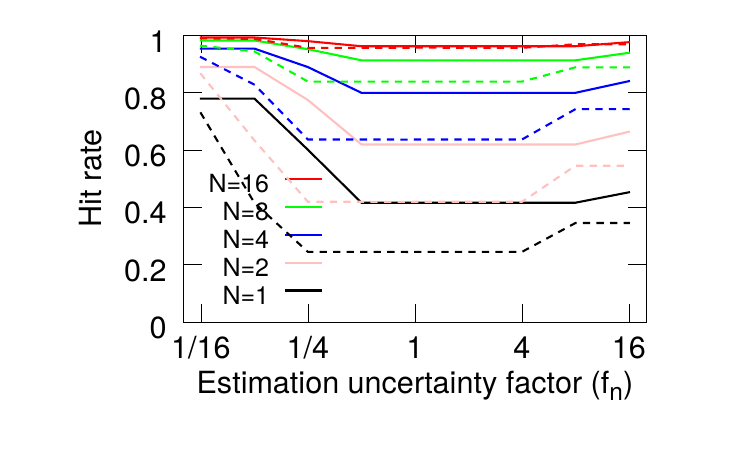}}
    \hspace{-4pt}
  \subfigure[Normalized average bandwidth\newline Constant bandwidth]{
    \includegraphics[trim = 6mm 5mm 6mm 0mm, width=0.31\textwidth]{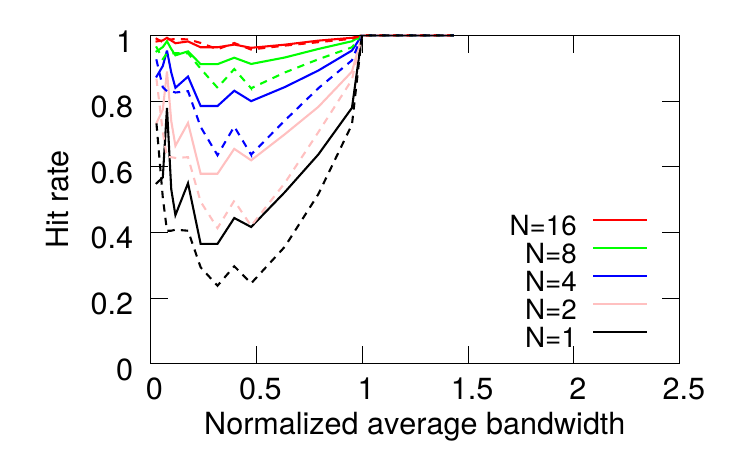}}\\
  \vspace{-10pt}
  \subfigure[Error factor $f_{\psi}$\newline National speedtest samples]{
    \includegraphics[trim = 8mm 12mm 8mm 0mm, width=0.32\textwidth]{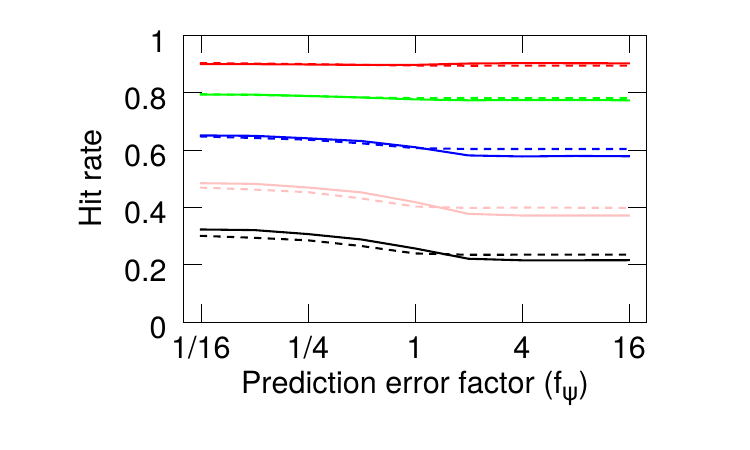}}
       \hspace{-4pt}
  \subfigure[Uncertainty factor $f_n$\newline National speedtest samples]{
    \includegraphics[trim = 8mm 12mm 8mm 0mm, width=0.32\textwidth]{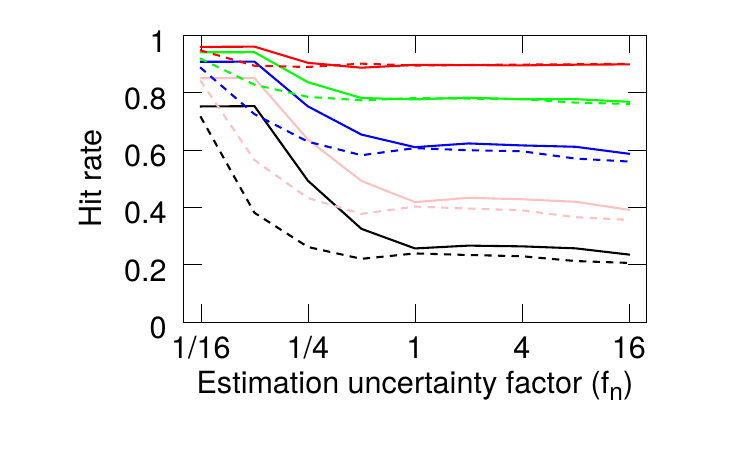}}
       \hspace{-4pt}
  \subfigure[Normalized average bandwidth\newline National speedtest samples]{
    \includegraphics[trim = 6mm 5mm 6mm 0mm, width=0.31\textwidth]{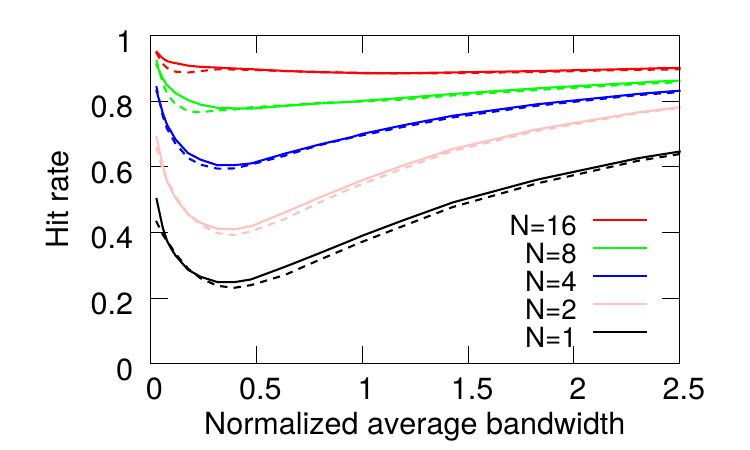}}
  \vspace{-14pt}
  \caption{\revAA{}{Impact of the prediction error factor $f_{\psi}$, the estimated uncertainty factor $f_n$, and the normalized average bandwidth.  Top row shows case when all clients have the fixed same bandwidth and the bottom row shows results using the national speed test samples.  All figures include results for both {\em static} videos (solid lines) and {\em explore} videos (dotted lines).}}
  \label{fig:simulations-44-and-45-all}
  \vspace{-10pt}
\end{figure}

\revAA{{\bf Impact of average bandwidth:}
In Figure~\ref{fig:simulations-44D} we show additional object hit rate results for the two
extreme cases of (a) constant bandwidth (identical for all clients) and
(b) the bandwidth distribution obtained by drawing random samples from measurements collected
by mobile 3G and 4G users of a dominant national speed testing service~\cite{anon16}.
When interpreting these results, it is important to note that clients sharing an edge-cache
(e.g., operated by a CDN or in cooperation with a CDN)
might be expected to experience more similar bandwidth conditions than in the speed testing data.  Also, with
the introduction of cap-based solutions~\cite{KrCH18},
and other streaming-aware network solutions,
used by different operators to stabilize HAS performance, improve QoE, and to reduce unnecessary bandwidth usage,
it seems likely that many networks in the future will provide fairly stable conditions for their streaming clients.
Therefore, we believe that likely bandwidth variations fall between these two extremes.}{{\bf Bandwidth variation baselines:} 
Figure~\ref{fig:simulations-44-and-45-all} shows
additional object hit rate results for the two
extreme cases of 
constant bandwidth (identical for all clients), and
the bandwidth distribution obtained by drawing random samples from measurements collected
by mobile 3G and 4G users of a dominant national speed testing service~\cite{anon16}.
When interpreting these results, it is important to note that clients sharing an edge-cache
(e.g., operated by a CDN or in cooperation with a CDN)
might be expected to experience more similar bandwidth conditions than in the speed testing data.  
Also, with the introduction of cap-based solutions~\cite{KrCH18},
and other streaming-aware network solutions,
used by different operators to stabilize HAS performance, improve QoE, and to reduce unnecessary bandwidth usage,
it seems likely that many networks in the future will provide fairly stable conditions for their streaming clients.
Therefore, we believe that likely bandwidth variations fall between these two extremes.}

\revtwo{}{{\bf Impact of head movement uncertainties and prediction accuracy:}}
\revAA{Figures~\ref{fig:simulations-44B} and~\ref{fig:simulations-44C}}{Figures~\ref{fig:simulations-44-and-45-all}(a), \ref{fig:simulations-44-and-45-all}(b), \ref{fig:simulations-44-and-45-all}(d), and~\ref{fig:simulations-44-and-45-all}(e)}
show results capturing the impact 
\revAA{of the}{of} 
prediction accuracy
(varying $f_{\psi}$)
and the concentration of the $P_n(n)$ distribution used 
\revAA{for the}{for} 
utility optimization
(varying $f_{n}$).
Note that the differences between the two video categories
are largest
for the constant bandwidth case 
\revAA{(or, more generally, are larger when the bandwidth variations are smaller).}{(or, more generally, larger when bandwidth variations are smaller).}
Although the differences 
\revAA{also appear}{appear} 
larger for smaller $N$,
it is also necessary to consider 
\revAA{the corresponding miss}{miss}
rates, as these determine
\revAA{the bandwidth}{bandwidth} 
costs.
For example, for $N=4$ and 
\revAA{}{the default}
constant bandwidth
\revAA{(Figures~\ref{fig:simulations-44B}(a) and~\ref{fig:simulations-44C}(a))}{(Figures~\ref{fig:simulations-44-and-45-all}(a) and~\ref{fig:simulations-44-and-45-all}(b))}
there are substantial regions where 
\revAA{the miss}{miss} 
rates (and hence bandwidth usage)
are almost twice as large for the {\em explore}
\revrev{video (dotted lines)}{video}
as for the {\em static}
\revrev{video (solid line).}{video.}

\revAA{Figures~\ref{fig:simulations-44B} and~\ref{fig:simulations-44C}}{Figures~\ref{fig:simulations-44-and-45-all}(a) and (d), and~\ref{fig:simulations-44-and-45-all}(b) and (e),}
clearly show that the
\revrev{initial hit rates}{hit rates for small $N$}
are much lower when the
prediction accuracy is poor and
\revrev{need to be taken into account in the optimizations,}{the estimated uncertainty is large,}
respectively,
but that the hit rates go up
\revrev{relatively quickly}{substantially as $N$ increases}
also for these cases.
These gains are especially visible when all clients have the same
\revrev{bandwidth (Figures~\ref{fig:simulations-44B}(a) and~\ref{fig:simulations-44C}(a)),}{bandwidth,}
illustrating that caching
\revfour{of HAS video (and in this case tiled 360$\degree$ video) is}{is}
most efficient when clients have similar
\revfour{bandwidth conditions}{bandwidth}
and request chunks (or tiles) of similar
\revfour{ quality. For example,
\revrev{}{when $f_{\psi}=2$,}
while the first {\em static} client seeing a non-empty cache (i.e., $N=1$)
\revrev{with the {\em static} video has}{has}
a hit rate below
\revrev{0.25 when $f_{\psi} \ge 2$,}{0.25,}
the hit rate quickly goes up to 0.61
\revrev{when $N=4$ for the same estimate error ($f_{\psi} = 2$),
  which itself is roughly equal the hit rate (0.62) seen by the first client when $f_{\psi}=0.5$.}{when $N$ increases to 4,
  \revrev{which is similar to the hit rate for}{a similar hit rate as for}
  $N=1$ and $f_{\psi}=0.5$.}}{quality.}

{\bf Impact of average bandwidth:}
\revAA{Figure~\ref{fig:simulations-44D} shows}{The results in Figures~\ref{fig:simulations-44-and-45-all}(c) and~\ref{fig:simulations-44-and-45-all}(f) show}
that our default case of a normalized bandwidth of 0.476 results in close to the worst-case hit rates, suggesting that the hit rates with tiled 360$\degree$ video could be
greater than suggested by
\revAA{Figures~\ref{fig:staticBW-categories}-\ref{fig:simulations-44C}.}{prior figures in this section.}
Also, when comparing 
\revAA{Figures~\ref{fig:simulations-44D}(a) and~\ref{fig:simulations-44D}(b)}{Figures~\ref{fig:simulations-44-and-45-all}(c) and~\ref{fig:simulations-44-and-45-all}(f)}
it should be noted that
owing to our choice of normalized units for bandwidth,
the hit rate is always one when all clients have the same (constant) bandwidth above one \revAA{(Figure~\ref{fig:simulations-44D}(a)),}{(Figure~\ref{fig:simulations-44-and-45-all}(c))}
whereas bandwidth variations in the national speedtest dataset 
\revAA{}{(Figure~\ref{fig:simulations-44-and-45-all}(f))}
result in significant periods of bandwidth below
  one even for average values substantially larger than one.
Again, in practice,
we expect clients
sharing the same cache
to see bandwidth variation 
between 
\revAA{the two extremes considered here,}{these two extremes,}
with operators likely to strive 
\revAA{towards providing}{towards} 
increasingly stable network conditions for streaming clients~\cite{KrCH18}.

\section{Related Work}\label{sec:related}

Broadly, the related work can be split into works that consider the head movements during
viewing of 360$\degree$
\revNot{videos, client-side techniques to provide the best possible QoE
  (e.g., through adaptive prefetching based on expected viewing directions),}{videos,}
  and caching of
HAS videos.
\revfour{To the best of our knowledge we provide the first study that considers the
  intersection of these three aspects.}{While some recent works have considered optimized
  cache management policies for 360$\degree$ videos~\cite{LLL+19, MNSP18,PaKo19}, none of these
  works provide a data-driven characterization of the caching opportunities that would
  be observed with traditional caching policies that simply cache the requested tiles when
  the clients apply adaptive prefetching techniques.}

{\bf Head-movement characterization:}
\revfour{A number of recent}{Some recent}
works have collected datasets
and characterized
the 360$\degree$ viewer behavior~\cite{CoSS17,FSMR18,DGC+18,LFL+17,vengat179,AAK+18,QHXG18}.
However, most of these datasets use relatively short video segments and do not capture
changes in behavior over time or across classes of videos.
The primary exception, and the work most closely related to ours,
is the work by Almquist et al.~\cite{AAK+18}, as we use their dataset.
In their work, they present a category-based characterization of the head movements over time,
and analyze how changes in viewing behavior
\revrev{depends}{depend}
on the time window considered,
but do not consider overlapping viewports of users watching the same video
or other similarity metrics of users' viewing directions.
\revAA{}{Coverage and overlap metrics, such as those introduced here, are valuable in identifying potential caching and bandwidth saving opportunities.}

{\bf Client-side techniques:}  Motivated by HMDs allowing different projection
and quality adaptive download techniques~\cite{ZHLL17},
\revAA{recent works have proposed various techniques}{various techniques have been proposed} 
to adaptively download different qualities for
different viewing directions~\cite{vengat181,vengat179, SoJR18, SoJR18b,ZhXL18,QHXG18}.
\revAA{These techniques}{These} 
typically combine user head movement tracking/prediction~\cite{vengat178,vengat179, XiZG18,QHXG18}
and bandwidth management~\cite{vengat181,vengat182}.
For example, Bao et al.~\cite{vengat179} 
\revAA{propose a motion-prediction-based transmission scheme based on viewing behavior data and show}{show} 
that view-dependent 360$\degree$ transmission schemes with motion prediction
can reduce bandwidth consumption by 45\% at the cost of only 
\revAA{very}{a} 
small performance degradation.
Similarly, Hosseini and Swaminathan~\cite{vengat181} present an adaptive tile-based streaming solution and show that
large bandwidth savings (72\%) can be achieved with only small quality degradation.
Graf et al.~\cite{GrTM17} studied the impact of projection techniques,
quantization parameters, and tile patterns on the playback experience and resource requirements.
Others have considered tradeoffs that try to address variations and uncertainties
\revAA{in both}{in} 
the user's bandwidth and viewing direction 
\revBB{simultaneously~\cite{AAK+18,SDL+18,QHXG18}.}{simultaneously~\cite{AAK+18,SDL+18,QHXG18,yuan2019spatial}.}
For example, Sun et al.~\cite{SDL+18} use simulations and experiments that capture the bandwidth variations,
Qian et al.~\cite{QHXG18} have implemented and tested a tile-based solution on a real network,
whereas Almquist et al.~\cite{AAK+18} note that HAS clients typically try to maintain a steady buffer (to protect against stalls)
and consider the optimized prefetch-aggressiveness tradeoff of such clients.
\revBB{}{Similarly, Yuan et al.~\cite{yuan2019spatial} present a buffer-based approach that tries to balance between the buffer length and video quality.}
In this paper, we leverage the optimization framework 
\revAA{provided by}{by}
Almquist et al.~\cite{AAK+18} to evaluate the impact prefetching optimizations
have on the caching performance of tiled 360$\degree$ video.
\revBB{}{Others have shown (in the context of regular HAS video) how different rate adaptive solutions can be combined to provide improved client-side performance~\cite{yuan2019ensemble}.}

Tile-based
\revfour{spatial \revthree{segmentation of a video}{video segmentation}}{segmentation}
has been used in
\revfour{several other}{other}
applications,
including to support pan/tilt/zoom interactions during live streaming of 
\revBB{high resolution}{high-resolution} 
videos~\cite{ref12},
\revthree{}{for interactive panoramic video~\cite{GRE+16},}
\revrev{interactable}{interactive}
4k video~\cite{ref13}, and to
allow users
\revfour{to navigate freely through}{free navigation in}
high resolution video feeds while minimizing
\revthree{the bandwidth}{bandwidth}
\revfour{usage~\cite{ref14}.
Other prior works have demonstrated interactive tiled streaming of ultra-high
resolution videos based on a user's region-of-interest~\cite{ref9,ref10}
and crowd-driven prefetch prediction~\cite{ref11}.}{usage~\cite{ref14}.}

\cutICPE{
An alternative to tile-based streaming is to create different versions for each potential viewing direction
and let the viewer adapt the version downloaded for each chunk.
For example, Kuzyakov and Pio~\cite{ref18}
create different smaller-sized versions in which each version has a specific area in high quality
and
\revrev{then gradually decreasing the quality}{with gradually decreasing quality}
away from this area.
\revfour{While some of the observations in this paper
\revthree{also are}{are}
applicable to this context,
we do not explicitly consider such techniques here.}{Some of the observations in this paper may be
  applicable to this other context also.}
}

\cutICPE{
\revthree{}{{\bf Bandwidth-interactivity tradeoffs:}
The tradeoffs between bandwidth constraints,
playback quality, and interactivity have also been considered in other contexts.
For example, Ma et al.~\cite{MaMF18} consider these tradeoffs in the context of interactive multiview streaming.
HAS also has been leveraged for bandwidth-aware support of other interactive services,
including interactive multiview streaming~\cite{ToFr17},
optimized stream bundles~\cite{CEKP17},
and to enhance parts of regular (linear) videos that the users show more interest in~\cite{GZH+18}.}
}

{\bf Caching \revrev{of}{for} HAS:}
Prior works have characterized the caching opportunities for HAS content in mobile networks~\cite{GHKL13},
evaluated the impact that cross traffic has on cache performance~\cite{BeER11}, identified HAS specific
instabilities and other tradeoffs associated with the use of caches combined with HAS~\cite{ KCE+13,LeDB14},
and proposed HAS-aware solutions to improve the client performance in such scenarios~\cite{KCE+13, LeDB14,TDS+17, TDS+16, MJMM18, LHZY15}.
Other works have considered various cache management problems in the context of
\revthree{adaptive streaming~\cite{ZWCK13}.}{HAS~\cite{ZWCK13,LTZ+18} and
optimized replication for interactive multiview streaming~\cite{RCCF14,ToCF16}.}

\revfour{However, none of the above caching papers consider tiled 360$\degree$ video.  In this paper, we present
the first characterization and analysis of
similarities in head movements between users watching the same video,
the users' viewport overlaps,
and their
implications on caching of tiled 360$\degree$ videos belonging to different categories.}{Most closely related
  to our work are perhaps recent works that present optimized cache management solutions for 360$\degree$ video~\cite{LLL+19, MNSP18,PaKo19}.
  These works formulate optimization problems related to the caching of tiled 360$\degree$ 
  videos~\cite{LLL+19, PaKo19} or try to learn probabilistic models of users FoV
  for each video so to improve cache performance~\cite{ MNSP18}.  However, none of the papers presents
  a data-driven characterization of the users' viewport overlaps and the
  bandwidth saving opportunities this provides basic caching policies.
  Here, we present the first such data-driven analysis of similarities in head movements between
  users watching the same video,
  the users' viewport overlaps,
  and their implications on caching of tiled 360$\degree$ videos belonging to different categories.}


\revBB{}{\section{Summary of Design Insights}\label{sec:insights}}

\revBB{}{{\bf Design recommendations:}
Our trace-based characterization and simulations provide {\em insights} that can be used in the design of more effective caching and preloading policies.  We next summarize some of these insights.
\begin{itemize}
\item {\bf Selective insertion policies:}
Our results highlight that video category has a major impact on probability of data reuse.  This observation suggests that selective edge-cache insertion policies~\cite{MaSi15,GaVa16,CaEa17,CaEa21} should take into account the category of video. We have also found that 
the cache hit rate often improves
over the duration of a video session, suggesting that such policies also may benefit from taking into account the timestamp of each chunk.
\item {\bf Chunk durations and buffer sizes:}  Other factors that improve cache performance are smaller chunk size and reduced uncertainties in head movements (e.g., due to improved client-side prediction).  While CDNs may not control these parameters (we expect most practical systems to use larger buffers to protect against unforeseen bandwidth variations and use larger chunks to allow more efficient encodings), we note that also these choices may impact the aggressiveness with which insertion policies may select to cache tiles.
\item {\bf Time-based prefetching of {\em static} videos:}  The initial exploration phase of {\em static} videos may need special consideration both with regards to insertion policies (due to hit probabilities mentioned above) and preloading of the start of a video.  {\em Moving} and {\em rides} videos provide the best opportunities to save bandwidth during preloading of the initial parts of a 360$\degree$ video.
\end{itemize}}

\revBB{More generally, our results can inform the design of new caching system policies tailored for 360$\degree$ video, and may also have implications for other contexts than caching. For example, our novel category-based characterization clearly highlights that there are substantial differences among the video categories
in the value of using the viewing directions of previous users for viewport prediction.
The results also clearly show that cache performance,
  and hence also likely user QoE, benefit from stable
  network conditions, motivating the use of cap-based
  network/server-side solutions or less greedy client-side
  solutions.}{{\bf Other system insights:} In addition to informing the design of new caching system policies, our findings may also have implications for other contexts than caching.
  \begin{itemize}
  \item {\bf Category-dependent predictors:} Our novel category-based characterization highlights substantial differences among the video categories in the value of using the viewing directions of previous users for viewport prediction.  This suggests that substantially different head-movement predictors may be needed for different video categories.
\item {\bf Tiling may have benefit also on a chunk level:}
Our observation that many chunks have a relatively small cover size
shows that a significant portion of the potential viewing area is not viewed during the playback of a chunk and suggests that tiles could indeed fruitfully be prioritized (by a client) on a per-chunk basis.
\item {\bf Caps-based network-side solutions and less aggressive client-side solutions may have additional advantages for 360$\degree$ video:}  Our results show that cache performance, and hence also likely user QoE, benefit from stable network conditions. This is something that can be provided through the use of cap-based network/server-side solutions or less greedy client-side solutions.
  \end{itemize}}

\section{Conclusions}\label{sec:conclusions}

This paper
presents novel trace-based analysis methods and
uses head-movement traces for
  different categories of 360$\degree$ videos, including {\em explore}, {\em static}, {\em moving}, {\em rides},
to characterize similarities in
the viewports of
users
watching the same video,
to study how the viewport overlaps and other related metrics
differ between the different
video categories, and to analyze and discuss how such similarities and differences impact the effectiveness
of caching tiled 360$\degree$ videos.  To the best of our knowledge, this is the first paper to provide such analysis.

Our results consistently highlight substantial differences between different video categories in the pairwise viewport overlaps observed
and their impact on the potential bandwidth savings from caching.
For example, 
\revBB{with the exception of}{except for} 
the initial 20-30 second exploration phase of {\em static} videos, the {\em static} videos provide the greatest caching opportunities.  However, during this initial phase, their pairwise viewport overlaps are almost as small as for the {\em explore} videos, which have the smallest overlaps among the categories considered here.  In contrast, {\em moving} and {\em rides} videos have a less pronounced exploration phase, 
\revBB{and overall}{and} 
often provide similar caching opportunities and performance as the {\em static} videos.
Our results also show that improved viewport prediction techniques~\cite{XiZG18} may not only help improve user QoE, through the use of more accurate prefetching, but may also help increase cache hit rates and reduce bandwidth requirements.

\revBB{}{ Based on our findings, we present design recommendations concerning caching and preloading policies.  We also describe implications of our results for other system design aspects.}

Finally, we note that the methodologies and metrics
  defined in this paper easily can be applied on other datasets,
  enabling others to compare with our results using alternative
  head-movement traces (e.g., different user groups, video categories, etc.), prediction algorithms, and network conditions.

{\bf Acknowledgements:}
This work was funded in part
by the Swedish Research Council (VR).

{
  \bibliographystyle{ACM-Reference-Format-Emir}
  \bibliography{caching360}
}

\end{document}